\RequirePackage[T1]{fontenc}
\documentclass[12pt]{article}

\usepackage[height=8.85in,width=6.45in]{geometry}

\usepackage[utf8]{inputenc}
\usepackage{amsmath}
\usepackage{amssymb}
\usepackage{mathtools}
\numberwithin{equation}{section}
\usepackage{slashed}
\usepackage{braket}
\usepackage[svgnames]{xcolor}
\usepackage[colorlinks,citecolor=DarkGreen,linkcolor=FireBrick]{hyperref}
\usepackage{cite}
\usepackage{graphicx}
\usepackage{tikz}
\usepackage{tikz-cd}
\usepackage{times}
\usepackage{courier}
\usepackage{bm}
\usepackage{subfig}

\usepackage[normalem]{ulem}

\usepackage{soul}

\usetikzlibrary{decorations.pathreplacing,decorations.markings}

\tikzset{
    partial ellipse/.style args={#1:#2:#3}{
        insert path={+ (#1:#3) arc (#1:#2:#3)}
    }
}

\tikzset{
  on each segment/.style={
    decorate,
    decoration={
      show path construction,
      moveto code={},
      lineto code={
        \path [#1]
        (\tikzinputsegmentfirst) -- (\tikzinputsegmentlast);
      },
      curveto code={
        \path [#1] (\tikzinputsegmentfirst)
        .. controls
        (\tikzinputsegmentsupporta) and (\tikzinputsegmentsupportb)
        ..
        (\tikzinputsegmentlast);
      },
      closepath code={
        \path [#1]
        (\tikzinputsegmentfirst) -- (\tikzinputsegmentlast);
      },
    },
  },
  mid arrow/.style={postaction={decorate,decoration={
        markings,
        mark=at position .5 with {\arrow[#1]{stealth}}
      }}},
}

\usetikzlibrary{shapes.multipart}
\usetikzlibrary{decorations.pathmorphing}
\tikzset{snake it/.style={decorate, decoration=snake}}

\usepackage{xcolor}
\usepackage{mdframed}

\renewenvironment{figure}[1][]{
  \begin{originalfigure}[#1]
    \begin{mdframed}[linecolor=black!0,backgroundcolor=black!1]
}{
    \end{mdframed}
  \end{originalfigure}
}

\def\cC{\mathcal{C}}

\definecolor{dgreen}{rgb}{0, 0.55, 0}

\def\CA{{\mathcal A}}
\def\CB{{\mathcal B}}
\def\CC{{\mathcal C}}

\def\CN{{\mathcal N}}
\def\CO{{\mathcal O}}
\def\CP{{\mathcal P}}

\def\CX{{\mathcal X}}

\def\CZ{{\mathcal Z}}
\newcommand{\Z}{\mathbb{Z}}
\newcommand{\ZZ}{\mathbb{Z}}
\newcommand{\Tr}{\mathrm{Tr}}

\usepackage{datetime}
\usepackage{dsfont}

\usetikzlibrary{shapes.geometric}

\def\cX{{\mathcal X}}
\def\tauYM{\tau_{\mathrm{YM}}}
\def\tauMax{\tau_{\mathrm{Max}}}
\def\cN{{\mathcal N}}
\def\cC{{\mathcal C}}
\def\cP{{\mathcal P}}

\def\cA{{\mathcal A}}

\def\no{\nonumber}

\newcommand{\be}{\begin{equation}}
\newcommand{\ee}{\end{equation}}
\newcommand{\bea}{\begin{eqnarray}}
\newcommand{\eea}{\end{eqnarray}}

\begin{document}

\begin{titlepage}

\begin{flushright}
~
\end{flushright}

\vskip 3cm

\begin{center}

{\Large \bfseries  Symmetry TFTs  and  Anomalies of Non-Invertible Symmetries}

\vskip 1cm

 Justin Kaidi$^{1}$, Emily Nardoni$^2$, Gabi Zafrir$^3$, and Yunqin Zheng$^{2,4}$
\vskip 1cm

\begin{tabular}{ll}
$^1$&Department of Physics, \\& University of Washington, Seattle, WA, 98195, USA
 \\
 $^2$&Kavli Institute for the Physics and Mathematics of the Universe, \\
 & University of Tokyo,  Kashiwa, Chiba 277-8583, Japan\\
 $^3$&Simons Center for Geometry and Physics, \\& Stony Brook University, Stony Brook, NY 11794-3636, USA\\
 $^4$&Institute for Solid State Physics, \\
  &University of Tokyo,  Kashiwa, Chiba 277-8581, Japan\\
 
\end{tabular}

\vskip 1cm

\end{center}

\noindent
It is known that the 't Hooft anomalies of invertible global symmetries can be characterized by an invertible TQFT in one higher dimension. 
The analogous statement remains to be understood for non-invertible symmetries. In this note we discuss how the linking invariants in a non-invertible TQFT known as the Symmetry TFT (SymTFT) can be used as a diagnostic for 't Hooft anomalies of non-invertible symmetries. When  the non-invertible symmetry is non-intrinsically non-invertible, and hence the SymTFT is a Dijkgraaf-Witten model, the linking invariants can be computed explicitly. We illustrate this proposal through the examples of the abelian Higgs model in 2d, as well as adjoint QCD and $\cN=4$ super Yang-Mills in 4d. We also comment on how the 't Hooft anomalies of non-invertible symmetries impose new constraints on the dynamics.

\end{titlepage}

\setcounter{tocdepth}{2}
\tableofcontents

\section{Introduction and summary}
\label{sec:intro}

Non-invertible symmetries have a long history of study in two dimensions \cite{verlinde1988fusion,Petkova:2000ip,Fuchs:2002cm,Bhardwaj:2017xup,Chang:2018iay,Lin:2022dhv,Komargodski:2020mxz,Tachikawa:2017gyf, Frohlich:2004ef, Frohlich:2006ch, Frohlich:2009gb, Carqueville:2012dk, Brunner:2013xna, Huang:2021zvu, Thorngren:2019iar, Thorngren:2021yso, Lootens:2021tet, Huang:2021nvb, Inamura:2022lun, Ji:2019jhk,Kong:2020cie,Ji:2021esj,Chatterjee:2022kxb, Chatterjee:2022tyg, Moradi:2022lqp}, but it was not until fairly recently that they were realized in higher-dimensional theories \cite{Kaidi:2021xfk,Choi:2021kmx, Koide:2021zxj,Choi:2022zal,Apruzzi:2021nmk,Arias-Tamargo:2022nlf,Hayashi:2022fkw,Roumpedakis:2022aik,Bhardwaj:2022yxj,Kaidi:2022uux,Choi:2022jqy,Cordova:2022ieu,Antinucci:2022eat,Bashmakov:2022jtl,Damia:2022rxw,Damia:2022bcd,Choi:2022rfe,Lu:2022ver,Bhardwaj:2022lsg,Bartsch:2022mpm,Lin:2022xod,Apruzzi:2022rei,GarciaEtxebarria:2022vzq, Benini:2022hzx, Wang:2021vki, Chen:2021xuc, DelZotto:2022ras,Bhardwaj:2022dyt,Brennan:2022tyl,Delmastro:2022pfo, Heckman:2022muc,Freed:2022qnc,Freed:2022iao,Niro:2022ctq,Kaidi:2022cpf,Mekareeya:2022spm,vanBeest:2022fss,Antinucci:2022vyk,Chen:2022cyw,Bashmakov:2022uek,Karasik:2022kkq,Cordova:2022fhg,Decoppet:2022dnz,GarciaEtxebarria:2022jky,Choi:2022fgx,Yokokura:2022alv,Bhardwaj:2022kot,Bhardwaj:2022maz,Bartsch:2022ytj,Hsin:2022heo,Heckman:2022xgu,Antinucci:2022cdi,Apte:2022xtu,Garcia-Valdecasas:2023mis,Delcamp:2023kew,Bhardwaj:2023zix}. One of the main goals in the study of non-invertible symmetries is to obtain a detailed understanding of their dynamical consequences. For standard, invertible symmetries, it is well known that `t Hooft anomalies provide important constraints on the low-energy physics of the theory, and one may optimistically hope that the notion of a `t Hooft anomaly can be extended to non-invertible symmetries as well. 
The purpose of this note is to discuss certain easy-to-compute quantities which can probe the existence of 't Hooft anomalies of non-invertible symmetries, and to illustrate them using examples in two and four dimensions.

\subsection{Anomalies of invertible symmetries}
\label{sec:intro1}

Let us begin by briefly reviewing the familiar case of 't Hooft anomalies for invertible symmetries; see e.g. \cite{Cordova:2019jnf, Cordova:2019bsd} for an overview.  Given a $d$-dimensional quantum field theory with an invertible global symmetry $G$ on a manifold $X_d$, we define the 't Hooft anomaly to be the obstruction to gauging $G$. 
To see how this obstruction arises, we couple the theory to a  background $G$ gauge field $A$. Under background field transformations, the partition function transforms as
\begin{equation}\label{eq:Def1Def2eq}
    Z[X_d, A] \to Z[X_d,A] \,e^{2\pi i \int_{X_{d}} f(A, g)}~,
\end{equation}
where $g$ is the gauge transformation parameter, and $f$ is some local functional of $g$ and $A$. Gauging $G$ means choosing a representative of $A$ in each gauge orbit (i.e. fixing a gauge) and then summing over such representatives.  When the phase $\mathrm{exp}\{{2\pi i \int_{X_{d}} f(A, g)}\}$ is non-trivial and cannot be cancelled by modifying the partition function by local counterterms built from background fields, summing over the representative $A$ in the gauge orbit is ambiguous, i.e. the resulting partition function depends on the choice of representatives. Hence the gauged theory is ill-defined, and there is an obstruction to gauging $G$.

In the modern understanding of anomalies (as explained in e.g. \cite{Freed:2014iua,Monnier:2019ytc}), the 't Hooft anomaly of $G$ is naturally described by an invertible field theory in one higher dimension, defined on a manifold $X_{d+1}$ whose boundary is $X_d$. This goes under the name of \emph{anomaly inflow} \cite{Callan:1984sa}. 
Concretely, the anomaly inflow theory is an invertible TQFT with action $2\pi \int_{X_{d+1}}\omega(A)$, such that under background gauge transformations it changes by $\omega(A)\to \omega(A) - df(A,g)$. Hence the combination 
\begin{equation}\label{eq:inflow}
    Z[X_d, A]\, e^{2\pi i \int_{X_{d+1}} \omega(A)}
\end{equation}
is gauge invariant. The possible inflow actions $\omega(A)$ are classified by bordisms equipped with a map to the classifying space of $G$,  and with a suitable spacetime structure.\footnote{Note that there are bordism invariants which cannot be written as the integral of a local functional of $A$, e.g. the Arf invariant. }  For convenience, we will refer to this invertible TQFT as the \textit{Anomaly TFT} (AnomTFT). Conversely, the AnomTFT can be taken to \textit{define} the 't Hooft anomaly of the invertible symmetry.

When the symmetry $G$ is a finite group, there is an alternative perspective. In this case, turning on a background gauge field $A$ amounts to inserting a network of $G$ symmetry defects, and gauge transformations amount to local deformations of the defect network. In 2d the defects are topological lines, and the local deformations of the topological lines are characterized by the $F$-symbols, as shown in Figure \ref{fig:Fsymbols}. Since gauging $G$ amounts to summing over all gauge field configurations up to gauge transformations, it is also equivalent to first fixing a triangulation of the spacetime manifold and then summing over all possible defect configurations on the dual lattice of the triangulation. When the partition function is not invariant under a local re-triangulation of the spacetime manifold (for instance by performing Pachner moves), then the resulting partition function is subject to an ambiguity, and hence there is an obstruction to gauging. Such an ambiguity arises when the $F$-symbols in Figure \ref{fig:Fsymbols} belong to a non-trivial group cohomology class in $H^3(G,U(1))$.

\begin{figure}[tbp]
\begin{center}
\begin{tikzpicture}[baseline=0,scale = 0.7, baseline=-10]
\draw [red, very thick] (-1,-1.5) to (1,1.5);
\draw [red, very thick] (0,0) to (1.1,-1.5);
\draw [red, very thick] (0.4,0.6) to (2,-1.5);
 \node[left] at (-1,-1.4) {$L_g$};
  \node[left] at (0.8,-1.4) {$L_h$};
  \node[right] at (1.9,-1.4) {$L_k$};
    \node[left] at (0.2,0.5) {$L_x$};
    \node[right] at (1,1.5) {$L_{\ell}$};
\end{tikzpicture}
$\,\,=\,\,  \sum_{L_y \in L_h \times L_k} (F_{g,h,k}^\ell)^x_y$
\begin{tikzpicture}[baseline=0,scale = 0.7, baseline=-10]
\draw [red, very thick] (-1,-1.5) to (1,1.5);
\draw [red, very thick] (-0.2,-1.5) to (0.8,0);
\draw [red, very thick] (0.4,0.6) to (2,-1.5);
 \node[left] at (-1,-1.4) {$L_g$};
  \node[right] at (0.1,-1.4) {$L_h$};
  \node[right] at (1.9,-1.4) {$L_k$};
      \node[right] at (0.6,0.3) {$L_y$};
    \node[right] at (1,1.5) {$L_{\ell}$};
\end{tikzpicture}
  \caption{Definition of the $F$-symbols, which generalize `t Hooft anomalies for non-invertible symmetries in $(1+1)$d. The sum is over all $L_y$ appearing in the fusion of $L_h$ with $L_k$. When the defects are invertible, the sum on the right hand side only contains one term. }
    \label{fig:Fsymbols}
    \end{center}
\end{figure}

\subsection{Anomalies of non-invertible symmetries}
\label{sec:intro2}

We now proceed to the case of non-invertible symmetries. For simplicity, we will assume that the non-invertible symmetries are finite. As for invertible symmetries,  it is possible to define gauging of non-invertible symmetries by fixing a triangulation of the spacetime manifold and summing over all possible non-invertible defect configurations.\footnote{The gauging of non-invertible symmetries in higher dimensions involves additional subtleties and a complete description remains unavailable at present. An example of such a subtlety is whether one should include all possible condensation defects in the sum above. Fortunately, the following discussion will not depend on these subtleties. }
Furthermore, non-invertible symmetries can also have non-trivial 't Hooft anomalies, in the sense of obstructions to gauging. The presence of non-trivial 't Hooft anomalies means that after summation the resulting partition function depends on the choice of triangulation (i.e. it is not ``gauge invariant''), and hence is ill-defined.

We may now ask about the anomaly inflow mechanism for non-invertible symmetries. 
It is natural to suspect that the anomalies for non-invertible symmetries can again be cancelled by coupling to a bulk invertible TQFT.  However, because the notion of background fields for non-invertible symmetries is still poorly understood, a systematic understanding of this AnomTFT is lacking. We will not shed light on this issue here. Instead, we point out that while the AnomTFT of non-invertible symmetries is poorly understood, the \textit{Symmetry TFT} (SymTFT)\cite{Freed:2012bs,2015arXiv150201690K, Freed:2018cec,Gaiotto:2020iye,Apruzzi:2021nmk,Apruzzi:2022dlm, Burbano:2021loy, Freed:2022qnc,Kaidi:2022cpf, Kitaev:2011dxc, 2017arXiv170401447K, 2021arXiv210403121K, Kong:2020cie,Freed:2022qnc,Freed:2022iao} is relatively well-understood (at least in some cases), and can be used as a partial diagnostic for the existence of `t Hooft anomalies.

Before proceeding, let us mention that there is already a definitive test for anomalies of non-invertible symmetries in 2d QFTs, introduced in \cite{Thorngren:2019iar,Thorngren:2021yso}. In particular, those references showed that the presence of an anomaly 
is equivalent to the lack of a so-called ``fiber functor,'' i.e. a module category with a single simple object. While this gives a complete characterization of anomalies in 2d, determining in practice whether a fiber functor exists is not always straightforward. Furthermore, the notion of fiber functor is more complicated in higher dimensions, where one has to consider  fusion higher-categories. Our construction in this note provides an alternative, occasionally more practical, but \textit{strictly weaker} characterization of anomalies of non-invertible symmetries.

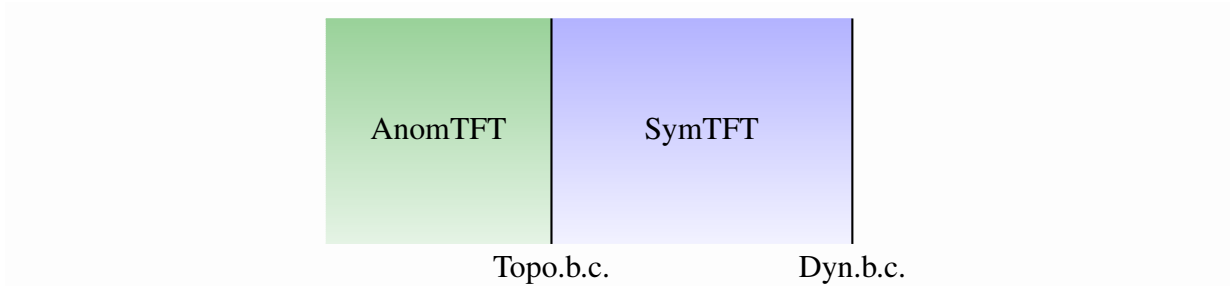
\begin{figure}[t]
	\centering
	\begin{tikzpicture}

	\shade[line width=2pt, top color=blue!30, bottom color=blue!5] 
	(3,0) to [out=90, in=-90]  (3,3)
	to [out=0,in=180] (4+3,3)
	to [out = -90, in =90] (4+3,0)
	to [out=180, in =0]  (0,0);
	
	\shade[line width=2pt, top color=dgreen!40, bottom color=dgreen!10] 
	(4-4,0) to [out=90, in=-90]  (4-4,3)
	to [out=0,in=180] (7-4,3)
	to [out = -90, in =90] (7-4,0)
	to [out=180, in =0]  (0,0);

	\draw[thick] (3,0) -- (3,3);
	\draw[thick] (4+3,0) -- (4+3,3);

	\node at (2+3,1.5) {\begin{tabular}{c}
	     SymTFT  
	\end{tabular}};
	\node[below] at (3,0) {Topo.b.c.};
	\node[below] at (4+3,0) {Dyn.b.c.}; 
	\node at (5.5-4,1.5) {\begin{tabular}{c}
	     AnomTFT 
	\end{tabular}};
	
	\end{tikzpicture}
	
	\caption{Any $d$-dimensional QFT  can be expanded into a $(d+1)$-dimensional slab with the SymTFT living inside of it, with a topological boundary condition on the left boundary and a dynamical boundary condition on the right boundary.  }
	\label{fig:SymTFTintro}
\end{figure}

\subsubsection{Symmetry TFT}
We now describe a \textit{sufficient} condition for the presence of an anomaly for a non-invertible symmetry, making use of the SymTFT. To set the stage, let us recall that for any $d$-dimensional QFT with non-invertible symmetry described by a higher fusion category $\CC$, one can expand the system into a $(d+1)$-dimensional slab with the SymTFT living inside it; see Figure \ref{fig:SymTFTintro}. The SymTFT is given by (the higher-categorical generalization of) the Turaev-Viro theory of $\CC$, whose topological operators are given by (the higher-categorical generalization of) the Drinfeld center of $\CC$, i.e. $\CZ(\CC)$. The left boundary is a topological Dirichlet boundary condition, while the right boundary is a dynamical boundary condition. Shrinking the slab by colliding the two boundaries reproduces the original $d$-dimensional QFT.  See e.g. \cite{Kaidi:2022cpf,Lin:2022dhv} for further details.

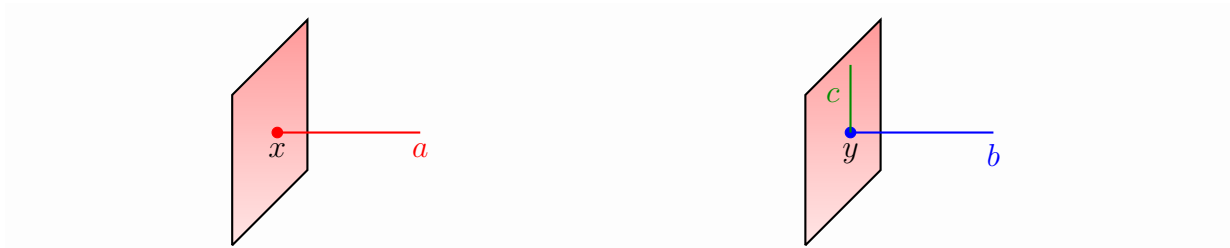
\begin{figure}[tbp]
\begin{center}
\begin{tikzpicture}

	\shade[line width=2pt, top color=red!40, bottom color=red!10] 
	(0,0)--(0,2)--(1,3)--(1,1)--(0,0);
	
	\draw[thick] (0,0)--(0,2)--(1,3)--(1,1)--(0,0);
	
	\filldraw[red, red] (0.6,1.5) circle (2pt);
	\draw[thick,red] (0.6,1.5)--(2.5,1.5); 
 \node[below, red] at (2.5,1.5) {$a$};
       \node[below] at (0.6,1.5) {$x$};
       
  \begin{scope}[xshift=3 in]
  \shade[line width=2pt, top color=red!40, bottom color=red!10] 
	(0,0)--(0,2)--(1,3)--(1,1)--(0,0);
	
	\draw[thick] (0,0)--(0,2)--(1,3)--(1,1)--(0,0);
	
	\filldraw[red, blue] (0.6,1.5) circle (2pt);
	\draw[thick,blue] (0.6,1.5)--(2.5,1.5); 
        \draw[thick, dgreen] (0.6,1.5) -- (0.6, 2.4);
   \node[below, blue] at (2.5,1.5) {$b$};
   \node[left, dgreen] at (0.6,2) {$c$};
   \node[below] at (0.6,1.5) {$y$};
  \end{scope}

	\end{tikzpicture}

\caption{Bulk topological operators terminating on the topological boundary. We refer to $a$ as ``terminable'' and $b$ as ``non-terminable.''}
\label{fig:oper}
\end{center}
\end{figure}

\begin{figure}[tbp]
\begin{center}
\begin{tikzpicture}

	\shade[line width=2pt, top color=red!40, bottom color=red!10] 
	(0,0)--(0,2)--(1,3)--(1,1)--(0,0);
	
	\draw[thick] (0,0)--(0,2)--(1,3)--(1,1)--(0,0);
	
	\filldraw[red, red] (0.6,1.5) circle (2pt);
	\draw[thick,red] (0.6,1.5)--(1.4,1.5); 
	\draw[thick,red] (1.6,1.5)--(2.5,1.5);
	
	\filldraw[red] (0.6,1) circle (2pt);
	\draw[thick,red] (0.6,1) to [out=-10, in=260]  (1.5,1.5); 
	\draw[thick,red] (1.5,1.5) to [out=90, in=180]  (2.5,2); 
  \draw[thick,-stealth] (0.5,1) to [out=170, in=270] (0.2,1.5);
  \node[right, red] at (2.5, 1.5) {$a$};
  \node[right, red] at (2.5, 2) {$a'$};
  
  \node[] at (5,1.5) {$\longrightarrow$};
  \begin{scope}[xshift=3 in]
  \shade[line width=2pt, top color=red!40, bottom color=red!10] 
	(0,0)--(0,2)--(1,3)--(1,1)--(0,0);
	
	\draw[thick] (0,0)--(0,2)--(1,3)--(1,1)--(0,0);
	
	\filldraw[red, red] (0.6,1.5) circle (2pt);
	\draw[thick,red] (0.6,1.5)--(2.5,1.5); 
	
	\filldraw[red, red] (0.6,1) circle (2pt);
	\draw[thick,red] (0.6,1) to [out=-10, in=260]  (1.5,1.4); 
	\draw[thick,red] (1.5,1.6) to [out=90, in=180]  (2.5,2); 
 \node[right, red] at (2.5, 1.5) {$a$};
  \node[right, red] at (2.5, 2) {$a'$};
  
  \end{scope}

	\end{tikzpicture}

\caption{
Two topological operators that are terminable on the topological boundary should have trivial linking invariant in the bulk. }
\label{fig:SymTFTarg}
\end{center}
\end{figure}
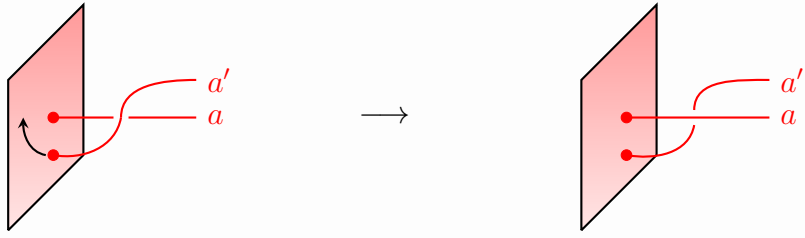

Conversely, given any $(d+1)$-dimensional TQFT in a slab with topological operators denoted by $\CZ$, together with a topological boundary condition (associated with a Lagrangian subalgebra $\CA$) on the left as well as a dynamical boundary condition on the right, shrinking the slab defines a $d$-dimensional QFT whose global symmetry is determined by $\CZ$ and the topological boundary condition $\cA$. To see this more concretely, let us place the topological operators in the bulk orthogonal to the topological boundary, as shown in Figure \ref{fig:oper}.

When the bulk $p$-dimensional topological operator labeled by $a$ belongs to the Lagrangian algebra $\CA$ associated with the topological boundary (red plane), $a$ can terminate on the topological boundary at a $(p-1)$-dimensional locus $x$; see the left panel of Figure \ref{fig:oper}.  This implies that the linking invariant between two bulk operators $a,a'\in \CA$ must be trivial. To see why, we consider the two configurations shown in Figure \ref{fig:SymTFTarg}, which differ by a braiding in the bulk. Since the boundary condition is topological, one can freely move the boundary locus of $a'$ around $a$, and hence the two configurations  are topologically equivalent. Thus the linking invariants between $a$ and $a'\in \CA$ must be trivial.

When the $p$-dimensional topological operator labeled by $b$ does not belong to $\CA$, the $(p-1)$-dimensional locus $y$ is further attached to a $p$-dimensional topological operator $c$  sitting within the topological boundary. Here $y$ is a morphism between the bulk operator $b$ and the boundary operator $c$, and we can denote the bulk operator $b$ as a pair $(c,y)$;\footnote{In general, this labeling is not unique. For a given $b\in \CB$, there may exist multiple $c\in \CC$ on the boundary such that they are connected by some morphisms. We will encounter such examples in the main text below. 
}  see the right panel of Figure \ref{fig:oper}. The topological operators $c$ are expected to form a higher fusion category $\CC$ describing the global symmetry of the $d$-dimensional QFT obtained by shrinking the slab. We will denote the set of bulk operators which can end on the boundary operator $c$ as $\CB_c$. We refer to the operators in $\CA$ as ``terminable,'' and to those in $\CB_c$ (for non-identity $c$) as ``non-terminable.''

\subsubsection{Probing the anomaly via linking invariants}
We would now like to consider a non-invertible symmetry whose defects are given by $\CC$. Below we will assume that $\CC$ is anomaly-free, meaning that it is gaugable. As reviewed above, gauging the symmetry means that we sum over all possible configurations of $\CC$ defects in the $d$-dimensional spacetime. We now discuss what this assumption implies in terms of the topological operators in the bulk SymTFT. 

Assuming that the symmetry is anomaly-free, we may gauge $\CC$ and then expand the $d$-dimensional 
$\CC$-gauged QFT to a $(d+1)$-dimensional slab. The SymTFT of the gauged theory is the same as the SymTFT of the ungauged theory, and in particular the topological operators within the slab are still given by $\CZ(\CC)$. However, the left topological boundary condition corresponding to the $\CC$-gauged theory is different from the one before. Because gauging means summing over all possible $\CC$ defect configurations, the left boundary becomes a condensate of the operators in $\CC$. Hence there should be at least one topological operator $b_c=(c,y_c)\in \CB_c$ for each $c\in \CC$ such that after gauging it becomes terminable. These operators will be within the Lagrangian subalgebra of the $\CC$-gauged topological boundary condition. By applying the discussion around Figure \ref{fig:SymTFTarg}, we see that such operators must have trivial braiding in the bulk.

We thus have the following main result,
\paragraph{Main result:}
\textit{If the non-invertible symmetry $\CC$ is anomaly-free (i.e. there is no obstruction to gauging), then for each $c\in \CC$ there must be a choice of morphisms $y_c$ denoted by $y_c^*$  such that the set of topological operators $\{b_c:=(c,y_c^*)\}$ in the SymTFT has trivial linking invariants, i.e. their correlation functions have trivial phases. Schematically, 
\bea
    \text{$\CC$ is anomaly-free } \,\,&\Longrightarrow&\,\, 
    \begin{matrix}\text{$\exists$ $b_c\in \CB_c$ such that the linkings between $b_{c_1}$ and $b_{c_2}$}\\
    \text{are trivial for every $c_1, c_2 \in \cC$}\end{matrix}~.
\eea
Said in another way, if one cannot find a choice of representatives $b_c\in \CB_c$ such that they have trivial linking  invariants, then the symmetry $\CC$ is anomalous,
\bea
   &\vphantom{.}&
   \begin{matrix}\text{$\nexists$ $b_c\in \CB_c$ such that the linkings between $b_{c_1}$ and $b_{c_2}$}\\
    \text{are trivial for every $c_1, c_2 \in \cC$}\end{matrix} \,\,\,\,\,\Longrightarrow\,\,\,\,\, \text{$\CC$ is anomalous}~. 
\eea
}

\bigskip
\noindent
Let us make a few comments before proceeding to examples:
\begin{enumerate}
    \item The only linking invariants relevant for this note are linking numbers. However, as reviewed in Appendix \ref{app:Link}, there can be multiple types of linking numbers involving different numbers of topological operators. 
    
    \item Our main result is only  a sufficient condition for a non-trivial 't Hooft anomaly. Indeed, the condition is \textit{not} necessary, i.e. even if the linking numbers among the $b_{c_i}$ are trivial, there can still be an  obstruction to gauging $\CC$. 
    
   As an example, consider a (non-invertible) one-form symmetry in 3d, whose defects are given by lines. We may now consider the linking numbers among lines in the 4d SymTFT. 
   Let us consider a link with $N$ components, of type $k$ (see \eqref{eq:typeklink}). Then the quantities $N$ and $k$ must satisfy
    \begin{equation}\label{eq:condition}
        2\cdot(N-k) + 1\cdot k = 4~.
    \end{equation}
    As reviewed in Appendix \ref{app:Link}, when $N=2$ there is only one type of linking number, with $k=1$; when $N>2$, the allowed linking numbers are labelled by $k=0, 1, ..., N-1$. None of these possibilities solves the constraint \eqref{eq:condition}, and we conclude that the anomaly of the 1-form symmetry is not captured by the linking numbers in the 4d SymTFT. However, interestingly, it is known that the anomaly of the one-form symmetry in this case is given by the spin of the defects in 3d \cite{Hsin:2018vcg}. 
    \item In two spacetime dimensions, our results are corollaries of the results in \cite{Thorngren:2019iar}; namely, the non-existence of a fiber functor is a sufficient and necessary condition for a nontrivial 't Hooft anomaly, whereas our condition is only sufficient.
    
    \item In the special case in which the non-invertible symmetry is \textit{non-intrinsically non-invertible} \cite{Kaidi:2022cpf, Kaidi:2022uux}, the associated SymTFT is a Dijkgraaf-Witten (DW) theory. Computing the linking invariants of operators in a DW theory is a standard exercise in any dimension. Hence in this case our main result provides a set of easily computable observables which probe the anomalies of non-invertible symmetries. 
    
    \item Using the gauged AnomTFT to probe anomalies has a long history. For example, in \cite{2012PhRvB..86k5109L} the anomaly of a $\Z_2$ symmetry in 2d was detected by computing the mutual braiding statistics between anyons (i.e. the linking number involving two magnetic lines) in the double semion TQFT in 3d. Here we point out that the gauged AnomTFT of the invertible symmetry is none other than the SymTFT, and that the linking invariants detecting anomalies can involve multiple operators and go beyond links involving two lines. 
    
Furthermore, the gauged AnomTFT arises naturally in the context of brane constructions in string theory via anomaly inflow \cite{Freed:1998tg,Harvey:1998bx,Bah:2019rgq,Bah:2020jas,Bah:2020uev}. In this context, the anomalous variation of the effective 10 or 11d action cancels the anomalies for degrees of freedom on the worldvolume of the branes, and the dimensional reduction of the topological terms in the 10 or 11d effective action yields the AnomTFT via the descent procedure.
The SymTFT can be seen to arise from the dimensional reduction of the same topological terms
\cite{Apruzzi:2021nmk}, and it was recently explained \cite{Apruzzi:2022rei} how to use this inflow perspective to obtain the SymTFT for non-invertible symmetries in the context of holography.

\end{enumerate}

\subsubsection{Simple examples}

It is useful to illustrate the above main result in two simple examples: namely the cases of $\CC= \text{Vec}_{\Z_2}^{\omega}$ with $\omega=0,1$ in 2d, i.e. \textit{invertible} $\Z_2$ symmetry in 2d without an anomaly ($\omega=0$) and with an anomaly ($\omega=1$).

\paragraph{$\omega=0$:} In this case the SymTFT is a $\Z_2$ gauge theory without DW twist. This SymTFT has four line operators denoted by $1, e, m,$ and $f:=e\times m$. We start with the topological boundary condition associated with the Lagrangian algebra $\CA=1\oplus e$. In this case the line $e$ is terminable, and the set of non-terminable lines is given by $ \{m, f\}$. In particular, both $m$ and $f$ are attached to a boundary line $\mathsf{m}\in \CC$, and we can write $m=(\mathsf{m}, x_m)$ and $f=(\mathsf{m}, x_f)$, where $x_m, x_f$ are appropriate morphisms between $m, f$, and $\mathsf{m}$, respectively.  The subset $\CB_{\mathsf{m}}$ thus contains both $m$ and $f$. To see whether $\mathsf{m}$ is anomalous, we choose an element in $\CB_{\mathsf{m}}$, i.e. either $m$ or $f$, and compute the linkings of it with itself. These are known to be trivial, and hence we don't find an obstruction to gauging. This is consistent with the known fact that for a $\Z_2$ gauge theory, $\mathsf{m}$ is anomaly-free.

\paragraph{$\omega=1$:} In this case the SymTFT is a $\Z_2$ twisted gauge theory, known as the double semion model. This theory has four line operators $1, b, s,$ and $\bar{s}:=b\times s$. We start with the topological boundary condition associated with the Lagrangian algebra $\CA=1\oplus b$. In this case $b$ is terminable and the set of non-terminable lines is $\{s, \bar{s}\}$. Both $s$ and $\bar{s}$ are attached to a boundary line $\mathsf{s}\in \CC$, and we can write $s=(\mathsf{s}, x_s)$ and $\bar{s}=(\mathsf{s}, x_{\bar{s}})$, where $x_{s}, x_{\bar{s}}$ are appropriate morphisms between $s, \bar{s},$ and $\mathsf{s}$ respectively. The subset $\CB_{\mathsf{s}}$ thus contains $\bar{s}$ and $s$. To see whether $\mathsf{s}$ is anomalous, we choose an element of $\CB_{\mathsf{s}}$, i.e. either $s$ or $\bar{s}$, and compute the linking invariants of that representative with itself. In the double semion theory, it is known that the Hopf linking between $s$ and itself, as well as the Hopf linking between $\bar{s}$ and itself, is non-trivial. In particular, unlinking produces a factor of $-1$. Hence by our main result, the anomaly of $\mathsf{s}$ must be non-trivial, as expected.

\bigskip 

In the remainder of this note, we will apply our main result to two families of examples, one in two dimensions (Section \ref{sec:Noninvsym2d})  and the other in four dimensions (Section \ref{sec:4d}). For simplicity, we will restrict to non-intrinsically non-invertible symmetries constructed via the strategy of \cite{Kaidi:2021xfk}, in which case the SymTFT is just a DW theory. In these cases the anomaly is in a sense ``obvious,'' since it follows from the self-anomaly of an invertible symmetry before gauging, but it is nevertheless a good illustration of our strategy. The computation of the linkings in the corresponding 3d and 5d DW theories is outlined in Appendices \ref{app:3dDW} and \ref{app:5d}, respectively.  We further use the anomaly of the non-invertible symmetries to make statements regarding the low-energy dynamics of the abelian Higgs models in 2d, together with adjoint QCD and $\CN=4$ super Yang-Mills (SYM) in 4d.

\section{Non-invertible symmetries and anomalies in 2d}
\label{sec:Noninvsym2d}

In this section we illustrate the discussion in the introduction by means of a simple 2d example.  Our starting point will be a 2d quantum field theory $\CX$ with an invertible $\Z_2^A\times \Z_2^B\times \Z_2^C$ zero-form global symmetry and 't Hooft anomalies specified by the 3d AnomTFT,
\begin{eqnarray}\label{eq:invTQFT}
\int_{X_3} \pi \left(  A B C +  A \beta A \right)~,
\end{eqnarray}
written in terms of the background fields $A, B, C$. 
The operation $\beta: H^1(X_3, \Z_2)\to H^2(X_3, \Z_2)$ is the Bockstein map associated with the central extension $1\to \Z_2\to \Z_4\to \Z_2\to 1$, and acts on $\Z_2$-valued cocycles as $\beta= \delta/2$. We will denote the collection of topological line defects of $\Z_2^A\times \Z_2^B\times \Z_2^C$ as $\CC$. 

In order to obtain a theory with a non-invertible symmetry, we  gauge $\Z_2^B\times \Z_2^C$, which is possible since neither of these symmetries has a self-anomaly, nor do they have a mixed anomaly with one another. By the general results of \cite{Kaidi:2021xfk}, this gauging changes $\Z_2^A$ into a non-invertible symmetry. It is natural to expect that the self-anomaly for the invertible $\Z_2^A$ symmetry before gauging leads to a self-anomaly for the non-invertible symmetry after gauging. We will now show that this is the case by relating it to non-trivial linking invariants in the SymTFT.

\subsection{SymTFT of invertible symmetries}
\label{sec:symtft2d}

\paragraph{The SymTFT:}
Let us begin by discussing the SymTFT for the theory $\CX$ with invertible symmetry. 
As reviewed in the introduction, a 2d QFT 
can be expanded into a 3d slab 
filled with the SymTFT. The SymTFT is a gauged version of the AnomTFT given in \eqref{eq:invTQFT}, i.e. it is a Dijkgraaf-Witten (DW) TQFT with DW twist term  
specified by the AnomTFT, 
\begin{eqnarray}\label{eq:2dSymTFT}
\int_{X_3} \pi \left(  \widehat{a} \delta a +  \widehat{b} \delta b +  \widehat{c} \delta c +  abc + \frac{1}{2} a\delta a \right) ~. 
\end{eqnarray}
Here all the dynamical fields $\widehat{a}, a, ...$ are $\Z_2$-valued 1-cochains. Integrating out $\widehat{a}$ enforces $a$ to be a $\Z_2$ cocycle, and likewise for $b$ and $c$. Such a DW theory has been studied in \cite{He:2016xpi, deWildPropitius:1995cf} and is reviewed in detail in Appendix \ref{app:3dDW}. For our purposes here, we note that this theory has 22 genuine line operators, among which eight are invertible and generated by the three independent invertible Wilson line operators
\begin{eqnarray}\label{eq:invop0}
U_{a}(M_1)= e^{i\pi \oint_{M_1} a}~, \hspace{1cm} U_{b}(M_1)= e^{i\pi \oint_{M_1} b}~, \hspace{1cm} U_{c}(M_1)= e^{i\pi \oint_{M_1} c}~.
\end{eqnarray}
The remaining 14 operators are non-invertible, each having quantum dimension 2. Three of them are given by the magnetic line operators 
\begin{equation}\label{eq:magopline}
    \begin{split}
        \widehat{U}_a(M_1)&\sim  \sum_{\phi_2, \phi_3\in C^0(M_1, \Z_2)} e^{i\pi \oint_{M_1} \widehat{a} + i\pi \oint_{M_1} (-\phi_2 c + \phi_3 b + \phi_2 \delta \phi_3)}~,\\
\widehat{U}_{b}(M_1)&\sim  \sum_{\phi_3, \phi_1\in C^0(M_1, \Z_2)} e^{i\pi \oint_{M_1} \widehat{b} + i\pi \oint_{M_1} (-\phi_3 a + \phi_1 c + \phi_3 \delta \phi_1)}~,\\
\widehat{U}_c(M_1)&\sim \sum_{\phi_1, \phi_2\in C^0(M_1, \Z_2)} e^{i\pi \oint_{M_1} \widehat{c} + i\pi \oint_{M_1} (-\phi_1 b + \phi_2 a + \phi_1 \delta \phi_2)}~,\\
    \end{split}
\end{equation}
and the others are obtained by fusion between these three and the invertible line operators. See  \eqref{eq:3dlines} for a complete list. Throughout this note, since we are mainly interested in the phases of partition and correlation functions, we will not specify the overall real positive  normalization factors and use $\sim$ to denote equality up to such an overall normalization. 
Note that in the definition of the magnetic line operators we have introduced a 1d TQFT on the worldline of the defect in order to maintain gauge invariance; since the TQFT is a $(0+1)$d $\Z_2$ gauge theory, the quantum dimension is 2.  It is this fact which leads to their non-invertibility. The total quantum dimension is $D_{\text{tot}}= \sqrt{8\cdot 1^2 + 14 \cdot 2^2}=8$.  
See Appendix \ref{app:3dDW} for more details.

\paragraph{Dirichlet boundary condition:}
To recover the theory $\cX$, the left topological boundary of the SymTFT should be taken to be the Dirichlet boundary condition for all three $\ZZ_2$ symmetries. This sets the dynamical fields $a,b,c$ equal to background fields $A,B,C$, respectively. In terms of boundary states, this can be written as
\begin{eqnarray}\label{eq:2dbra}
\bra{D_{a,b,c}} = \sum_{a,b,c\in C^1(X_2, \Z_2)} \bra{a,b,c}\, \delta(a-A) \delta(b-B)\delta(c-C)~.
\end{eqnarray}
This Dirichlet boundary condition means that all the invertible lines in \eqref{eq:invop0} become trivial on the boundary, and can also terminate perpendicularly on the boundary. We denote the set of all invertible lines as $\CA_{\text{inv}}$, where the subscript indicates that the topological boundary condition gives rise to invertible symmetries. The sum of the quantum dimension of all the objects in $\CA_{\text{inv}}$ is $D_{\text{inv}}=8$, matching the total quantum dimension $D_{\text{inv}}=D_{\text{tot}}$.

To see how this boundary condition gives rise to an invertible $\Z_2^A\times \Z_2^B\times \Z_2^C$ symmetry, we consider placing each of the 22 bulk line operators orthogonal to the boundary as shown in Figure \ref{fig:oper}. Without loss of generality, we may set all the background fields $A,B,C=0$.  All invertible lines belong to $\CA_{\text{inv}}$ by definition, and hence can terminate on the boundary, whereas all of the non-invertible lines cannot end. For example, when the non-invertible line $\widehat{U}_a$ touches the Dirichlet boundary, both $b$ and $c$ become trivial (and consequently $\phi_2$ and $\phi_3$ also become trivial) but the magnetic portion $e^{i\pi \oint_{M_1}\widehat{a}}$ is not trivialized on the boundary, and instead extends along the boundary; see the right panel of Figure \ref{fig:oper}. In other words, there are eight boundary lines collectively denoted as $\CC_{\text{inv}}$, generated by
\begin{eqnarray}
    \widehat{U}_a(M_1|_\partial)= e^{i\pi\oint_{M_1|_{\partial}} \widehat{a}}~, \hspace{1cm} \widehat{U}_b(M_1|_\partial)= e^{i\pi\oint_{M_1|_{\partial}} \widehat{b}}~, \hspace{1cm} \widehat{U}_c(M_1|_\partial)= e^{i\pi\oint_{M_1|_{\partial}} \widehat{c}}~,
\end{eqnarray}
where $M_1|_{\partial}$ represents a line on the boundary. As all of these boundary lines are invertible and satisfy $\Z_2$ fusion rules, upon shrinking the slab they generate a $\Z_2^A\times \Z_2^B\times \Z_2^C$ invertible global symmetry. We denote the  boundary symmetry generators by $U_A, U_B,$ and $U_C$, respectively.

\paragraph{Dynamical boundary condition:}
On the other hand, the dynamical boundary condition on the right  captures the dynamics of the 2d theory, and in particular is non-topological. In terms of boundary states, it can be written as
\begin{eqnarray}\label{eq:2dket}
\ket{\CX} = \sum_{a,b,c\in C^1(X_2, \ZZ_2)} Z_{\CX}[X_2; a,b,c]\ket{a,b,c} ~. 
\end{eqnarray}
Shrinking the 3d slab amounts to taking the inner product between the bra \eqref{eq:2dbra} and the ket \eqref{eq:2dket}, which recovers the 2d partition function $Z_\CX[X_2; A,B,C]$.

\subsection{Anomaly of invertible symmetry from linking invariants}
\label{sec:AnomTFT2d}

Before discussing non-invertible symmetries, let us review how the anomaly \eqref{eq:invTQFT} of the invertible symmetry $\Z_2^A\times \Z_2^B\times \Z_2^C$  can be probed via linking invariants in the SymTFT, following the general discussion in the introduction.

Note that both $\widehat{U}_{a}$ and $\widehat{U}_{a} U_{a}$ are non-terminable, ending on the boundary line $\widehat{U}_{a}|_{\partial}$. Thus these two lines live in the same subset $\CB_{\widehat{a}}$. In a similar way, the 14 non-invertible lines in the bulk can be grouped into 7 subsets, 
\begin{align}
    \CB_{\widehat{a}}&= \{\widehat{U}_{a},\, \widehat{U}_{a} U_{a}\}~, & \CB_{\widehat{b}}&= \{\widehat{U}_{b},\, \widehat{U}_{b} U_{b}\}~, & \CB_{\widehat{c}}&= \{\widehat{U}_{c},\, \widehat{U}_{c} U_{c}\}~,\nonumber \\
    \CB_{\widehat{ab}}&= \{\widehat{U}_{ab},\, \widehat{U}_{ab} U_{a}\}~, & \CB_{\widehat{ac}}&= \{\widehat{U}_{ac},\, \widehat{U}_{ac} U_{a}\}~, & \CB_{\widehat{bc}}&= \{\widehat{U}_{bc},\, \widehat{U}_{bc} U_{b}\}~,\nonumber \\
    & & \CB_{\widehat{abc}}&= \{\widehat{U}_{abc},\, \widehat{U}_{abc} U_{a}\}~. & &
\end{align}
Following the discussion in the introduction, in order to probe the anomaly, we must check whether it is possible to choose one representative from each subset such that all of the linking invariants are trivial. If for every choice of representatives the linking invariants are non-trivial, then the anomaly must be non-trivial as well.

As reviewed in Appendix \ref{app:Link}, there are only two types of linking invariants to consider in this context: the linking number involving two loops, and the type-0 linking number involving three loops. The former can be probed by the Hopf link, and the latter by the Borromean rings.  We first consider the  linking configuration involving two identical representatives $\widehat{U}_a(M_1)U^{p_a}_{a}(M_1) $, where $p_a=0,1$ denotes the two choices of representatives in $\CB_{\widehat{a}}$. This  is expected to probe the self-anomaly for $\ZZ_2^A$. A straightforward computation described in Appendix \ref{app:3dDW} gives
\begin{eqnarray}\label{eq:3dbilink}
\begin{split}
    \braket{\widehat{U}_a(M_1)U^{p_a}_{a}(M_1) \widehat{U}_a(M_1')U^{p_a}_{a}(M_1')} &\sim (-1)^{\text{Link}(M_1,M_1')}~, \hspace{1cm} p_a=0,1\,,
\end{split}
\end{eqnarray}
where $\sim$ means equal up to a real positive normalization. 
We see that no matter which representative of $\mathcal{B}_{\widehat a}$ we choose, the Hopf link configuration gives a non-trivial sign. By the main result in the introduction, we conclude that the {invertible} $\Z_2^A$ symmetry is anomalous. Indeed, from the calculation of linking invariants in Appendix \ref{app:3dDW},  this non-trivial linking invariant is seen to follow directly from the Dijkgraaf-Witten term $\frac{1}{2}a\delta a$ in (\ref{eq:2dSymTFT}), which in turn follows from the anomaly $A\beta A$ in \eqref{eq:invTQFT}. Hence the non-trivial linking invariant detects the anomaly $A\beta A$.

We may also consider linking configurations involving three components. We begin by choosing representatives from the subsets $\CB_{\widehat{a}}, \CB_{\widehat{b}}$, and $\CB_{\widehat{c}}$, which we denote by $\widehat{U}_a U_{a}^{p_a}, \widehat{U}_b U_{b}^{p_b}$, and $\widehat{U}_c U_{c}^{p_c}$. Their correlation function is
\begin{eqnarray}\label{eq:3dtrilink}
\begin{split}
    &\braket{\widehat{U}_a(M_1) U_{a}^{p_a}(M_1) \widehat{U}_b(M_1') U_{b}^{p_b}(M_1')\widehat{U}_c(M_1'') U_{c}^{p_c}(M_1'') } \\&\sim  (-1)^{\text{Link}(M_1,M_1',M_1'')_{0}} (-1)^{(p_a+p_b)\text{Link}(M_1, M_1') +(p_a+p_c)\text{Link}(M_1, M_1'')  + (p_b+p_c)\text{Link}(M_1', M_1'')}~. 
\end{split}
\end{eqnarray}
If we take $M_1, M_1', M_1''$ to form the Borromean rings, for which the linking number between any pair of loops vanish and $(-1)^{\text{Link}(M_1,M_1',M_1'')_{0}} =-1$, then the above correlation function is non-trivial for any $p_a, p_b, p_c$. 
By the main result in the introduction, we then find that the invertible $\Z_2^A\times \Z_2^B\times \Z_2^C$ symmetry is anomalous. Indeed, since the non-trivial type-0 linking number between three loops follows from the $abc$ term in the DW theory \eqref{eq:2dSymTFT}, which in turn follows from the anomaly $ABC$ in \eqref{eq:invTQFT}, this configuration detects the anomaly $ABC$.

\subsection{Non-invertible symmetries from gauging}
\label{sec:noninvsymmfromgauging}

We have just seen how the non-trivial `t Hooft anomalies of invertible symmetries can be detected using the linkings of operators in the SymTFT. 
We now turn to the case of non-invertible symmetries, for which the AnomTFT is not well-understood. As explained above, it is possible to obtain a theory $\widehat \CX$ with non-invertible symmetries by starting with the theory $\CX$ and gauging the $\Z_2^B \times \Z_2^C$ symmetry. Let us briefly review the properties of these non-invertible symmetries, before discussing their anomalies. 
To begin, note that the partition function of $\widehat \CX$ is 
\begin{eqnarray}\label{eq:hatX}
Z_{\widehat{\CX}}[B,C] = \frac{1}{|H^0(X_2, \Z_2)|^2} \sum_{b,c\in H^1(X_2, \Z_2)} Z_{\CX}[b,c] \,e^{i \pi \int_{X_2} bC + cB}~,
\end{eqnarray}
where we have turned off the $\Z_2^A$ background field in the partition function. Our goal is to show that the topological defect for $\Z_2^A$ becomes a non-invertible topological defect.

To see this, we define two topological manipulations: a gauging of the $\Z_2^B\times \Z_2^C$ symmetry denoted by $\sigma$, and a stacking with a 2d $\Z_2^B\times \Z_2^C$ SPT denoted by $\tau$. Concretely,
\begin{eqnarray}\label{eq:topomani}
\begin{split}
    Z_{\sigma\CX}[B,C] &= \frac{1}{|H^0(X_2, \Z_2)|^2} \sum_{b,c\in H^1(X_2, \Z_2)} Z_{\CX}[b,c]\, e^{i \pi \int_{X_2} bC + cB}~,\\
    Z_{\tau\CX}[B,C] &= Z_{\CX}[B,C]\, e^{i\pi \int_{X_2} BC}~.
\end{split}
\end{eqnarray}
Note that the right-hand side of \eqref{eq:hatX} is precisely the partition function of $\sigma\CX$. To see the non-invertible symmetry of $\widehat{\CX}$, we start with $\widehat{\CX}$ and perform a $\tau\sigma\tau$ transformation followed by a $\Z_2^A$ transformation $g$. It is straightforward to check that $\widehat{\CX}$ is invariant under $g\tau\sigma\tau$ (up to an Euler counterterm):
\begin{eqnarray}
\begin{split}
Z_{g\tau\sigma\tau\widehat{\CX}}[B,C] &= \frac{1}{|H^0(X_2, \Z_2)|^4} \sum_{b,c,\widetilde{b}, \widetilde{c}\in H^1(X_2, \Z_2)} Z_{\CX}[b,c]\, e^{i \pi \int_{X_2} b\widetilde{c} + c\widetilde{b} + \widetilde{b} \widetilde{c} + \widetilde{b}C + \widetilde{c} B + BC + bc}\\
&= \frac{|H^1(X_2, \Z_2)|}{|H^0(X_2, \Z_2)|^4} \sum_{b,c\in H^1(X_2, \Z_2)} Z_{\CX}[b,c]\, e^{i \pi \int_{X_2} (b+B)(c+C) + BC + bc}\\
&= \chi^{-1}[X_2]  \frac{1}{|H^0(X_2, \Z_2)|^2} \sum_{b,c\in H^1(X_2, \Z_2)} Z_{\CX}[b,c]\, e^{i \pi \int_{X_2} bC + cB} = \chi^{-1}[X_2] Z_{\widehat{\CX}}[B,C]~. \\
\end{split}
\end{eqnarray}
In the first equality, we have used the mixed anomaly \eqref{eq:invTQFT}, which implies that under a global $\Z_2^A$ transformation $g$ the partition function of $\CX$ acquires a phase $Z_{\CX}[b,c] \to Z_{\CX}[b,c] e^{i \pi \int_{X_2} bc}$. In the last line we have used the definition of the Euler counterterm $\chi[X_2]:= |H^0(X_2, \Z_2)|^2/|H^1(X_2, \Z_2)|$, assuming $X_2$ is a closed manifold. Ignoring the Euler counterterm, we then have
\begin{eqnarray}
g\tau\sigma\tau\widehat{\CX} = \widehat{\CX}
\end{eqnarray}
and hence we see that $g\tau\sigma\tau$ is a symmetry of $\CX$.
Since the symmetry involves gauging, i.e. a $\sigma$ operation, the symmetry is non-invertible, as can be confirmed by explicitly calculating the fusion rules following \cite{Kaidi:2021xfk, Choi:2021kmx}. Moreover, since the operation $g\tau\sigma\tau$ is obtained by dressing the $\Z_2^A$ transformation $g$ with a twisted gauging $\tau\sigma\tau$, we conclude that the invertible $\Z_2^A$ symmetry in $\CX$ becomes a non-invertible symmetry implementing $g\tau\sigma\tau$ upon gauging $\Z_2^B\times \Z_2^C$.

\begin{figure}[tbp]
\begin{center}

\begin{tikzpicture}[scale=0.4]

\shade[top color=gray!40, bottom color=gray!10] (0.03,-1) 
to [out=0, in=180] (-6.43,-1) 
to [out=270, in=90] (-6.43,1)
to [out=180, in=0] (0.03,1)
to [out=90, in=270] (0.03,-1) ;

\shade[top color=gray!40, bottom color=gray!10,rotate = 90]  (-1,0) coordinate (-left) 
to [out=260, in=60] (-3,-2) 
to [out=240, in=110] (-3,-4) 
to [out=290,in=180] (0,-7) 
to [out=0,in=250] (3,-4) 
to [out=70,in=300] (3,-2) 
to [out=120,in=280] (1,0)  coordinate (-right);

\draw[rotate = 90]  (-1,0) coordinate (-left) 
to [out=260, in=60] (-3,-2) 
to [out=240, in=110] (-3,-4) 
to [out=290,in=180] (0,-7) 
to [out=0,in=250] (3,-4) 
to [out=70,in=300] (3,-2) 
to [out=120,in=280] (1,0)  coordinate (-right);

\shade[top color=gray!40, bottom color=gray!10,rotate = 270,yshift=-2.5in]  (-1,0) coordinate (-left) 
to [out=260, in=60] (-3,-2) 
to [out=240, in=110] (-3,-4) 
to [out=290,in=180] (0,-7) 
to [out=0,in=250] (3,-4) 
to [out=70,in=300] (3,-2) 
to [out=120,in=280] (1,0)  coordinate (-right);

\draw[rotate = 270,yshift=-2.5in]  (-1,0) coordinate (-left) 
to [out=260, in=60] (-3,-2) 
to [out=240, in=110] (-3,-4) 
to [out=290,in=180] (0,-7) 
to [out=0,in=250] (3,-4) 
to [out=70,in=300] (3,-2) 
to [out=120,in=280] (1,0)  coordinate (-right);

\draw[] (0,-1)--(-6.4,-1);
\draw[] (0,1)--(-6.4,1);

\pgfgettransformentries{\tmpa}{\tmpb}{\tmp}{\tmp}{\tmp}{\tmp}
\pgfmathsetmacro{\myrot}{-atan2(\tmpb,\tmpa)}
\draw[rotate around={\myrot:(0,-2.5)},yshift=1.1in,xshift=1.4in] (-1.2,-2.4) to[bend right]  (1.2,-2.4);
\draw[fill=white,rotate around={\myrot:(0,-2.5)},yshift=1.1in,xshift=1.4in] (-1,-2.5) to[bend right] (1,-2.5) 
to[bend right] (-1,-2.5);

\draw[rotate around={\myrot:(0,-2.5)},yshift=0.5in,xshift=-4in] (-1.2,-2.4) to[bend right]  (1.2,-2.4);
\draw[fill=white,rotate around={\myrot:(0,-2.5)},yshift=0.5in,xshift=-4in] (-1,-2.5) to[bend right] (1,-2.5) 
to[bend right] (-1,-2.5);

\draw[rotate around={\myrot:(0,-2.5)},yshift=1.4in,xshift=-3.8in] (-1.2,-2.4) to[bend right]  (1.2,-2.4);
\draw[fill=white,rotate around={\myrot:(0,-2.5)},yshift=1.4in,xshift=-3.8in] (-1,-2.5) to[bend right] (1,-2.5) 
to[bend right] (-1,-2.5);

\draw[red, thick] (-3.2,-1) arc(-90:90:0.5cm and 1cm);
\draw[red, dotted, thick] (-3.2,-1) arc(270:90:0.5cm and 1cm);
\node[below] at (-3.2,-1) {$M_1|_0$};

\node[below] at (-12.5,4) {$X_2^{<0}$};
\node[below] at (6.5,4) {$X_2^{\geq 0}$};
\end{tikzpicture}

\caption{Decomposition of $X_2$ along a neck. The interface is located at $x=0$.}
\label{fig:X2}
\end{center}
\end{figure}
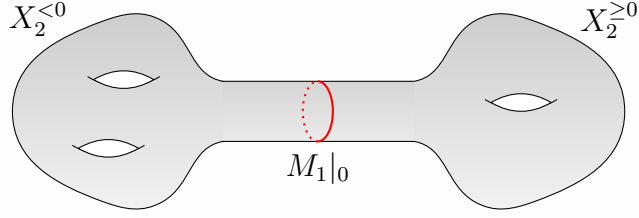

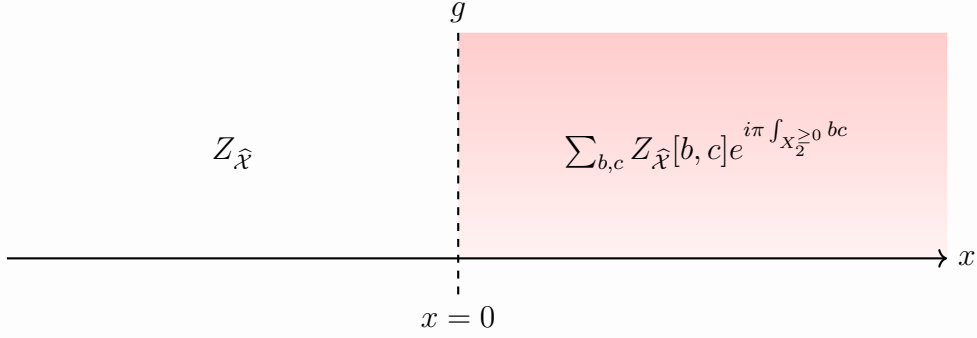
\begin{figure}
    \centering
    \begin{tikzpicture}
      \shade[top color=red!20, bottom color=red!5]  (6,0) 
to [out=0, in=180] (12.5,0) 
to [out=90, in=270] (12.5,3) 
to [out=180,in=0] (6,3) 
to [out=270,in=90] (6,0);

    \draw[thick, ->] (0,0) -- (12.5,0);
    \draw[thick, dashed] (6,3) -- (6,-0.5); 
    \node[above] at (6,3) {$g$};
    \node[] at (3,1.4) {$Z_{\widehat \CX}$};
    \node[] at (9.3,1.5) { $ \sum_{b,c}Z_{\widehat \CX}[b,c] e^{i \pi \int_{X_2^{\geq 0}} b c }$};
    \node[below] at (6,-0.5) {$x=0$};
    \node[right] at (12.5,0) {$x$};

    \end{tikzpicture}
    \caption{The $g\tau\sigma\tau$ defect is obtained by placing $\Z_2^A$ defect $g$ at $M_1|_0$ and performing $\tau\sigma\tau$ on half of the spacetime $X_2^{\geq 0}$. }
    \label{fig:gTST}
\end{figure}

Let us construct the non-invertible operator implementing $g\tau\sigma\tau$ explicitly. We start by dividing the spacetime into two parts, as shown in Figure \ref{fig:X2}. The left part $X_2^{< 0}$ and right part $X_2^{> 0}$ share a common boundary $M_1|_0$. The subscript/superscript indicates the value of the local coordinate around the neck.  The defect implementing $g\tau\sigma\tau$ is obtained by placing the $\Z_2^A$ defect $g(M_1|_0)$ at $M_1|_0$ and further acting by $\tau\sigma\tau$ on only half of the spacetime, say $X_2^{> 0}$, with Dirichlet boundary conditions at $M_1|_0$ as shown in Figure \ref{fig:gTST}. For simplicity, we have turned off the background fields $B,C$. Note that the $g(M_1|_0)$ defect alone is not gauge invariant, but rather only the combination 
\begin{eqnarray}
g(M_1|_0)\cdot \exp\left( i\pi \int_{X_2^{\geq 0}} bc\right)~
\end{eqnarray}
is.
 To interpret this as a line operator, we introduce a 1d TQFT  supported on $X_2^{\geq 0}$ to cancel the bulk $X_2^{\geq 0}$ dependence. Supposing that the gauge transformations are $b\to b+ \delta \beta$ and $c\to c+ \delta \gamma$, the 1d TQFT can be chosen to be $i \pi (- \phi_2 c + \phi_3 b + \phi_2 \delta \phi_3)$, where $\phi_{2,3}$ are scalars supported only on the line $\gamma|_0$ with the gauge transformation  $\phi_{2}\to \phi_2 + \beta$ and $\phi_3\to \phi_3+ \gamma$. Thus the combination
\begin{equation}\label{eq:noninvdefect2d}
\CN_A(M_1|_0)=\frac{1}{|C^0(M_1|_0, \Z_2)|} \sum_{\phi_2, \phi_3\in C^0(M_1|_0, \Z_2)}g(M_1|_0)\cdot \exp\left( i \pi \int_{M_1|_0} - \phi_2 c + \phi_3 b + \phi_2 \delta \phi_3\right) 
\end{equation}
is a genuine topological line operator, generating the transformation $g\tau\sigma\tau$. Following the prescription in \cite{Choi:2021kmx, Choi:2022zal} (and  treated more carefully in \cite{Kaidi:2022cpf}) we find the fusion rules
\begin{eqnarray}\label{eq:fusion2d}
\begin{split}
    \CN_A \times \CN_A&= 1+ U_b + U_c + U_b U_c~,\\
    \CN_A \times U_b &=   U_b \times \CN_A=  \CN_A~,\\
    \CN_A \times U_c &= U_c \times \CN_A=  \CN_A~,\\
    U_b \times U_b &= U_c \times U_c = 1~,
\end{split}
\end{eqnarray}
where $U_{a,b,c}=e^{i\pi \oint_{M_1|_0} a,b,c}$ are the generators for the quantum $\Z_2^A\times \Z_2^B\times \Z_2^C$ symmetry. 
These fusion rules coincide with those of the  $\Z_2\times \Z_2$ Tambara-Yamagami fusion category. Note that the discussion in this section does not depend on the existence of the self-anomaly of $\Z_2^A$ (i.e. the term $\pi A \beta A$) because we turned off its background field $A$ throughout.

\subsection{Non-invertible defects from SymTFT}
\label{sec:defectsfromSymTFT}

We next review how non-invertible symmetry defects are realized in the SymTFT. 
Because the theory $\widehat{\CX}$ is obtained by gauging a non-anomalous finite symmetry in $\CX$,  the SymTFT is the same for both $\CX$ and $\widehat{\CX}$. Hence the SymTFT is still given by the DW theory in \eqref{eq:2dSymTFT}; the only difference is the topological boundary condition on the left. For $\CX$, the relevant topological boundary condition was the Dirichlet boundary condition such that all of $U_a, U_b, U_c$ could terminate. For $\widehat{\CX}$, the relevant boundary condition is given by Neumann boundary conditions for $b$ and $c$ and Dirichlet boundary conditions for $a$, or equivalently the Dirichlet boundary condition for $a, \widehat{b}$, and $\widehat{c}$.  In terms of boundary states, this is
\begin{eqnarray}\label{eq:newbdy}
\bra{D_{a}, N_{b,c}} = \sum_{a,b,c\in C^1(X_2, \Z_2)} \bra{a,b,c} \delta(a-A) \,e^{i \pi \int_{X_2} bC + cB} ~. 
\end{eqnarray}
Under this new boundary condition, the lines $U_a, \widehat{U}_b, \widehat{U}_c$, and the composite $\widehat{U}_{bc}$ can terminate.\footnote{To see this, we note that the Dirichlet boundary condition for $a$ implies $\phi_1=0$ on the boundary. Hence all three terms in the 1d TQFT of $\widehat{U}_b$ and $\widehat{U}_c$ are trivial. Together with the Dirichlet boundary condition for $\widehat{b}, \widehat{c}$, we conclude that $\widehat{U}_b$, $\widehat{U}_c$ are terminable on the new boundary. } We thus have $\CA_{\text{noninv}}= 1\oplus U_a \oplus \widehat{U}_b \oplus \widehat{U}_c \oplus \widehat{U}_{bc}$. Indeed, the total quantum dimension of $\CA_{\text{noninv}}$ is $1+1+2+2+2=8$, matching the total quantum dimension $D_{\text{tot}}$ as required for a Lagrangian algebra. As another consistency check, all the operators in $\CA_{\text{noninv}}$ have trivial linking invariants in the bulk.

There are 5 lines belonging to $\CA_{\text{noninv}}$, and the remaining 17 lines belong to the complement of $\CA_{\text{noninv}}$. This includes 6 invertible lines and 11 non-invertible lines. 
We now discuss how these lines connect with boundary lines. As usual, it is useful to organize the bulk lines into subsets labelled by the boundary lines they can end on. We first list the results,
\begin{eqnarray}\label{eq:bdyop}
    \begin{tabular}{|c|c|c|}
    \hline
       subset  & bulk line &  boundary line\\
       \hline
       $\CB_b$  & $U_b, U_{ab}, \widehat{U}_{b} U_b, \widehat{U}_{bc} U_b$ & $U_b|_{\partial}$ \\
       $\CB_c$ & $U_c, U_{ac}, \widehat{U}_{c} U_c, \widehat{U}_{bc} U_c$ & $U_c|_{\partial}$\\
       $\CB_{bc}$ & $U_{bc}, U_{abc}$ & $U_{bc}|_{\partial}$\\
       $\CB_{\widehat{a}}$ & $\widehat{U}_a, \widehat{U}_{ab}, \widehat{U}_{ac}, \widehat{U}_{abc}, \widehat{U}_a U_a, \widehat{U}_{ab} U_a, \widehat{U}_{ac} U_a, \widehat{U}_{abc} U_a$ & $\widehat{U}_{a}|_{\partial}$\\
       \hline
    \end{tabular}
\end{eqnarray}
We now explain the above results by means of a few representative examples. First consider the bulk line $U_b$. Since $b$ obeys Neumann boundary conditions, it cannot terminate on the topological boundary. Instead, when placed orthogonal to the boundary as shown in Figure \ref{fig:oper}, it bends at the junction and becomes a boundary line $U_b|_{\partial}$, which is the 
 $\Z_2^B$ symmetry defect of the 2d QFT.

A slightly more non-trivial example is the bulk line $\widehat{U}_{a}$. Because $b$ and $c$ obey Neumann boundary conditions,  the 1d TQFT survives in the boundary line $\widehat{U}_a|_{\partial}$. Hence it becomes a non-invertible symmetry defect of the 2d QFT $\widehat{\CX}$. Note that the 1d TQFT is precisely the one appearing in the definition of the non-invertible defect $\CN_A$, c.f. \eqref{eq:noninvdefect2d}.

Here, we also see an example where the same bulk line can terminate on different boundary lines. To see this, we consider the bulk line $\widehat{U}_{bc}U_b$. Since $\widehat{U}_{bc}$ is terminable, only $U_b|_{\partial}$ survives on the boundary. On the other hand, the bulk line $\widehat{U}_{bc}U_c$ is attached to $U_c|_{\partial}$ on the boundary. Note that the two bulk lines $\widehat{U}_{bc}U_b$ and $\widehat{U}_{bc}U_c$ are actually the same on a closed circle, and hence the same bulk line can terminate on two different boundary lines. 

We find that there are five line operators, $1, U_b|_{\partial}, U_c|_{\partial}, U_{bc}|_{\partial}$, and $\widehat{U}_{a}|_{\partial}$ 
on the boundary. These operators are in obvious one-to-one correspondence with the symmetry operators in Section \ref{sec:noninvsymmfromgauging}; concretely, we have the map
\begin{eqnarray}
    \begin{tabular}{|c|c|}
    \hline
        operators in 2d QFT $\widehat{\CX}$ in Section \ref{sec:noninvsymmfromgauging} & operators on the top.bdy. in  \eqref{eq:bdyop} \\
        \hline
        $U_b$ & $U_b|_{\partial}$\\
        $U_c$ & $U_c|_{\partial}$\\
        $U_{bc}$ & $U_{bc}|_{\partial}$\\
        $\CN_A$ & $\widehat{U}_{a}|_{\partial}$\\
        \hline
    \end{tabular}
\end{eqnarray}
The fusion rules among $U_b|_{\partial}, U_c|_{\partial}, U_{bc}|_{\partial},$ and $\widehat{U}_{a}|_{\partial}$ can also be straightforwardly computed, and coincide with those in \eqref{eq:fusion2d}.

\subsection{Anomalies of the non-invertible symmetries}
\label{sec:anomnoninvsym}

We may now finally turn to the question of anomalies of the non-invertible symmetry of $\widehat{\CX}$. Since there is currently no notion of a background field for non-invertible symmetries, the meaning of the AnomTFT is unclear.
In Section \ref{sec:AnomTFT2d}
we reviewed how the SymTFT can be used to probe the anomalies of invertible symmetries, following the general discussion in the introduction. In this section, we will use the SymTFT to probe the anomalies of the non-invertible symmetry reviewed in Section \ref{sec:noninvsymmfromgauging} and \ref{sec:defectsfromSymTFT}.

As before, the idea is to pick representatives from each subset $\CB_b, \CB_c, \CB_{bc}$, and $\CB_{\widehat{a}}$, and then to compute the linking invariants among them. Since the original, invertible symmetry $U_a|_\partial$ had a self-anomaly before gauging, it is natural to expect that the corresponding non-invertible symmetry $\CN_{A} = \widehat U_a|_\partial$ also has a self-anomaly. To detect this, we choose a representative from the subset $\CB_{\widehat{a}}$ and compute the linking number between the representative and itself. In fact, it is straightforward to check that for any choice of representative, the linking number is always non-trivial,
\begin{eqnarray}\label{eq:3dbilink2}
\begin{split}
    \braket{ \CO(M_1) \CO(M'_1) } &\sim (-1)^{\text{Link}(M_1,M_1')}~, \hspace{1cm} \forall ~~ \CO\in \CB_{\widehat{a}}~, 
\end{split}
\end{eqnarray}
which can (for instance) be detected by a Hopf link.
In particular, the self-linkings of $\CO=\widehat{U}_a$ and $\widehat{U}_a U_a$ were already computed to detect the invertible $\Z_2^A$ self-anomaly in Section \ref{sec:AnomTFT2d}. Using the main result in the introduction, we thus conclude that the non-invertible symmetry $\CN_A$ in the 2d QFT $\widehat{\CX}$ enjoys a non-trivial 't Hooft anomaly.

We give two remarks before moving on to an example:
\begin{enumerate}
    \item The merit of using linking invariants to quantify 't Hooft anomalies is that the linking invariants (especially when the SymTFT is Dijkgraaf-Witten) are relatively easy to compute, and can be easily generalized to higher dimensions.
    \item In \cite{Choi:2021kmx, Choi:2022zal,Apte:2022xtu}, a powerful sufficient condition for a non-invertible symmetry implementing $g\tau\sigma\tau$ to have non-trivial anomaly was proposed. Concretely, suppose a QFT $\widehat{\CX}$ has an invertible global symmetry $G$, and furthermore that $\widehat{\CX}$ is invariant under a sequence of topological manipulations $f(\sigma, \tau, \rho)$, where $\rho$ rescales the background fields, e.g. $B\to k B$ for certain $k$.  Then in general $\widehat{\CX}$ has a non-invertible symmetry implementing $f(\sigma, \tau, \rho)$. The authors of \cite{Choi:2021kmx, Choi:2022zal} showed that $f(\sigma, \tau, \rho)$ is anomalous if one cannot find a $G$-SPT such that it is invariant under $f(\sigma, \tau, \rho)$. However, this condition is not necessary. Namely, the existence of an $f(\sigma, \tau, \rho)$-invariant $G$-SPT does not imply that $f(\sigma, \tau, \rho)$ is anomaly-free. Indeed, in our example of $G=\Z_2\times \Z_2$, there is a 2d $\Z_2\times \Z_2$ SPT which is invariant under $g\tau\sigma\tau$,
    whose topological action is
\begin{eqnarray}\label{eq:2dSPT}
\pi \int_{X_2} BC~.
\end{eqnarray}
 Here $g$ acts trivially on \eqref{eq:2dSPT}. Hence the condition in \cite{Choi:2021kmx, Choi:2022zal,Apte:2022xtu} would not detect the anomaly, whereas our current condition does (although as emphasized in the introduction, our condition is still only sufficient, and not necessary). The non-trivial linking invariant \eqref{eq:3dbilink2} dictates additional structures for the junction between the non-invertible defects, and these are not satisfied by \eqref{eq:2dSPT} \cite{Thorngren:2019iar}.

\end{enumerate}

\subsection{Application: Abelian Higgs Model}

We close this section by giving a  concrete example of the discussion above. We will take the theory $\CX$ to be the 2d Abelian Higgs Model (AHM) with $N_f=2$ complex scalars charged under the dynamical gauge group $U(1)$, and with a  non-trivial theta term $\theta=\pi$. The action is 
\begin{eqnarray}
S_{\CX} = \int_{X_2} \sum_{i=1}^2 |D_a \phi_i|^2 + m^2 \sum_{i=1}^2 |\phi_i|^2 + \lambda \left(\sum_{i=1}^2 |\phi_i|^2 \right)^2 + \frac{1}{2e^2} da * da + \frac{\pi}{2\pi} da~. 
\end{eqnarray}
Let us discuss the symmetries and anomalies of this theory.

\paragraph{Invertible symmetries and anomalies:}

The AHM has an $SU(2)/\Z_2= SO(3)$ global symmetry; the $\Z_2$ quotient is included since the $\Z_2$ normal subgroup of $SU(2)$ actually belongs to the $U(1)$ gauge group. Below, we will focus on only the $\Z_2^{x}\times \Z_2^{y}$ subgroup of $SO(3)$. This means that we are allowed to turn on interactions that explicitly break $SO(3)$ while preserving its $\Z_2^{x}\times \Z_2^{y}$ subgroup. To show how the $\Z_2^{x}\times \Z_2^{y}$ subgroup acts on the scalars, we consider the scalar bilinears, 
\begin{eqnarray}
n^\alpha= \sum_{i,j=1}^2 \phi^\dagger_i \tau^{\alpha}_{ij} \phi_j~, \hspace{1cm} \alpha=1,2,3\,,
\end{eqnarray}
which transform as vectors under $SO(3)$ and which are $U(1)$ gauge invariant.\footnote{Here $\tau^\alpha$ are the Pauli matrices, not to be confused with the topological manipulation $\tau$ in \eqref{eq:topomani}. }
Then the $\Z_2^{x}\times \Z_2^{y}$ symmetry acts on $n^\alpha$ as 
\begin{eqnarray}
\begin{split}
    \Z_2^x:\hspace{1cm} &(n^1, n^2, n^3) \to (n^1, -n^2, -n^3)~,\\
    \Z_2^y:\hspace{1cm} &(n^1, n^2, n^3) \to (-n^1, n^2, -n^3)~.\\
\end{split}
\end{eqnarray}

Apart from the flavor rotation symmetry, there is also a charge conjugation symmetry $\Z_2^{C}$ acting on the scalars and $U(1)$ gauge field as \cite{Sulejmanpasic:2018upi,Wan:2018zql, Komargodski:2017dmc}
\begin{eqnarray}
\Z_2^C: \hspace{1cm} \phi_i\to (i \tau^2 \phi)_i~, \hspace{1cm} a\to -a~.
\end{eqnarray}
Note that although $\Z_2^C$ acts on the scalar as $C^2 \phi_i = - \phi_i$, and hence $C^2=(-1)^{2j}$ where $j$ is the $SU(2)$ isospin, all the gauge invariant local operators have integer $SU(2)$ isospin, and therefore $C^2=1$ on all gauge invariant operators. The $\Z_2^C$ symmetry acts on the $SO(3)$ vector $n^\alpha$ as 
\begin{eqnarray}
\Z_2^C: \hspace{1cm} (n^1, n^2, n^3) \to (-n^1, n^2, -n^3)~,
\end{eqnarray}
which coincides with the action of $\Z_2^y$ on the scalar bilinears. However, note that $\Z_2^C$ additionally acts on the $U(1)$ gauge field.

As we have mentioned, we will allow ourselves to add interactions to the action such as
\begin{eqnarray}\label{eq:SO3toZ2Z2}
V=\int_{X_3} \xi_1 (n^1)^2 + \xi_2 (n^2)^2 + \xi_3 (n^3)^2 + \xi_{123} n^1 n^2 n^3 + \dots~
\end{eqnarray}
which preserve the $\Z_2^C\times \Z_2^x\times \Z_2^y$ symmetry while explicitly breaking the $SO(3)$ symmetry.

The anomaly of the $\Z_2^C\times \Z_2^x\times \Z_2^y$ symmetry of the AHM has been studied in detail in \cite{Metlitski:2017fmd, Komargodski:2017dmc, Wan:2018zql}. In the presence of the $SO(3)$ symmetry, the anomaly is found to be
\begin{eqnarray}\label{eq:SO3Z2Z2anom}
\int_{X_3} \pi C w_2^{SO(3)} + \pi C \beta C\,,
\end{eqnarray}
where $C$ is the background field of $\Z_2^C$. 
The mixed anomaly $\pi C w_2^{SO(3)}$ is interpreted as a Lieb-Schultz-Mattis (LSM) type anomaly, since $\Z_2^C$ can be interpreted as the $\Z_2$ reduction of the translation symmetry when the AHM is realized as the low-energy description of an antiferromagnetic spin chain in the UV. The self-anomaly $\pi C \beta C$ was referred to as an ``emergent anomaly'' in \cite{Metlitski:2017fmd}, 
since it is beyond LSM-type anomalies; indeed, the translation symmetry $\Z$ of the UV spin chain clearly does not have an anomaly, and the anomaly only appears when $\Z\to \Z_2$ at low energies.\footnote{To see that translation itself does not have a self-anomaly, we consider deforming the antiferromagnetic spin chain Hamiltonian $H_{\text{AFM}}= \sum_{i} \sigma^x_{i}\sigma^x_{i+1}+ \sigma^y_{i}\sigma^y_{i+1}+\sigma^z_{i}\sigma^z_{i+1}$  by a transverse field term $-h \sum_{i} \sigma^x_{i}$ with large $h$. This term preserves translation symmetry and drives the theory to a trivially gapped phase. } 
We should note that recently the $\Z_2^C$ in the low-energy theory has been dubbed an \emph{emanant symmetry} (to be distinguished from an emergent symmetry) since it descends from a UV symmetry \cite{Cheng:2022sgb}; the UV spin-chain manifestation of this self-anomaly was identified there as well.\footnote{We thank Shu-Heng Shao for discussions on this point.}   
When we explicitly break the $SO(3)$ symmetry of the AHM by adding the potential \eqref{eq:SO3toZ2Z2}, the anomaly \eqref{eq:SO3Z2Z2anom} reduces to 
\begin{eqnarray}\label{eq:Z2Z2Z2anom}
\int_{X_3}\pi C A^x A^y + \pi C \beta C
~,\end{eqnarray}
where $A^x$ and $A^y$ are the background fields of $\Z_2^x$ and $\Z_2^y$ respectively. 
This anomaly is precisely the one in \eqref{eq:invTQFT} which we have studied extensively throughout this section.

\paragraph{Dynamical constraints on the $\Z_2^x\times \Z_2^y$ gauged AHM:}

Because our discussions in  Sections \ref{sec:AnomTFT2d} through \ref{sec:anomnoninvsym} are based purely on the symmetries and anomalies of the theory, we may immediately apply our results to the AHM. In particular, if we gauge the $\Z_2^x\times \Z_2^y$ symmetry of the perturbed AHM (whose action is $S_{\CX} + V$) the resulting theory will have a non-invertible charge conjugation symmetry. In addition, from the results of Section \ref{sec:anomnoninvsym}, this non-invertible charge conjugation symmetry is anomalous, and measurable via the linking number between two non-invertible charge conjugation defects in the SymTFT.

The presence of the anomalous non-invertible symmetry has the following dynamical consequences: 
\begin{enumerate}
    \item Firstly, independent of the anomaly, the non-invertible symmetry itself forbids some terms that might otherwise have been radiatively generated. If one were not aware of the existence of the non-invertible symmetry, then after gauging $\Z_2^x\times \Z_2^y$ 
    one would naively conclude that $\Z_2^C$ was explicitly broken, and that there was no obstruction for  terms such as $\eta_1 n^1+ \eta_3 n^3 + ...$ charged under $\Z_2^C$ to be radiatively generated. In other words, not including such terms in the action would be ``unnatural.'' 
    However, when the non-invertible symmetry is accounted for, such operators are not allowed, and naturalness is restored.
    \item Secondly, the presence of the anomaly forbids the system from flowing to a trivially gapped phase, no matter what perturbation is turned on (as long as it is uncharged under the non-invertible symmetry, and hence under $\Z_2^C$ before gauging). 
\end{enumerate}

\section{Non-invertible symmetries and anomalies in 4d}
\label{sec:4d}

We now repeat the discussion of the previous section in a very similar four-dimensional setup. Our starting point will be a 4d spin quantum field theory $\CX$ with an invertible $\Z_{2MN}^{(0)}\times \Z_N^{(1)}$ global symmetry, where the superscripts indicate that they are respectively zero- and one-form symmetries. We will denote the background fields as $A^{(1)}$ and $B^{(2)}$ respectively. We further assume that the theory has a 't Hooft anomaly specified by the 5d AnomTFT, 
\begin{eqnarray}\label{eq:AnomTFT5d}
\begin{split}
    \int_{X_5} \left(\frac{2\pi}{2N} A^{(1)} \CP(B^{(2)}) + \frac{\pi(N^2-1)}{6N} A^{(1)} \beta A^{(1)} \beta A^{(1)}\right)~,& \hspace{1cm} N\in 2\Z\\
    \int_{X_5} \left(\frac{2\pi}{N}  \frac{1+N}{2}A^{(1)} B^{(2)}B^{(2)} + \frac{\pi(N^2-1)}{6N} A^{(1)} \beta A^{(1)} \beta A^{(1)}\right)~,& \hspace{1cm} N\in 2\Z+1\\
\end{split}
\end{eqnarray}
which is an invertible TQFT. Here $\cP$ is the Pontryagin square operation, while $\beta$ is the Bockstein map associated with the exact sequence $1\to \Z_{2MN} \to \Z_{(2MN)^2} \to \Z_{2MN}\to 1$, acting on $\Z_{2MN}$-valued cocycles as $\beta=\delta/(2MN)$. The coefficients are chosen to match with our two examples: (1) adjoint QCD and (2) $\CN=4$ SYM. The zero-form symmetry being  $\Z_{2MN}$ is a simplifying assumption, and one can in principle discuss more general symmetries and anomalies. 

In order to obtain a theory with non-invertible symmetry, we now gauge the $\Z_N^{(1)}$ one-form symmetry, which is possible since it does not have a self-anomaly. This gauging changes $\ZZ_{2MN}^{(0)}$ into a non-invertible symmetry. As in the two-dimensional case studied in the previous section, the self-anomaly of $\ZZ_{2MN}^{(0)}$ before gauging will lead to an anomaly for the non-invertible symmetry after gauging. Our analysis will again proceed by studying the structure of the SymTFT.

\subsection{SymTFT of invertible symmetries}

\paragraph{The SymTFT:}
Let us begin by discussing the SymTFT for the theory $\CX$ with invertible symmetry. As reviewed in the introduction, a 4d QFT can be expanded into a 5d slab in which the SymTFT lives. The SymTFT is a gauged version of the AnomTFT given in 
\eqref{eq:AnomTFT5d}, i.e. it is a DW TQFT. The SymTFT for even $N$ is  
\begin{eqnarray}\label{eq:5dSymTFT}
\int_{X_5} \left( \frac{2\pi}{2MN} \widehat{a}^{(3)} \delta a^{(1)} + \frac{2\pi}{N} \widehat{b}^{(2)}\delta b^{(2)} + \frac{2\pi}{2N} a^{(1)} b^{(2)} b^{(2)} + \frac{\pi(N^2-1)}{6 N} a^{(1)} \frac{\delta a^{(1)}}{2 MN } \frac{\delta a^{(1)}}{2 MN}\right)~.
\end{eqnarray}
Here all the fields $a^{(1)}, \widehat{a}^{(3)}, b^{(2)}$, and $\widehat{b}^{(2)}$ are dynamical, with the superscripts indicating their form degrees. Note that we suppressed the higher cup products in the SymTFT.\footnote{Throughout this work we suppress $\cup_1$ terms and ignore the fact that cochains are not super-commutative. To justify our treatment here, one may replace all cochains by differential forms by adjusting their normalizations, e.g. treating $\frac{2\pi}{4}a^{(1)}$ as a $2\pi$-periodic one-form. In the differential form formalism, all fields are super-commutative. }  For odd $N$, the third term in \eqref{eq:5dSymTFT} should be replaced by $\frac{2\pi}{N} \frac{1+N}{2} a^{(1)} b^{(2)} b^{(2)} $. As the discussions of even and odd $N$ are similar, we will focus on the case of even $N$ below.

The spectrum of topological operators in the DW theory \eqref{eq:5dSymTFT} is studied in Appendix \ref{app:5d}, and we simply summarize the results here. First there are Wilson lines and Wilson surfaces, 
\begin{eqnarray}
    U_a(M_1)= e^{\frac{2\pi i}{2 MN }\oint_{M_1} a^{(1)}}~, \hspace{1cm} U_{b}(M_2) = e^{ \frac{2\pi i}{N} \oint_{M_2} b^{(2)}}~, 
\end{eqnarray}
which satisfy the obvious invertible fusion rules $U_a^{2MN}=1$ and $U_b^{N}=1$. There are also non-invertible magnetic surface and 3-volume operators, 
\begin{eqnarray}\label{eq:magneticop5d0}
\begin{split}
    \widehat{U}_{a}(M_3) &\sim \sum_{\phi^{(1)}\in C^1(M_3, \Z_{N})} e^{\frac{2\pi i}{2 MN }\oint_{M_3} \left( \widehat{a}^{(3)} - M\phi^{(1)} \delta \phi^{(1)} + 2 M\phi^{(1)} b^{(2)} \right)}~,\\
    \widehat{U}_{b}(M_2) &\sim \sum_{\substack{\phi^{(0)}\in C^0(M_2, \Z_{2 MN})\\ \phi^{(1)}\in C^1(M_2, \Z_{N})}} e^{{2\pi i \over N}\oint_{M_2} \left( \widehat{b}^{(2)} - \phi^{(0)}b^{(2)} -  \phi^{(1)} a^{(1)} + \phi^{(1)} \delta \phi^{(0)} \right)}~,\\
\end{split}
\end{eqnarray}
where the TQFTs attached to the naive magnetic operators $e^{\frac{2\pi i}{2 MN }\oint_{M_3} \widehat{a}^{(3)}}$ and $e^{\frac{2\pi i}{N}\oint_{M_2} \widehat{b}^{(2)}}$ are needed to cancel the gauge non-invariance under $a^{(1)}\to a^{(1)}+\delta \alpha^{(0)}$ and $b^{(2)}\to b^{(2)}+\delta \beta^{(1)}$. 
The fields $\phi^{(0)}$ and $\phi^{(1)}$ live only on the worldvolumes of the topological operators and transform as $\phi^{(0)}\to \phi^{(0)}+ \alpha^{(0)}$ and $\phi^{(1)}\to \phi^{(1)}+ \beta^{(1)}$. The TQFTs on the operator worldvolumes render the operators non-invertible. The non-invertibility can also be seen from the fusion rules 
\begin{eqnarray}
\begin{split}
    \widehat{U}_a(M_3)\times \overline{\widehat{U}}_a(M_3) &\sim \sum_{\phi^{(1)}\in H^1(M_3, \Z_N)} e^{{2\pi i \over N}Q(\text{PD}(\phi^{(1)}))} e^{{2 \pi i \over N}\oint_{\text{PD}(\phi^{(1)})} b^{(2)}}~,\\
    \widehat{U}_b(M_2)\times \overline{\widehat{U}}_b(M_2)  &\sim \sum_{\substack{\phi^{(0)}\in H^0(M_2, \Z_{2MN})\\ \phi^{(1)}\in H^1(M_2, \Z_{N})}} e^{{2\pi i \over N} \oint_{\text{PD}(\phi^{(0)})} b^{(2)} + {2\pi i \over N} \oint_{\text{PD}(\phi^{(1)})} a^{(1)}}~,
\end{split}
\end{eqnarray}
where $Q(\text{PD}(\phi^{(1)})):= \frac{1}{N}\int_{M_3} \phi^{(1)} \delta \phi^{(1)}$ is the triple intersection number in $M_3$, and $\text{PD}(\phi)$ is the Poincar{\'e} dual of $\phi$.

\paragraph{Dirichlet boundary condition:}
To obtain the theory $\CX$, the topological boundary of the SymTFT should be the Dirichlet boundary condition for both $a^{(1)}$ and $b^{(2)}$. This sets the dynamical fields $a^{(1)}$ and $b^{(2)}$ to background fields $A^{(1)}$ and $B^{(2)}$, respectively. The corresponding boundary state is 
\begin{eqnarray}\label{eq:4dbra}
    \bra{D_{a^{(1)}, b^{(2)}}} = \sum_{\substack{a^{(1)}\in C^1(X_4, \Z_{2MN})\\ b^{(2)}\in C^2(X_4, \Z_{N})}} \bra{a^{(1)}, b^{(2)}} \,\delta(a^{(1)}-A^{(1)}) \delta(b^{(2)}- B^{(2)})~.
\end{eqnarray}
This Dirichlet boundary condition means that all the invertible line and surface operators in the bulk become trivial on the boundary, and can also terminate perpendicularly on the boundary. The set of all invertible operators is denoted by $\CA_{\text{inv}}$. To see how the boundary condition gives rise to invertible $\Z_{2MN}^{(0)}\times \Z_N^{(1)}$ symmetries, we consider placing the bulk operators orthogonal to the boundary as shown in Figure \ref{fig:oper}. Since both $a^{(1)}$ and $b^{(2)}$ become trivial on the boundary, the TQFTs on both operator volumes are trivialized, and consequently the boundary operators are invertible, i.e. 
\begin{eqnarray}\label{eq:bdyoperators5d}
    \widehat{U}_{a}(M_3|_\partial) = e^{\frac{2\pi i}{2MN}\oint_{M_3|_{\partial}} \widehat{a}^{(3)} }~, \hspace{1cm} \widehat{U}_{b}(M_2|_\partial) = e^{\frac{2\pi i}{N}\oint_{M_2|_{\partial}} \widehat{b}^{(2)} }~.
\end{eqnarray}
We denote the collection of the above boundary operators as well as the condensation defects constructed out of them as $\CC_{\text{inv}}$.\footnote{Although $\widehat{U}_{a}(M_3|_\partial)$ and $\widehat{U}_{b}(M_3|_\partial)$ are both invertible operators, the condensation defects constructed from them can be non-invertible. The subscript on $\cC_\text{inv}$ means that the  simple defects (which are not the sum of other defects of the same dimension) in $\cC_\text{inv}$  that are not condensation defects are all invertible. } 
Since $\widehat{U}_a$ and $\widehat{U}_b$ are 
respectively codimension-one and -two operators on the 4d boundary, they generate a $\Z_{2MN}^{(0)}\times \Z_N^{(1)}$ invertible symmetry after shrinking the slab.

\paragraph{Dynamical boundary condition:} 
The dynamical boundary condition on the right captures the dynamics of the 4d theory, and in particular is non-topological. The corresponding boundary state is given by
\begin{eqnarray}\label{eq:4dket}
    \ket{\CX}= \sum_{\substack{a^{(1)}\in C^1(X_4, \Z_{2MN})\\ b^{(2)}\in C^2(X_4, \Z_{N})}} Z_{\CX}[X_4; a^{(1)}, b^{(2)}] \ket{a^{(1)}, b^{(2)}}~.
\end{eqnarray}
Shrinking the 5d slab and taking the inner product between the two boundary states \eqref{eq:4dbra} and \eqref{eq:4dket} reproduces the 4d partition function $Z_{\CX}[X_4; A^{(1)}, B^{(2)}]$.

\subsection{Anomaly of invertible symmetry from linking invariants}

Proceeding in parallel with Section \ref{sec:Noninvsym2d}, we now review how to use the linking invariants to diagnose the anomaly of the invertible $\Z_{2MN}^{(0)}\times \Z_N^{(1)}$ symmetry. We begin by grouping the bulk defects into subsets, such that all defects in a given subset can end on the same boundary operator. For convenience, we denote the condensation of an operator $U$ on the manifold $\Sigma$ as $\text{Cond}_{\Sigma}(U)$. We will suppress the $\Sigma$ when the manifold dependence is not emphasized.  Then the operators in the SymTFT can be organized into the following subsets
\begin{eqnarray}\label{eq:equiv5d}
    \CB_{\widehat{a}} = \{ \widehat{U}_a,\, \widehat{U}_{a} \text{Cond}(U_{a}),\, \widehat{U}_{a} \text{Cond}(U_{b}),\, ...\}~, \hspace{1cm} \CB_{\widehat{b}} = \{  \widehat{U}_{b} U_b^p,\, \widehat{U}_{b} \text{Cond}(U_{a}),\, ...\}~,
\end{eqnarray}
where the $...$ represents stacking $\widehat{U}_{a,b}$ with other condensation operators---with or without discrete torsions---and $p=0, ..., N-1$.
For simplicity we will not discuss the condensation defects of non-invertible operators, and consequently will not discuss the condensation defects of boundary operators \eqref{eq:bdyoperators5d}. Following the discussion in the introduction, in order to probe the anomaly, we must check whether it is possible to choose one representative from each subset such that all of the linking invariants are trivial. If such a choice of representatives does not exist, then the anomaly for $\CC_{\text{inv}}$ is non-trivial.

For simplicity, we will consider only operators of spherical topology, which means that all the condensation operators in each subset are trivialized. Hence the two subsets in \eqref{eq:equiv5d} simplify significantly, and we have $\CB_{\widehat{a}} = \{ \widehat{U}_a\}$ and $\CB_{\widehat{b}} = \{ \widehat{U}_b U_b^p,\, p=0, ..., N-1\}$. We first check the linking invariants among two components. As discussed in Appendix \ref{app:Link}, the only possible linking invariants are between two surface operators. It is easy to check that there exists a representative $\widehat{U}_b$ such that the correlation function $\braket{\widehat{U}_b(M_2) \widehat{U}_b(M_2')}$ has a trivial phase when both $M_2$ and $M_2'$ are $S^2$. This is consistent with the fact that there is no anomaly of the form $B^{(2)}\beta B^{(2)}$.

We next consider the linking invariants among three components. There are three types of such linking invariants referred to as type 0, type 1, and type 2 in Appendix \ref{app:Link}, and we will find that the type 0 and type 2 invariants are non-trivial in the present case. 
The type 0 linking involves one 3-volume operator and two surface operators, and we may compute the following correlation function
\begin{eqnarray}
    \braket{\widehat{U}_a(M_3) \widehat{U}_b(M_2') U^p_b(M_2')\widehat{U}_b(M_2'')U^p_b(M_2'')}\sim  e^{-\frac{2\pi i}{N} \text{Link}(M_3, M'_2, M''_2)_{0}}\,,
\end{eqnarray}
where we have assumed that $M_2'$ and $M_2''$ are not Hopf linked. We see that the right-hand side is nontrivial, and is independent of the choice of representative. This non-trivial type 0 linking invariant captures the $A^{(1)}B^{(2)}B^{(2)}$ mixed anomaly in \eqref{eq:AnomTFT5d}.

On the other hand, the type 2 linking involves three 2-volume operators, and hence there is only one choice of correlation function, 
\begin{eqnarray}\label{eq:trip2main}
    \braket{\widehat{U}_a(M_3) \widehat{U}_a (M'_3) \widehat{U}_a(M''_3)} \sim e^{-{ i \pi (N^2 - 1) \over 4 N^3 M^2}\text{Link}(M_3, M'_3, M''_3)_{2}}~.
\end{eqnarray}
This captures the non-trivial $A^{(1)}\beta A^{(1)}\beta A^{(1)}$ anomaly in \eqref{eq:AnomTFT5d}.

\subsection{Non-invertible symmetries from gauging}
\label{sec:4dNonsymfromgauging}

We have just seen how the non-trivial 't Hooft anomalies of invertible symmetries can be detected via the linkings of operators in the SymTFT. We now turn  to the case of non-invertible symmetries. We consider non-invertible symmetries which are obtained by gauging the $\Z_N^{(1)}$ one-form symmetry of the QFT $\CX$. Let us denote the  $\Z_N^{(1)}$-gauged QFT by $\widehat{\CX}$, with partition function given by 
\begin{eqnarray}
    Z_{\widehat{\CX}}[B^{(2)}] = \frac{|H^0(X_4, \Z_N)|}{|H^1(X_4, \Z_N)|} \sum_{b^{(2)}\in H^2(X_4, \Z_N)} Z_{\CX}[b^{(2)}] e^{\frac{2\pi i}{N}\int_{X_4} b^{(2)}B^{(2)}}~.
\end{eqnarray}
Here we have turned off the $\Z_{2MN}^{(0)}$ background field in the partition function. The fact that $\widehat{\CX}$ has a non-invertible symmetry has been discussed in \cite{Kaidi:2021xfk, Choi:2021kmx,Bhardwaj:2022yxj,Kaidi:2022cpf}, and we now briefly review this fact here.

We first define two topological manipulations: gauging $\Z_N^{(1)}$ and stacking with a $\Z_N^{(1)}$ SPT, denoted by $\sigma$ and $\tau$ respectively. Concretely, we have
\begin{eqnarray}
\begin{split}
    Z_{\sigma\CX}[B^{(2)}] &= \frac{|H^0(X_4, \Z_N)|}{|H^1(X_4, \Z_N)|} \sum_{b^{(2)}\in H^2(X_4, \Z_N)} Z_{\CX}[b^{(2)}] e^{\frac{2\pi i}{N}\int_{X_4} b^{(2)}B^{(2)}}~,\\
    Z_{\tau\CX}[B^{(2)}] &= 
    \begin{cases}
        Z_{\CX}[B^{(2)}] e^{\frac{2\pi i}{2N}\int_{X_4} \CP(B^{(2)})}~, & N\in 2\Z\\
        Z_{\CX}[B^{(2)}] e^{\frac{2\pi i}{N}\frac{1+N}{2}\int_{X_4} B^{(2)}B^{(2)}}~, & N\in 2\Z+1\\
    \end{cases}~.
\end{split}
\end{eqnarray}
To see the non-invertible symmetry of $\widehat{\CX}$, we note that it is invariant under $gC\tau\sigma\tau$, where $g$ is a global $\Z_{2MN}^{(0)}$ transformation and $C$ is the charge conjugation operation. To see this concretely, for $N\in 2\Z$ we have
\begin{eqnarray}
\begin{split}
    Z_{gC\tau\sigma\tau\widehat{\CX}}[B^{(2)}]& = \frac{|H^0(X_4, \Z_N)|^2}{|H^1(X_4, \Z_N)|^2}\sum_{b^{(2)}, \widetilde{b}^{(2)}} Z_{\CX}[b^{(2)}] e^{\frac{2\pi i}{N} \int_{X_4} b^{(2)} \widetilde{b}^{(2)} + \frac{\CP(\widetilde{b}^{(2)})}{2}- \widetilde{b}^{(2)} B^{(2)} + \frac{\CP(B^{(2)})}{2} +\frac{\CP(b^{(2)})}{2} } \\
    &=\chi^{\frac{1}{2}}[X_4]\, \frac{|H^0(X_4, \Z_N)|}{|H^1(X_4, \Z_N)|} \sum_{b^{(2)}} Z_{\CX}[b^{(2)}] e^{\frac{2\pi i}{N} \int_{X_4} b^{(2)} B^{(2)}} = \chi^{\frac{1}{2}}[X_4] \,Z_{\widehat{\CX}}[B^{(2)}]~,
\end{split}
\end{eqnarray}
where the Euler counterterm is given by $\chi[X_4]:= \frac{|H^0(X_4, \Z_N)|^2 |H^2(X_4, \Z_N)|}{|H^1(X_4, \Z_N)|^2}$. 
In short, we have 
\begin{eqnarray}
    gC\tau\sigma\tau \widehat{\CX}= \widehat{\CX}~.
\end{eqnarray}
The defect implementing $gC\tau\sigma\tau$ can be constructed from half-space gauging, following \cite{Choi:2021kmx,Kaidi:2022cpf}. Another equivalent construction is to start with the 3d $\Z_{2MN}^{(0)}$ defect, and decorate on the defect worldvolume some TQFT to cancel the worldvolumn $\Z_N^{(1)}$ anomaly \cite{Kaidi:2021xfk}. Both methods give rise to the following non-invertible 3-volume defect,
\begin{eqnarray}\label{eq:4dnondefect}
   \CN_{A}(M_3)\sim \sum_{\phi^{(1)}\in C^1(M_3, \Z_N)} g(M_3) \cdot \exp\left( \frac{2\pi i}{2N} \int_{M_3} - \phi^{(1)} \delta \phi^{(1)} + 2 \phi^{(1)} b^{(2)}\right)\,,
\end{eqnarray}
where the gauge transformation of $\phi^{(1)}$ is $\phi^{(1)}\to \phi^{(1)}+ \beta^{(1)}$. The fusion rules are 
\begin{eqnarray}
    \begin{split}
        \CN_A(M_3)\times \overline{\CN}_A(M_3) &\sim \sum_{M_2\in H_2(M_3, \Z_N)} (-1)^{Q(M_2)} U_b(M_2)~, \\
        \CN_A(M_3)\times U_{b}(M_2)&= \CN_A(M_3)~,
    \end{split}
\end{eqnarray}
where $U_b(M_2)=e^{\frac{2\pi i}{N}\int_{M_2}b^{(2)}}$ is the defect for the quantum $\Z_N^{(1)}$ symmetry in $\widehat{\CX}$, $Q(M_2):= \frac{1}{N}\int_{M_3} \text{PD}(M_2) \delta \text{PD}(M_2)$ is the triple intersection number in $M_3$, and $\text{PD}(M_2)$ is the Poincar{\'e} dual of $M_2$.

We conclude that $\widehat{\CX}$ has a non-invertible symmetry whose defects satisfy non-invertible fusion rules. This symmetry follows from $\Z_{2MN}^{(0)}$ in $\CX$. The above derivation is insensitive to the presence of the self-anomaly $A^{(1)}\beta A^{(1)} \beta A^{(1)}$, as we turned off the background field $A^{(1)}$ throughout.

\subsection{Non-invertible defects from SymTFT}

We now review how the non-invertible symmetry defects are realized in the SymTFT. Gauging only changes the topological boundary condition,  and the SymTFT of $\widehat{\CX}$ remains the same as $\CX$, i.e. it is still given by \eqref{eq:5dSymTFT}. For $\CX$, we saw that the topological boundary condition was the Dirichlet boundary condition for $a^{(1)}$ and $b^{(2)}$. After gauging $\Z_N^{(1)}$, the relevant topological boundary condition becomes the Neumann boundary condition for $b^{(2)}$ and the Dirichlet boundary condition for $a^{(1)}$, or equivalently the Dirichlet boundary condition for both $a^{(1)}$ and $\widehat{b}^{(2)}$. In terms of boundary states, this is
\begin{eqnarray}
    \bra{D_{a^{(1)}}, N_{b^{(2)}}} = \sum_{a^{(1)}, b^{(2)}} \bra{a^{(1)}, b^{(2)}}\, \delta(a^{(1)}-A^{(1)}) e^{\frac{2\pi i}{N}\int_{X_4} b^{(2)} B^{(2)}}~.
\end{eqnarray}

What bulk operators are terminable on the new topological boundary? Clearly the topological line $U_a$ can still terminate. Moreover, due to the Dirichlet boundary condition for $\widehat{b}^{(2)}$, the non-invertible surface $\widehat{U}_{b}$ can terminate as well. Indeed, since $a|_{\partial}=0$ we have $\phi^{(0)}|_{\partial}=0$,  and the TQFT on its worldvolume is trivialized. We denote the collection of bulk operators that are terminable as $\CA_{\text{noninv}}=\{U_a, \widehat{U}_b, ...\}$, where $...$ represents the operators constructed from $U_a, \widehat{U}_b$ such as their condensation defects.

By definition, all other bulk operators besides those in $\CA_{\text{noninv}}$ should be attached to boundary operators when intersecting the new topological boundary. Under this new boundary condition, the topological surface $U_b$ would be attached to  a boundary surface operator $U_b|_{\partial}$. The topological 3-volume operator $\widehat{U}_a$ remains non-invertible when moved to the boundary and becomes a boundary operator $\widehat{U}_a|_{\partial}$---in particular, the worldvolume TQFT is not trivialized. The boundary operators are collectively denoted as $\CC_{\text{noninv}}= \{U_b^p|_{\partial}, \widehat{U}_a^q|_{\partial}, ...\}$ where $...$ are again various condensation defects.  The bulk operators can be organized into the following subsets,
\begin{equation}
    \CB_{b}= \{U_b,\, U_b \widehat{U}_b^p,\, U_b \text{Cond}(U_a), ...\}~, \hspace{0.8cm} \CB_{\widehat{a}}= \{\widehat{U}_a,\, \widehat{U}_a \text{Cond}(U_a),\, \widehat{U}_a \text{Cond}(\widehat{U}_b), ... \}~.
\end{equation}
Assuming all defects to be of spherical topology, the above subsets simplify significantly to 
\begin{eqnarray}
    \CC_{\text{noninv}}= \{U_b^p|_{\partial}, \widehat{U}_a^q|_{\partial}\}~,\hspace{1cm}  \CB_{b}=\{U_b, U_b \widehat{U}_b^p\}~, \hspace{1cm} \CB_{\widehat{a}}= \{ \widehat{U}_a \}~. 
\end{eqnarray}

We finally note that the invertible boundary operator $U_b|_\partial$ and the non-invertible boundary operator $\widehat{U}_a|_{\partial}$ are precisely the generators of the invertible $\Z_N^{(1)}$ one-form symmetry and the non-invertible zero-form symmetry of the 4d QFT $\widehat{\CX}$ after shrinking the slab, i.e. we have
\begin{eqnarray}
    \begin{tabular}{|c|c|}
    \hline
        operators in 4d QFT $\widehat{\CX}$ in Section \ref{sec:4dNonsymfromgauging} & operators on the top. bdy. $\CC_{\text{noninv}}$ \\
        \hline
        $U_b$ & $U_b|_{\partial}$\\
        $\CN_A$ & $\widehat{U}_{a}|_{\partial}$\\
        \hline
    \end{tabular}
\end{eqnarray}

\subsection{Anomalies of non-invertible symmetries}
\label{sec:4dnonanom}

We finally turn to the question of anomalies of non-invertible symmetry of $\widehat{\CX}$, using the linking invariants among the operators in the SymTFT. Since the QFT $\CX$ before gauging has a $\Z_{2MN}^{(0)}$ self-anomaly and the $\Z_{2MN}^{(0)}$ generators become the non-invertible defects of the QFT $\widehat{\CX}$ after gauging, it is natural to expect that the non-invertible symmetry also has a non-trivial 't Hooft anomaly. To see this, we pick a representative in $\CB_{\widehat{a}}$ and compute the type 2 linking number. Since we assume spherical topology, there is only one choice of representative $\widehat{U}_a$, and indeed the linking number is non-trivial, 
\begin{eqnarray}\label{eq:trip2main2}
    \braket{\widehat{U}_a(M_3) \widehat{U}_a (M'_3) \widehat{U}_a(M''_3)} \sim e^{-{ i \pi (N^2 - 1) \over 4 N^3 M^2}\text{Link}(M_3, M'_3, M''_3)_{2}}~.
\end{eqnarray}
This non-trivial  linking invariant diagnoses the non-trivial 't Hooft anomaly of the non-invertible symmetry $\CN_A$ in $\widehat{\CX}$.

The two remarks at the end of Section \ref{sec:anomnoninvsym} still apply. Namely, the advantage of this approach is that linking invariants for the DW theories are easy to compute, though a full understanding of the anomaly requires knowledge of the fusion  higher-category. See  \cite{Bhardwaj:2022kot, Decoppet:2022dnz} for recent developments in higher categories from the generalized symmetry point of view. Moreover, the anomaly of the non-invertible symmetry detected here cannot be probed using the method of \cite{Choi:2021kmx, Choi:2022zal}, although our condition is still only a sufficient condition for the anomaly to be non-trivial.

\subsection{Application 1: Adjoint QCD}

We  close this section with two examples. We first take $\CX$ to be $SU(N_c)$ gauge theory with $N_f$ Weyl fermions in the adjoint representation of $SU(N_c)$. The action is
\begin{eqnarray}\label{eq:4dadjqcd}
    S_{\CX}= \int_{X_4} 
 \left( - \frac{1}{2g^2} \Tr f\wedge \star f + \sum_{i=1}^{N_f} i \overline{\psi}_i \slashed D_{a} \psi_i \right)~,
\end{eqnarray}
where $D_a$ is the covariant derivative with adjoint indices (suppressed).

\paragraph{Invertible symmetries and anomalies:}
Because the fermions are in the adjoint representation, this theory has an electric $\Z_{N_c}^{(1)}$ one-form symmetry. 
Classically, there is also a $U(1)$ zero-form symmetry which acts on the fermion by a phase, $\psi_i \to e^{i \alpha} \psi_i$. It is well-known that the $U(1)$ is broken by $SU(N_c)$ instantons down to $\Z_{2N_f N_c}^{(0)}$. Thus the total symmetry is $\Z_{N_c}^{(1)}\times \Z_{2N_f N_c}^{(0)}$. This symmetry has a non-trivial 't Hooft anomaly \cite{Cordova:2018acb, Delmastro:2022pfo}, specified by the 5d AnomTFT, 
\begin{eqnarray}\label{eq:4dadjanom}
    \begin{split}
    \int_{X_5} \left(\frac{2\pi}{2N_c} A^{(1)} \CP(B^{(2)}) + \frac{\pi(N_c^2-1)}{6N_c} A^{(1)} \beta A^{(1)} \beta A^{(1)}\right)~,& \hspace{1cm} N_c\in 2\Z\\
    \int_{X_5} \left(\frac{2\pi}{N_c} \frac{1+N_c}{2}  A^{(1)} B^{(2)}B^{(2)} + \frac{\pi(N_c^2-1)}{6N_c} A^{(1)} \beta A^{(1)} \beta A^{(1)}\right)~,& \hspace{1cm} N_c\in 2\Z+1\\
\end{split}
\end{eqnarray}
where $A^{(1)}$ and $B^{(2)}$ are the $\Z_{2N_f N_c}^{(0)}$ and $\Z_{N_c}^{(1)}$ background fields respectively. Thus, the  anomaly of adjoint QCD is precisely the anomaly \eqref{eq:5dSymTFT} we have been discussing throughout this section, upon replacing $(N, M)\to (N_c, N_f)$. All of the discussions so far thus apply straightforwardly. 

We will now discuss some of the dynamical implications of the anomaly. Before doing so, let us comment that the theory \eqref{eq:4dadjqcd} actually has a larger symmetry than just $\Z_{N_c}^{(1)}\times \Z_{2N_f N_c}^{(0)}$; for example, there is a flavor rotation symmetry $SU(N_f)$. We will not require that these additional symmetries be preserved below, and in particular we will allow for perturbations that explicitly break $SU(N_f)$ as long as $\Z_{N_c}^{(1)}\times \Z_{2N_f N_c}^{(0)}$ is preserved.

The mixed anomaly \eqref{eq:4dadjanom} has an immediate dynamical consequence: adding any $\Z_{N_c}^{(1)}\times \Z_{2N_f N_c}^{(0)}$ preserving deformation does not drive the theory \eqref{eq:4dadjqcd} to a trivially gapped phase. For example, assuming that $SU(N_c)$ Yang-Mills without a theta term has a mass gap, the fermion bilinear mass term $m \sum_{i=1}^{N_f}\epsilon^{\alpha\beta}{\psi}_{i \alpha} \psi_{i 
\beta}$ with $m>0$ drives the theory to a trivially gapped phase. Here $\alpha,\beta=1,2$ are the Lorentz spinor indices, and  pairing the two spinors via the epsilon tensor ensures a Lorentz singlet. However, this term explicitly breaks $\Z_{2N_fN_c}^{(0)}$ chiral symmetry.

\paragraph{Dynamical constraints of $PSU(N_c)$ adjoint QCD:}
Since $\Z_{N_c}^{(1)}$ does not have a self-anomaly, it can be gauged resulting in $PSU(N_c)$ adjoint QCD. This theory has a $\Z_{N_c}^{(1)}$ quantum one-form symmetry, whose defect is
\begin{eqnarray}\label{eq:quantumsym}
    U_b(M_2)= e^{\frac{2\pi i}{N_c}\oint_{M_2} b^{(2)}}= e^{\frac{2\pi i}{N_c}\oint_{M_2} w_2^{PSU(N_c)}}~.
\end{eqnarray}
The general results reviewed in this section show that the theory has a non-invertible codimension-one defect
\begin{eqnarray}
    \CN_{A}(M_3)\sim \sum_{\phi^{(1)}\in C^1(M_3, \Z_{N_c})} g(M_3) \cdot \exp\left( \frac{2\pi i}{2N_c} \int_{M_3} - \phi^{(1)} \delta \phi^{(1)} + 2 \phi^{(1)} w_2^{PSU(N_c)}\right)
\end{eqnarray}
implementing a twisted gauging $g\tau\sigma\tau$, where $g$ is a $\Z_{2N_f N_c}^{(0)}$ global transformation, $\tau$ is a stacking with a $\Z_{N_c}^{(1)}$ invertible phase, and $\sigma$ is a gauging of $\Z_{N_c}^{(1)}$. Moreover, the non-invertible defect $\CN_{A}$ has a self-anomaly, diagnosed by a non-trivial triple linking invariant in the bulk.

As is the case for invertible symmetries, the presence of an anomalous non-invertible symmetry has non-trivial dynamical implications. Indeed, suppose that we start with $SU(N_c)$ adjoint QCD and add $\Z_{N_c}^{(1)}\times \Z_{2N_f N_c}^{(0)}$ symmetric deformations,  then gauge $\Z_{N_c}^{(1)}$ to obtain $PSU(N_c)$ adjoint QCD. We would like to ask whether there is a symmetric perturbation which triggers a flow from the gauged theory to a trivially gapped phase.

Without knowledge of the non-invertible symmetry, one may have naively concluded that the $PSU(N_c)$ theory arising from gauging of the $\Z_{N_c}^{(1)}$ symmetry of the $SU(N_c)$ theory has only a $\Z_{2N_f}^{(0)}\times {\Z}_{N_c}^{(1)}$ symmetry, where the first component is the quotient subgroup of $\Z_{2N_f N_c}^{(0)}$ that is free of mixed anomaly, while the second component is the quantum symmetry in \eqref{eq:quantumsym}. The one-form symmetry is anomaly-free, while the zero-form symmetry has an anomaly 
\begin{eqnarray}
   \int_{X_5} \frac{\pi(N_c^2-1)}{6} \widetilde{A}^{(1)} \widetilde{\beta}\widetilde{A}^{(1)} \widetilde{\beta}\widetilde{A}^{(1)} 
\end{eqnarray}
with $A^{(1)}= N_c \widetilde{A}^{(1)}$ and $\widetilde{A}^{(1)}$ the $\Z_{2N_f}^{(0)}$ background field. Here $\widetilde{\beta}= \delta/(2N_f)$ is the Bockstein map.  The above anomaly vanishes for certain $(N_c, N_f)$, e.g. $N_c=5$ and any $N_f$, and thus one may naively conclude that when $N_c=5$ and  $N_f$ is arbitrary, the theory is free of anomalies and can be deformed to a trivially gapped phase.

However, with our current knowledge of the non-invertible symmetry, we now know that this is not true. The gauged theory has a non-invertible symmetry that suffers from a self-anomaly for \textit{any} $N_c>1$ and $N_f\geq 1$, and thus the theory cannot be deformed to a trivially gapped phase.

\paragraph{Low-energy dynamics for $N_f=1$:}
Let us comment on some special features when $N_f=1$. 
$SU(N_c)$ massless adjoint QCD  with $N_f=1$ has enhanced supersymmetry---namely, it becomes $\CN=1$ $SU(N_c)$ SYM. The low-energy dynamics of this theory are known: the $\Z_{2N_c}^{(0)}\times \Z_{N_c}^{(1)}$ global symmetry is spontaneously broken to $\Z_{2}^{(0)}\times \Z_{N_c}^{(1)}$ 
 by a gaugino bilinear condensation, where $\Z_{2}^{(0)}$ is  fermion parity. As a consequence, there are $N_c$ supersymmetric vacua. Each vacuum is trivially gapped, and the domain walls between two different vacua support a Chern-Simons theory \cite{Gaiotto:2017yup} needed to match the mixed anomaly between $\Z_{2N_c}^{(0)}$ and $\Z_{N_c}^{(1)}$ in \eqref{eq:4dadjanom}. Note that the Chern-Simons theory couples to the bulk only through the $\Z_{N_c}^{(1)}$ background field (rather than a dynamical field), and hence the domain wall still has invertible fusion rules.

There has been some confusion in the literature about how the Chern-Simons theory on the domain wall can match the $\Z_{2N_c}^{(0)}$ self-anomaly in \eqref{eq:4dadjanom}. In \cite{Delmastro:2022pfo}, it was pointed out that non-invertible defects and the junctions of the domain walls are potentially important to the resolution of this confusion. In the remainder of this section, we make this observation more concrete using the SymTFT.

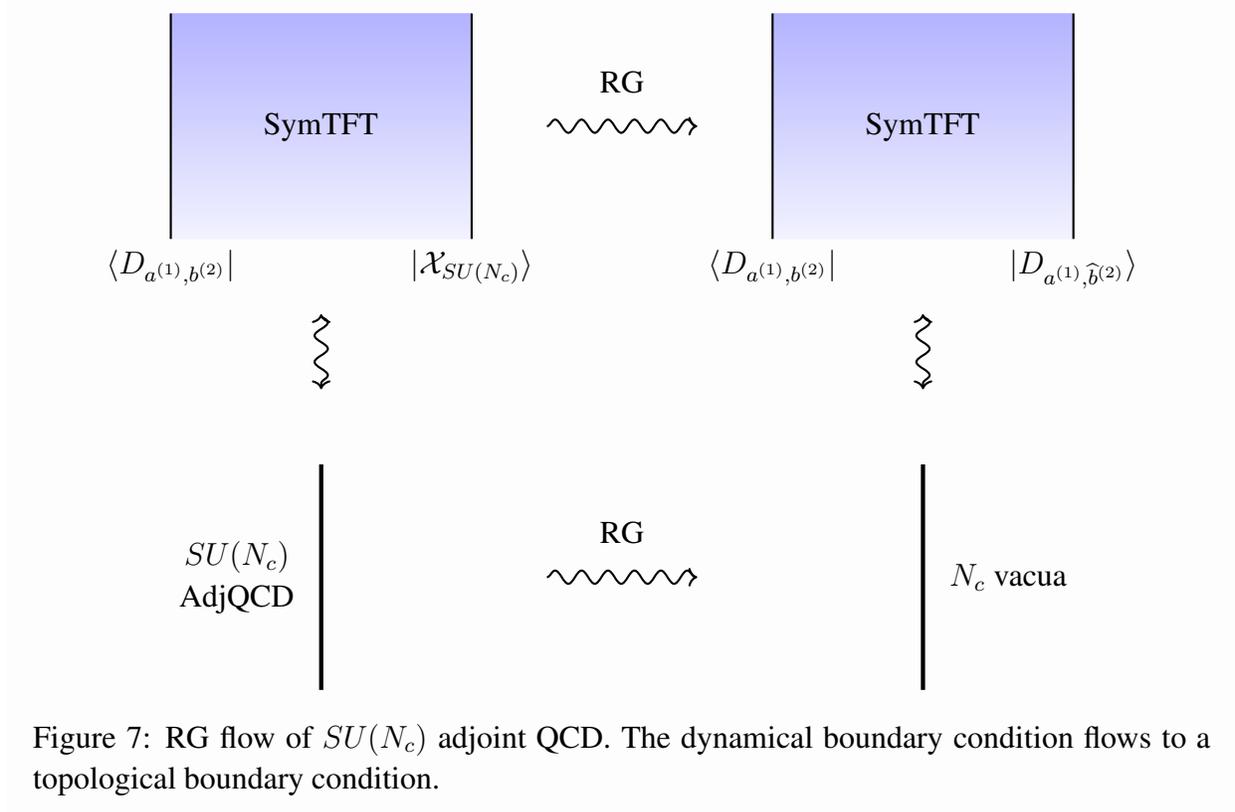
\begin{figure}[t]
	\centering
	\begin{tikzpicture}

	\shade[line width=2pt, top color=blue!30, bottom color=blue!5] 
	(0,0) to [out=90, in=-90]  (0,3)
	to [out=0,in=180] (4,3)
	to [out = -90, in =90] (4,0)
	to [out=180, in =0]  (0,0);

	\draw[thick] (0,0) -- (0,3);
	\draw[thick] (4,0) -- (4,3);

	\node at (2,1.5) {\begin{tabular}{c}
	     SymTFT  
	\end{tabular}};
	\node[below] at (0,0) {$\bra{D_{a^{(1)},b^{(2)}}}$};
	\node[below] at (4,0) {$\ket{\CX_{SU(N_c)}}$};

    \draw[snake it, thick, ->] (5,1.5) -- (7, 1.5);
    \node[above] at (6,1.8) {RG};

    \shade[line width=2pt, top color=blue!30, bottom color=blue!5] 
	(0+8,0) to [out=90, in=-90]  (0+8,3)
	to [out=0,in=180] (4+8,3)
	to [out = -90, in =90] (4+8,0)
	to [out=180, in =0]  (0+8,0);

 \draw[thick] (0+8,0) -- (0+8,3);
	\draw[thick] (4+8,0) -- (4+8,3);

 \node at (2+8,1.5) {\begin{tabular}{c}
	     SymTFT  
	\end{tabular}};
	\node[below] at (0+8,0) {$\bra{D_{a^{(1)},b^{(2)}}}$};
	\node[below] at (4+8,0) {$\ket{D_{a^{(1)},\widehat{b}^{(2)}}}$}; 

 \draw[thick, snake it, <->] (2,-1) -- (2, -2);

 \draw[ultra thick] (2, -3) -- (2, -6);

 \draw[snake it, thick, ->] (5,1.5-6) -- (7, 1.5-6);

 \draw[thick, snake it, <->] (2+8,-1) -- (2+8, -2);

 \draw[ultra thick] (2+8, -3) -- (2+8, -6);

 \node[left] at (2,-4.5) {\begin{tabular}{c}
      $SU(N_c)$  \\
      AdjQCD
 \end{tabular} };

 \node[right] at (2+8,-4.5) {\begin{tabular}{c}
      $N_c$ vacua  
 \end{tabular} };
 
\node[above] at (6,1.8-6) {RG};
 
	\end{tikzpicture}
	
	\caption{RG flow of $SU(N_c)$ adjoint QCD. The dynamical boundary condition flows to a topological boundary condition.  }
	\label{fig:RGadjQCD}
\end{figure}

At low energies, the theory flows to a gapped RG fixed point, and hence the dynamical boundary condition on the right boundary in the SymTFT flows to a topological boundary condition; see Figure \ref{fig:RGadjQCD}. Because it is known that the  $\Z_{2N_c}^{(0)}$ zero-form symmetry is spontaneously broken to $\Z_2^{(0)}$ while the $\Z_{N_c}^{(1)}$ one-form symmetry remains unbroken, the topological boundary condition should be such that $U_a$ is terminable while $U_b$ is not. Indeed, in this case $U_a$ is terminable on both left and right boundaries, and shrinking the slab gives rise to a topological local order parameter of $\Z_{2N_c}^{(0)}$ labeling the distinct vacua. Likewise, since $U_b$ is not terminable on the right boundary, shrinking the slab does not produce a topological order parameter for $\Z_{N_c}^{(1)}$, and hence it is not spontaneously broken.

The domain wall supporting a Chern-Simons theory in the SymTFT setup comes from the operator $\widehat{U}_a|_{\partial}$ supported purely on the left topological boundary. Note that it is an invertible defect on the boundary. The Chern-Simons theory on the domain wall follows from the TQFT on the worldvolume of $\widehat{U}_a$, where the dynamical field $b^{(2)}$ is replaced by the background field $B^{(2)}$.  Using the discussion in Section \ref{sec:4dnonanom}, the fact that the domain walls and their junctions saturate the $\Z_{2N_c}^{(0)}$ self-anomaly in the UV is related to the fact that the bulk operators in the SymTFT $\widehat{U}_a$ have non-trivial triple linking invariants. Therefore the right panel in Figure \ref{fig:RGadjQCD} reproduces all the known IR features of adjoint QCD and manifestly matches the anomaly since the SymTFT and the topological boundary condition on the left boundary are the same as in the UV. It would be interesting to study the junctions between domain walls or $\widehat{U}_{a}|_{\partial}$ in detail. 

After gauging $\Z_{N_c}^{(1)}$, the left topological boundary condition is changed to $\bra{D_{a^{(1)}, \widehat{b}^{(2)}}}$, while the right boundary condition is unchanged. At low energies, this in particular means that the $\Z_{N_c}^{(1)}$ one-form symmetry can also be spontaneously broken. Indeed, the vacua in this case are $\Z_{N_c}$ discrete gauge theories with $b^{(2)} b^{(2)}$ couplings, where the $\Z_{N_c}^{(1)}$ one form symmetry is (partially) spontaneously broken.

\subsection{Application 2: $\mathcal{N}=4$ SYM}
\label{sec:N=4}

We next consider the example of $\mathcal{N}=4$ SYM. This theory possesses an $SL(2,\mathbb{Z})$ duality group relating equivalent but different-looking $\mathcal{N}=4$ theories. The dual theories generally differ in both the value of the complexified coupling constant $\tau_{\mathrm{YM}}$ and the gauge group. There are however special gauge groups and values of $\tau_{\mathrm{YM}}$ for which the theory is mapped to itself under some discrete subgroup of $SL(2,\mathbb{Z})$. In such cases this discrete subgroup becomes an invertible symmetry of the theory. On the other hand, when $\tau_{\mathrm{YM}}$ is fixed under a subgroup of $SL(2,\mathbb{Z})$ but the gauge group changes, we can in general combine the transformation with a gauging of the one-form symmetry in order to obtain a non-invertible symmetry.  For more information, we refer the reader to \cite{Kaidi:2022uux,Bashmakov:2022uek}. As in the previous examples, we will focus on non-invertible symmetries which are non-instrinsic. In the current context, this means that there exists a different global form, i.e. a theory with the same gauge algebra but either a different gauge group or a different invertible phase (or both), in which the symmetry becomes invertible.

We now ask when these non-invertible symmetries are anomalous.
It is convenient to begin by considering the anomaly of the $SL(2,\mathbb{Z})$ duality group of Maxwell theory. This anomaly was originally identified in \cite{Witten:1995gf}, and further studied in \cite{Kravec:2014aza,Seiberg:2018ntt,Hsieh:2019iba}. In the original reference \cite{Witten:1995gf}, it was discovered that the partition function is not invariant under $SL(2,\mathbb{Z})$ on a curved manifold, and instead acquires a factor depending on the Euler characteristic $\chi(X_4)$ and signature $\mathsf{sig}(X_4)$ of the 4d manifold $X_4$ as follows, 
\bea
Z(\tauMax+1) &=& Z(\tauMax)~, 
\no\\
Z(-1/\tauMax) &=& \tauMax^{{1\over 4}\left(\chi(X_4) - \mathsf{sig}(X_4)\right)}\,\overline{\tau}_{\mathrm{Max}}^{{1\over 4}\left(\chi(X_4) + \mathsf{sig}(X_4)\right)}\,Z(\tauMax)~.
\eea
This signals a mixed anomaly between gravity and the $SL(2,\mathbb{Z})$ duality symmetries at certain values of the coupling. 

We first consider $\tauMax = i$, where we would expect a $\ZZ_4$ symmetry generated by $\mathsf{S}: \tauMax\mapsto -1/\tauMax$. According to the above results, under the modular $\mathsf{S}$ transformation the partition function transforms as  
\bea
 \mathsf{S}: \hspace{0.3 in} Z(i) \hspace{0.2in} \longrightarrow \hspace{0.2 in}i^{-{1\over 2}{\mathsf{sig}(X_4)}} Z(i)~.
\eea
Rokhlin's theorem states that $\mathsf{sig}(X_4) \in 16 \ZZ$ on spin manifolds, so in this case the overall factor is trivial. Hence we do not expect a mixed anomaly on spin manifolds. On the other hand, for $\tauMax = e^{2\pi i /3}$ we would expect a $\ZZ_3$ symmetry generated by $\mathsf{ST}$. In this case we find that 
\bea
 \mathsf{ST}: \hspace{0.3 in}Z(e^{2\pi i \over 3}) \hspace{0.2in} \longrightarrow \hspace{0.2 in} e^{-{\pi i \over 3} \mathsf{sig}(X_4)} Z(e^{2\pi i \over 3})~
\eea
where $\mathsf{T}$ acts as $\mathsf{T}:\tauMax \mapsto \tauMax+1$. The overall factor here is non-trivial (even on spin manifolds) and hence we conclude that there is a mixed anomaly.

We have just seen that by coupling to non-trivial gravitational backgrounds (i.e. $\mathsf{sig}(X_4)$), one can detect a mixed anomaly between (subgroups of) $SL(2,\ZZ)$ and gravity at $\tauMax = e^{2\pi i /3}$. For the rest of the discussion here, we shall concentrate on this value of $\tauMax$. We can now also allow for non-trivial backgrounds of $SL(2,\ZZ)$, which will allow us to detect self-anomalies of the $\mathbb{Z}_3$ symmetry. In general, such anomalies must take values in $\Omega^{\text{Spin}}_5 (B\mathbb{Z}_3) = \mathbb{Z}_{9}$, and we now ask which element of $\mathbb{Z}_{9}$ Maxwell theory realizes. 

This problem was considered in \cite{Hsieh:2019iba}, which claimed that the self-anomaly of Maxwell theory is given by $56$ times that of a 4d Weyl fermion. The strategy in that work was to make use of a 6d interpretation, in terms of the compactification of the 6d rank-1 E-string SCFT. This theory has a moduli space consisting of two branches referred to as the tensor branch and Higgs branch. On a generic point of the former one gets a free $(1,0)$ tensor multiplet, while on a generic point of the latter one gets $29$ free hypermultiplets. When reduced to 4d on a torus, the former gives Maxwell theory plus $2$ Weyl fermions and some scalars, while the latter gives $58$ free Weyl fermions plus some scalars. As we can continuously move from the tensor branch to the Higgs branch, and as anomalies should be invariant under continuous deformations, we see that the anomaly for the Maxwell theory should be equal to that of $56$ Weyl fermions. If we assume that the minimal anomaly is realized by a single Weyl fermion, then we conclude that the duality anomaly of Maxwell theory realizes the element $56 = 2 \in \ZZ_{9}$.

One subtlety in the above derivation is how the $SL(2,\mathbb{Z})$ acts on the fermions. In particular, the 6d picture leads to  $\mathsf{S}^4$ being $(-1)^\mathsf{F}$, i.e. fermion parity, instead of just $1$. As such, we should really be considering $[\mathrm{Spin}(3,1)\times Mp(2,\mathbb{Z})]/\mathbb{Z}^{F}_2$ instead of $\mathrm{Spin}(3,1)\times SL(2,\mathbb{Z})$, where $Mp(2,\mathbb{Z})$ is a double cover of $SL(2,\mathbb{Z})$ such that $\mathsf{S}^4 \neq 1$, and $\mathbb{Z}^{F}_2$ is the combination of the fermion parity of $\mathrm{Spin}(3,1)$ and $\mathsf{S}^4$ in $Mp(2,\mathbb{Z})$. A simplifying feature is that, as we are only concerned with the $\ZZ_{3}$ subgroup at $\tauMax = e^{2\pi i /3}$, we can take the two groups to be $[\mathrm{Spin}(3,1)\times \mathbb{Z}_6]/\mathbb{Z}^{F}_2$ and $\mathrm{Spin}(3,1)\times \mathbb{Z}_3$, respectively. These groups are actually equivalent, since starting from the former we can just combine the generator of $\mathbb{Z}_3$ with fermion parity to get $\mathbb{Z}_6$. Furthermore, as we expect fermion parity to be non-anomalous, we expect the self-anomalies of the two symmetries to be the same. Indeed, the cobordism groups with both structure groups, evaluated in \cite{Hsieh:2018ifc}, are equal to $\mathbb{Z}_{9}$. From now on we shall concentrate on the $\mathbb{Z}_6$ version that includes fermion parity, though we expect the results to also hold for the $\mathbb{Z}_3$ version.

Having discussed the anomaly for Maxwell theory, we are now ready to consider the case of $\mathcal{N}=4$ SYM. To understand this case, we shall use the fact that $SL(2,\mathbb{Z})$ (and fermion parity) do not act on the moduli space and hence are preserved at a generic point on this space. At a generic point, the theory looks like a theory of $r$ free $\mathcal{N}=4$ vector multiplets, with $r$ being the rank of the gauge group. Each vector multiplet contains a copy of Maxwell theory, four Weyl fermions, and six scalars, the latter of which will play no role in our analysis. As we have not broken the $SL(2,\ZZ)$ by moving out on the moduli space, the anomaly of the $\mathcal{N}=4$ SCFT should be equal to the anomaly at a generic point in this space. From our previous results, we conclude that the total anomaly is 
\bea
\label{eq:N4modularanom}
I^{\mathrm{rank}\,\,r\,\,\text{SYM}}= (2+4) \,r I^{\text{4d fermion}} = 6\, r I^{\text{4d fermion}}~,
\eea
again in terms of the anomaly of a single Weyl fermion. The anomaly identified is thus a self-anomaly for the $\mathbb{Z}_6$ zero-form modular symmetry. Note that as the anomaly in $[\mathrm{Spin}(3,1)\times \mathbb{Z}_6]/\mathbb{Z}^{F}_2$ is mod $9$, this anomaly actually only depends on $r$ mod $3$.

For the $\mathfrak{a}$-type cases we can arrive at the same result using the 6d approach of \cite{Hsieh:2019iba}. Here we start with the rank-$(r+1)$ E-string theory and use the fact that on the tensor branch it reduces to the type $\mathfrak{a}_r$ $(2,0)$ theory plus a decoupled $(1,0)$ tensor, while on the Higgs branch it reduces to $30r-1$ free hypers. The reduction of the former now gives $\mathcal{N}=4$ SYM with gauge algebra $\mathfrak{a}_r$ plus a decoupled $\mathcal{N}=2$ $U(1)$ gauge theory and equating the anomaly on the two sides yields the desired result.

To summarize our discussion so far, we have found that at the value of the complex coupling $\tauYM=e^{\frac{2\pi i}{3}}$, the modular $\mathsf{ST}$ symmetry can have a self-anomaly, determined by the rank $r$ of the gauge group mod 3. When the gauge group in question is left invariant under the modular $\mathsf{ST}$ symmetry, then this is a standard `t Hooft anomaly in an invertible symmetry. On the other hand, when the gauge group is \textit{not} left invariant under $\mathsf{ST}$, then the modular transformation must be dressed with appropriate discrete gaugings to make it a non-invertible symmetry, and in the case of non-intrinsic non-invertible symmetries the anomaly of the invertible symmetry implies an anomaly for the non-invertible symmetry. 

As concrete examples, let us consider the $\cN=4$ SYM theories with gauge algebras $\mathfrak{a}_1$, $\mathfrak{a}_2$, and $\mathfrak{e}_6$, all of which were analyzed in \cite{Kaidi:2022uux}. In the case of $\mathfrak{a}_1$, there are three global forms $SU(2)$, $SO(3)_+$, and $SO(3)_-$ (up to stacking with SPT phases), and none of them have an invertible $\mathsf{ST}$ symmetry at $\tauYM=e^{\frac{2\pi i}{3}}$. Thus all of the corresponding non-invertible $\mathsf{ST}$ symmetries are intrinsically non-invertible, and we will not say anything about them here. On the other hand, for $\mathfrak{a}_2$, there are four global forms $SU(3)$, $PSU(3)_0$, $PSU(3)_1$, and $PSU(3)_2$ (again up to SPT phases), and it turns out that the  $PSU(3)_1$ theory has an invertible $\mathsf{ST}$ symmetry. Since the rank $r=2$ is non-zero mod 3, we conclude that this $\mathsf{ST}$ symmetry has a `t Hooft anomaly, and by our general discussions in this paper that the non-invertible symmetries in the other global variants are also anomalous. Finally, in the case of $\mathfrak{e}_6$, there are again four global forms up to SPTs, one of which has an invertible $\mathsf{ST}$ symmetry. However, in this case the rank $r=6$ is zero mod 3, and hence we expect no anomalies for this symmetry.

Let us close with a bit more detail on the case of $PSU(3)_1$, which has a $\Z_6$ self-anomaly given by $6r=3\mod 9$ copies of the anomaly of the 4d fermion. Note that the 5d anomaly inflow action of the 4d fermion can be inferred from the $U(1)$ inflow action by restricting $U(1)$ to $\Z_6$ subgroup. Three copies of the 4d fermions have the $U(1)$ anomaly  
$\frac{3}{24\pi^2} \mathsf{A}d\mathsf{A}d\mathsf{A}$. Substituting $\mathsf{A}=\frac{2\pi}{6}A^{(1)}$, we then get the $\Z_6$ self-anomaly $\frac{2\pi}{12}A^{(1)}\beta A^{(1)} \beta A^{(1)}$. 
Combining this with the mixed anomaly with the $\Z_3$ one-form symmetry, the total anomaly is given by\footnote{The $A^{(1)} B^{(2)} B^{(2)}$ anomaly can be checked by turning off $A^{(1)}$ and computing how the partition function transforms under a global $\Z_{6}^{(0)}$ transformation. In principle, another mixed anomaly of the form $B^{(2)} A^{(1)} \beta A^{(1)} $ can also appear. Although it is interesting to explicitly check it, we will assume that such term vanishes. If it does not vanish, there exists a choice of symmetry fractionalization \cite{Delmastro:2022pfo} by shifting $B^{(2)}\to B^{(2)}+ k \beta A^{(2)} \mod 3$ for some $k$ such that this term is absorbed. As a consequence, the coefficient of the self anomaly $A^{(2)}\beta A^{(1)}\beta A^{(2)}$ will be modified. }
\begin{eqnarray}
    \int_{X_5} \frac{4\pi}{3} A^{(1)}B^{(2)}B^{(2)}+ \frac{2\pi}{12}A^{(1)}\beta A^{(1)} \beta A^{(1)}
\end{eqnarray}
where $\beta=\delta/6$, the field $A^{(1)}$ is the $\Z_6^{(0)}$ one-form gauge field, and $B^{(2)}$ is the $\Z_3^{(1)}$ two-form gauge field. We can further gauge the $\Z_3^{(1)}$ one-form symmetry, mapping $PSU(3)_1$ SYM to $SU(3)$ SYM (coupled to a non-trivial $\Z_3^{(1)}$ SPT). From the results in this section,\footnote{Although we focused on $\Z_N^{(1)}$ one-form symmetry for even $N$, parallel discussions apply for odd $N$ as well.} we conclude that $SU(3)$ SYM has an anomalous non-invertible symmetry.

\section*{Acknowledgements}

We thank Yichul Choi, Linhao Li, Kantaro Ohmori, Sakura Schafer-Nameki, Yuji Tachikawa, and Zheyan Wan for discussions. We thank Shu-Heng Shao and Yuji Tachikawa for comments on a draft. 
JK and GZ thank Kavli IPMU for their generous hospitality during the inception of this work. EN and YZ are partially supported by WPI Initiative, MEXT, Japan at IPMU, the University of Tokyo. GZ is partially supported by the Simons Foundation grant 815892.

\appendix

\newpage
\section{Linking numbers}
\label{app:Link}

In this appendix we discuss the linking numbers relevant to the main text.

\paragraph{Two component links:} 
Given two closed manifolds $M_1^{(p)}$ and $M_2^{(q)}$, we aim to compute the linking number between them. We denote their Seifert surfaces as $N_1^{(p+1)}$ and $N_2^{(q+1)}$ respectively, which means that $\partial N_1^{(p+1)}=M_1^{(p)}$ and $\partial N_2^{(q+1)}=M_2^{(q)}$. The linking number between $M_1^{(p)}$ and $M_2^{(q)}$ in spacetime $S^d$ is given by 
\begin{equation}
   \text{Link}(M_1^{(p)}, M_2^{(q)}):= \int_{S^d} \text{PD}(N_1^{(p+1)}) d \text{PD}(N_2^{(q+1)}) = \text{Int}(N_1^{(p+1)}, M_2^{(q)})
\end{equation}
where $\text{PD}$ is the Poincar{\'e} dual and $\text{Int}(\cdot, \cdot)$ counts the number of (oriented) intersection points between the two arguments.  Note that for the integration to be non-trivial, we need to match the degrees
\begin{equation}\label{eq:hopfdim}
    p+1 + q = d~.
\end{equation}
As an example, one can consider $p=q=1$ and $d=3$. In this case, the two lines form a standard link in three-dimensions. In this case it is easy to see that a non-trivial Hopf link has linking number 1, while an L4a1 link (also known as Solomon's knot) has linking number 2.\footnote{See \url{http://katlas.math.toronto.edu/wiki/L4a1}.} 

One may wonder why we don't define another type of linking number by $\int_{S^d} \text{PD}(N_1^{(p+1)})  \text{PD}(N_2^{(q+1)}) = \text{Int}'(N_1^{(p+1)}, N_2^{(q+1)})$ with $p+q+2=d$. The reason is that this intersection number is unstable: it is possible to move the boundaries $M_1^{(p)}$ and $M_2^{(q)}$ without crossing each other such that the intersection number is zero.

\begin{figure}[t]
	\centering
	\begin{tikzpicture}
	
\draw[very thick, red, -] (0,0) [partial ellipse=55:-210:2cm and 0.6cm];
\draw[very thick, red, -] (0,0) [partial ellipse=65:115:2cm and 0.6cm];
\draw[very thick, red, -] (0,0) [partial ellipse=125:150:2cm and 0.6cm];
\node[above, red] at (0,0.6) {$M_1^{(1)}$};

\draw[very thick, orange, -] (0,0) [partial ellipse=110:250:2.5cm and 2.5cm];
\draw[very thick, orange, -] (0,0) [partial ellipse=-100:100:2.5cm and 2.5cm];
\node[right, orange] at (2.5,0) {$M_2^{(1)}$};

\draw[very thick, blue, -] (0,0) [partial ellipse=-5:50:1cm and 3cm];
\draw[very thick, blue, -] (0,0) [partial ellipse=60:185:1cm and 3cm];
\draw[very thick, blue, -] (0,0) [partial ellipse=-15:-50:1cm and 3cm];
\draw[very thick, blue, -] (0,0) [partial ellipse=-60:-165:1cm and 3cm];
\node[above, blue] at (0,3) {$M_3^{(1)}$};

	\end{tikzpicture}
	
	\caption{Borromean rings linking between $M_1^{(1)}, M_2^{(1)}$ and $M_3^{(1)}$ . }
	\label{fig:Borr}
\end{figure}
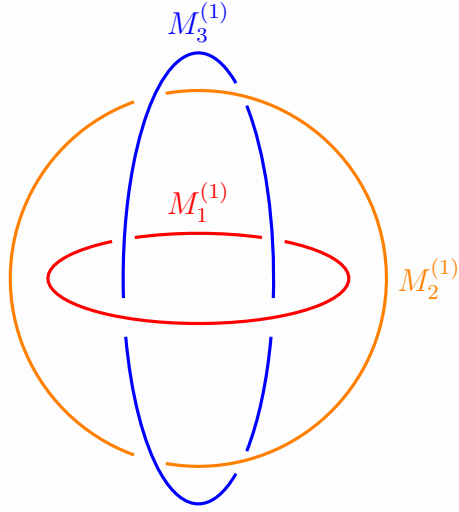

\paragraph{Three component links:} 
Unlike the the case with two components, there are multiple types of linking numbers involving three components. Suppose the three components are $M_1^{(p)}, M_2^{(q)}$, and $M_3^{(r)}$ respectively. We denote their Seifert surfaces by $N_1^{(p+1)}, N_2^{(q+1)}$, and $N_3^{(r+1)}$. There are three types of linking numbers between $M_1^{(p)}, M_2^{(q)}$, and $M_3^{(r)}$ in spacetime $S^d$, denoted by type 0, type 1, and type 2 respectively. The type 0 linking number is given by
\begin{equation}
\begin{split}
   \text{Link}(M_1^{(p)}, M_2^{(q)}, M_3^{(r)})_0 &:= \int_{S^d} \text{PD}(N_1^{(p+1)}) \text{PD}(N_2^{(q+1)}) \text{PD}(N_3^{(r+1)})
   \\&= \text{Int}(N_1^{(p+1)}, N_2^{(q+1)}, N_3^{(r+1)})
\end{split}
\end{equation}
where the dimensions should satisfy 
\begin{equation}
    p+1+q+1+r+1 =2d~.
\end{equation}
The type 1 linking number is given by 
\begin{equation}
\begin{split}
   \text{Link}(M_1^{(p)}, M_2^{(q)}, M_3^{(r)})_1&:= \int_{S^d} \text{PD}(N_1^{(p+1)}) \text{PD}(N_2^{(q+1)}) d\text{PD}(N_3^{(r+1)})
   \\
   &= \text{Int}(N_1^{(p+1)}, N_2^{(q+1)}, M_3^{(r)})
\end{split}
\end{equation}
where the dimensions should satisfy 
\begin{equation}
    p+1+q+1+r =2d~.
\end{equation}
Finally, the type 2 linking number is given by 
\begin{equation}
\begin{split}
   \text{Link}(M_1^{(p)}, M_2^{(q)}, M_3^{(r)})_2&:=
   \int_{S^d} \text{PD}(N_1^{(p+1)}) d\text{PD}(N_2^{(q+1)}) d\text{PD}(N_3^{(r+1)})\\ &= \text{Int}(N_1^{(p+1)}, M_2^{(q)}, M_3^{(r)})
\end{split}
\end{equation}
where the dimensions should satisfy 
\begin{equation}
    p+1+q+r =2d~.
\end{equation}
As an example, for $p=q=r=1$ and $d=3$, the Borromean rings have type 0 linking number 1 because the three Seifert surfaces of the three loops intersect at one point; see Figure \ref{fig:Borr}. Moreover, for $p=q=r=2$ and $d=4$, the linking configuration with a non-trivial type 1 linking number is associated with a 3-loop braiding process \cite{2014PhRvL.113h0403W, Wang:2014oya}.

\paragraph{$N$ component links:} 
By generalizing the discussion for triple links, it is clear that there are $N$ types of $N$-component linking numbers for $N>2$. Suppose the $N$ topological operators are of dimension $p_i$ for  $i=1, ..., N$.  We can define the $N$-component link of type $k$ by having $N-k$ Seifert surfaces and $k$ operators intersect at a point,
\begin{equation}
    \begin{split}
        &\text{Link}(M_1^{(p_1)}, ...., M_N^{(p_N)})_k:=
        \\
        &\hspace{0.2 in}\int_{S^d} \text{PD}(N_1^{(p_1+1)}) ... \text{PD}(N_{N-k}^{(p_{N-k}+1)}) d\text{PD}(N_{N-k+1}^{(p_{N-k+1}+1)})... d\text{PD}(N_{N}^{(p_{N}+1)})\\
        &\hspace{0.5 in}= \text{Int}(N_1^{(p_1+1)}, ..., N_{N-k}^{(p_{N-k}+1)}, M_{N-k+1}^{(p_{N-k+1})}, ..., M_N^{(p_N)})
    \end{split}
\end{equation}
where the various dimensions should satisfy
\begin{equation}\label{eq:typeklink}
    \sum_{i=1}^{N-k} (p_i+1) + \sum_{j=N-k+1}^{N} p_j = (N-1)d~.
\end{equation}

\paragraph{Linking invariants between lines in $d=3$:}
From the results above, we find that the linking invariants between lines in $d=3$ are of two types: the  linking number involving two loops probed by the Hopf link, and the type 0 linking number involving three loops probed by the Borromean rings. To see that there are no other linking numbers with number of components $N\geq 3$, we use \eqref{eq:typeklink} and substitute $p_i=p_j=1$,  
\begin{eqnarray}
    (N-k)\cdot (1+1)+ k \cdot 1 = (N-1)3 \hspace{0.5cm} \Rightarrow \hspace{0.5cm} N=3-k~.
\end{eqnarray}
Hence the only solution for $N\geq 3$ is $N=3$ and $k=0$.

\paragraph{Linking invariants among surfaces and 3-volumes in $d=5$:} From the results above, we find that the linking invariants involving surfaces and 3-volumes in $d=5$ are of the following types: the linking number involving two 2d operators, the type 0 linking number between one 3d operator and two 2d operators, the type 1 linking number between two 3d operators and a 2d operator,  the type 2 linking number between three 3d operators, the type 1 linking number between four 3d operators, and the type 0 linking number between five 3d operators.

\section{$\Z_2^A\times \Z_2^B\times \Z_2^C$ Dijkgraaf-Witten model in 3d}
\label{app:3dDW}

In this appendix we study the properties of the Dijkgraaf-Witten model \eqref{eq:2dSymTFT}. For convenience we reproduce the action of this theory, 
\begin{eqnarray}\label{eq:2dSymTFTapp}
\int_{X_3} \pi \left(  \widehat{a} \delta a +  \widehat{b} \delta b +  \widehat{c} \delta c +  abc + \frac{1}{2} a\delta a \right)~.
\end{eqnarray}
All of the gauge fields are $\Z_2$ cochains, i.e. elements of $C^1(X_3, \Z_2)$. In this appendix we will assume that $X_3$ does not have a boundary. The action is invariant under the gauge transformations \cite{He:2016xpi}
\begin{eqnarray}\label{eq:3dDWgaugetran}
\begin{split}
&a\to a+ \delta \alpha, \hspace{1cm} b\to b+\delta \beta, \hspace{1cm} c\to c+\delta \gamma~,\\
&\widehat{a}\to \widehat{a} + \delta \widehat{\alpha} + \beta c - \gamma b + \beta \delta\gamma~,\\
&\widehat{b}\to \widehat{b} + \delta \widehat{\beta} +  \gamma a - \alpha c + \gamma \delta \alpha~,\\
&\widehat{c}\to \widehat{c} + \delta \widehat{\gamma} + \alpha b - \beta a + \alpha \delta \beta~.
\end{split}
\end{eqnarray}
Note that the twist term $abc$ is not gauge invariant under the gauge transformations of $a,b,c$. The gauge transformations of $\widehat{a}, \widehat{b}$, and $\widehat{c}$ are then fixed by requiring gauge invariance of the full action.

\subsection{Line operators and fusion rules}

The gauge invariant line operators of the theory are as follows. We first have the obvious invertible Wilson loops,
\begin{eqnarray}\label{eq:invop}
U_{a}(M_1)= e^{i\pi \oint_{M_1} a}~, \hspace{1cm} U_{b}(M_1)= e^{i\pi \oint_{M_1} b}~, \hspace{1cm} U_{c}(M_1)= e^{i\pi \oint_{M_1} c}~,
\end{eqnarray}
which together with their linear combinations yield eight invertible lines. The magnetic operators are more interesting. Without the DW twist terms, the naive magnetic lines  $e^{i\pi \oint \widehat{a}}$, $e^{i\pi \oint \widehat{b}}$, and $e^{i\pi \oint \widehat{c}}$ are gauge invariant. However, in the presence of the DW twist terms these lines are not gauge invariant. To achieve gauge invariance, one needs to couple them to appropriate 1d TQFTs.  The gauge invariant operators are as follows, 
\begin{eqnarray}\label{eq:noninvop}
\begin{split}
\widehat{U}_a(M_1)&\sim  \sum_{\phi_2, \phi_3\in C^0(M_1, \Z_2)} e^{i\pi \oint_{M_1} \widehat{a} + i\pi \oint_{M_1} (-\phi_2 c + \phi_3 b + \phi_2 \delta \phi_3)}~,\\
\widehat{U}_{b}(M_1)&\sim \sum_{\phi_3, \phi_1\in C^0(M_1, \Z_2)} e^{i\pi \oint_{M_1} \widehat{b} + i\pi \oint_{M_1} (-\phi_3 a + \phi_1 c + \phi_3 \delta \phi_1)}~,\\
\widehat{U}_c(M_1)&\sim \sum_{\phi_1, \phi_2\in C^0(M_1, \Z_2)} e^{i\pi \oint_{M_1} \widehat{c} + i\pi \oint_{M_1} (-\phi_1 b + \phi_2 a + \phi_1 \delta \phi_2)}~,\\
\widehat{U}_{ab}(M_1)&\sim \sum_{\phi_{12}, \phi_3\in C^0(M_1, \Z_2)} e^{i\pi \oint_{M_1} (\widehat{a}+ \widehat{b}) + i\pi \oint_{M_1} (-\phi_3 (a-b) + \phi_{12} c + \phi_3 \delta \phi_{12})}~,\\
\widehat{U}_{bc}(M_1)&\sim \sum_{\phi_{23}, \phi_1\in C^0(M_1, \Z_2)} e^{i\pi \oint_{M_1} (\widehat{b}+ \widehat{c}) + i\pi \oint_{M_1} (-\phi_1 (b-c) + \phi_{23} a + \phi_1 \delta \phi_{23})}~,\\
\widehat{U}_{ca}(M_1)&\sim \sum_{\phi_{31}, \phi_2\in C^0(M_1, \Z_2)} e^{i\pi \oint_{M_1} (\widehat{c}+ \widehat{a}) + i\pi \oint_{M_1} (-\phi_2 (c-a) + \phi_{31} b + \phi_2 \delta \phi_{31})}~,\\
\widehat{U}_{abc}(M_1)&\sim \sum_{\phi_{12}, \phi_{23}\in C^0(M_1, \Z_2)} e^{i\pi \oint_{M_1} (\widehat{a}+ \widehat{b}+\widehat{c}) + i\pi \oint_{M_1} (-\phi_{12} (b-c) + \phi_{23} (a-b) + \phi_{12} \delta \phi_{23})}~,
\end{split}
\end{eqnarray}
where the gauge transformations of the scalars on the lines are given by
\begin{eqnarray}
\phi_i\to \phi_i+ \alpha_i~, \hspace{1cm} \phi_{ij}\to \phi_{ij}+ \alpha_i-\alpha_j~, \hspace{1cm} i,j=1,2,3~
\end{eqnarray}
and $\alpha_{1,2,3}=\alpha,\beta,\gamma$ respectively.\footnote{The last operator can be naively written as $$\widehat{U}_{abc}(M_1)\sim   \sum_{\phi_{1}, \phi_{2}, \phi_3\in C^0(M_1, \Z_2)} e^{i\pi \oint_{M_1} (\widehat{a}+ \widehat{b}+\widehat{c}) + i\pi \oint_{M_1} (\phi_2-\phi_3)a + (\phi_3-\phi_1)b + (\phi_1-\phi_2)c + \phi_1 \delta \phi_2 + \phi_2 \delta \phi_3 +\phi_3 \delta \phi_1}~.$$ One can then define $\phi_{ij}=\phi_i-\phi_j$. Note that $\phi_{ij}$ are not completely independent---they are required to sum to zero.  By using $\phi_2\delta \phi_2 = 0$, it is possible to rewrite  the last three terms purely in terms of $\phi_{ij}$'s $$\phi_{12}\delta \phi_{23}= \phi_1 \delta \phi_2 + \phi_2 \delta \phi_3 + \phi_3\delta \phi_1~. $$}

The magnetic lines in \eqref{eq:noninvop} can also fuse with the invertible lines \eqref{eq:invop} to give new lines. 
For instance, fusing $U_b$ with $\widehat{U}_a$ amounts to a global shift $\phi_3\to \phi_3+1$ in the sum. Since $\phi_3$ is summed over, the result is simply $\widehat{U}_a$ itself. This shows the fusion rule $\widehat{U}_a\times U_b= \widehat{U}_a$. By the same reasoning, one can also derive $\widehat{U}_a\times U_c= \widehat{U}_a$. However, the above argument does not apply to fusing $U_a$ with $\widehat{U}_a$, and instead $U_a \widehat{U}_a$ is a new operator. Likewise there are 7 new operators involving the magnetic operators, 
\begin{eqnarray}\label{eq:3dlines}
\begin{split}
&\widehat{U}_a U_a~, \hspace{1cm} \widehat{U}_b U_b~, \hspace{1cm}\widehat{U}_c U_c~,\hspace{1cm} \widehat{U}_{ab} U_a = \widehat{U}_{ab} U_b~, \hspace{1cm} \widehat{U}_{bc} U_b = \widehat{U}_{bc} U_c~, \\ &\widehat{U}_{ca} U_c = \widehat{U}_{ca} U_a~, \hspace{1cm} \widehat{U}_{abc} U_a = \widehat{U}_{abc} U_b = \widehat{U}_{abc} U_c~.
\end{split}
\end{eqnarray}
Together with the 8 invertible lines in \eqref{eq:invop}, there are a total of $8+7\times 2= 22$ line operators in the theory. The fact that there are 22 lines in a closely related theory was already known in \cite{deWildPropitius:1995cf,Wang:2014oya,He:2016xpi}.

We finally consider the fusion rules between the magnetic line operators. Let us first compute the fusion $\widehat{U}_a\times \widehat{U}_a$,\footnote{Further specifying the overall normalization of $\widehat{U}_a$ to be $1/|C^0(M_1, \Z_2)|$, we actually find the fusion rule $\widehat{U}_a\times \widehat{U}_a=1+ U_{b}+ U_{c}+ U_{bc}$, i.e. the first and last expression are actually equal. }
\begin{eqnarray}\label{eq:AAfusionrule}
\begin{split}
\widehat{U}_a\times \widehat{U}_a &\sim  \sum_{\substack{\phi_2, \phi_3, \widetilde{\phi}_2, \widetilde{\phi}_3\\ \in C^0(M_1, \Z_2)}} e^{i\pi \oint_{M_1} (-\phi_2 c + \phi_3 b + \phi_2 \delta \phi_3 - \widetilde{\phi}_2c + \widetilde\phi_3 b + \widetilde\phi_2 \delta  \widetilde\phi_3 )  }\\
&\sim \sum_{\substack{\phi_2', \phi_3', \widetilde{\phi}_2, \widetilde{\phi}_3\\ \in C^0(M_1, \Z_2)}} e^{i\pi \oint_{M_1} (-\phi_2' c + \phi_3' b + \phi_2' \delta \phi_3' + \widetilde\phi_2 \delta  \phi_3' + \phi_2' \widetilde{\phi}_3)   }\\
&\sim  \sum_{\substack{\phi_2', \phi_3'\in Z^0(M_1, \Z_2)}} e^{i\pi \oint_{M_1} (-\phi_2' c + \phi_3' b)   } \sim 1+ U_{b}+ U_{c}+ U_{bc}~.\\
\end{split}
\end{eqnarray}
In the first equality, we used the fact that $\widehat{a}$ is a $\Z_2$ cochain, and hence that the two factors $e^{i\pi \oint \widehat{a}}$ cancel against each other. Only the contributions from the attached 1d TQFT survive. Note that since the $\phi_i$ only live on the individual lines, we should distinguish the scalars for each line separately. 
In the second equality, we introduced $\phi_2':= \phi_2-\widetilde{\phi}_2, \phi_3':= \phi_3-\widetilde{\phi}_3$, and as a consequence 
 the fields $\widetilde{\phi}_{2,3}$ became Lagrangian multiplers. In the third equality we integrated over $\widetilde{\phi}_{2,3}$, which enforced that $\phi_{2,3}'$ were $\Z_2$-valued 0-cocycles.  In other words, they are constants on $M_1$.  Finally, summing over $\phi_{2,3}'\in Z^0(M_1, \Z_2)$ simply amounts to summing over $\phi_{2,3}'\in \{0,1\}$, and we obtain four terms as shown above. As a consistency check, one can stack $U_{b}$ on both sides, and ones finds that  both sides are indeed invariant.

Similar manipulations can be used to find the other fusion rules. Below, we enumerate the fusion rules up to permutations of $a,b,c$,
\begin{eqnarray}
\begin{split}
\widehat{U}_a\times \widehat{U}_a  &\sim 1+ U_{b}+ U_{c}+ U_{bc}~,\\
\widehat{U}_a\times \widehat{U}_b & \sim \widehat{U}_{ab} + \widehat{U}_{ab} U_a \sim  \widehat{U}_{ab} + \widehat{U}_{ab} U_b~,\\
\widehat{U}_a \times \widehat{U}_{ab} &\sim \widehat{U}_{b} + \widehat{U}_{b} U_b~,\\
\widehat{U}_c \times \widehat{U}_{ab} &\sim  \widehat{U}_{abc} + \widehat{U}_{abc} U_a\sim  \widehat{U}_{abc} + \widehat{U}_{abc} U_b\sim \widehat{U}_{abc} + \widehat{U}_{abc} U_c~.\\
\end{split}
\end{eqnarray}
Specifying the normalization, one finds that all $\sim$ can be replaced by $=$. 
We thus find that all magnetic operators $\widehat{U}_{i}, \widehat{U}_{ij}, \widehat{U}_{ijk}$ have quantum dimension 2 and  are non-invertible.

\subsection{Linking numbers between lines}

We now compute the linking numbers from the correlation functions of the lines discussed above. 
A useful simplifying assumption is to take all the lines to be unknotted circles $S^1$, and to take the spacetime to be of trivial topology $X_3=S^3$. This allows us to avoid issues of loops wrapping spacetime cycles; indeed, if  a loop were to wrap a non-contractable spacetime cycle, this could lead to  additional phase factors and contaminate the linking numbers.

The non-trivial linking numbers can be organized into the following classes: 
\begin{enumerate}
    \item Linking number between an electric and magnetic line, 
\begin{eqnarray}\label{eq:AAhatlink}
    \braket{U_i(M_1) \widehat{U}_j(M'_1)}\sim  (-1)^{\text{Link}(M_1,M'_1) }\delta_{ij}
    \end{eqnarray}
    for $i,j\in \{a,b,c\}$. For instance, $M_1, M_1'$ forming a Hopf link can be used to probe this linking number. This linking invariant is a consequence of the BF couplings in \eqref{eq:2dSymTFTapp}.  
    \item  Linking number between two magnetic lines,
    \begin{eqnarray}\label{eq:AhatAhatlink}
    \braket{\widehat{U}_i(M_1) \widehat{U}_j(M'_1)}\sim 
    \begin{cases}
     (-1)^{\text{Link}(M_1,M'_1)} & i=j=a\\
    1 & \text{otherwise}
    \end{cases}
    \end{eqnarray}
    Once again, $M_1, M_1'$ forming a Hopf link can be used to probe this linking number.
    This is a consequence of the DW twist  $\frac{1}{2}a\delta a$  in \eqref{eq:2dSymTFTapp}.  
    \item  Linking number of type 0 between three magnetic lines,
    \begin{eqnarray}\label{eq:ABClink}
    \braket{\widehat{U}_a(M_1) \widehat{U}_{b}(M'_1) \widehat{U}_c(M''_1)}\sim (-1)^{\text{Link}(M_1,M'_1,M''_1)_{0}}~.
    \end{eqnarray}
   The lines $M_1, M_1', M_1''$ forming the Borromean rings can be used to probe this linking number.
    This is a consequence of the DW twist $abc$  in \eqref{eq:2dSymTFTapp}.  
\end{enumerate}
All other linking invariants are either trivial or combinations of the above ones. For example, the Hopf linkings between electric lines $\braket{{U}_i(M_1) {U}_j(M'_1)}$ are all trivial.

The computations are straightforward. Below we derive \eqref{eq:ABClink} in detail; all other invariants can be likewise derived. We first compute the partition function without any line insertions. 
\begin{eqnarray}
\begin{split}
Z(S^3)&\sim  \sum_{\substack{a,b,c,\widehat{a}, \widehat{b}, \widehat{c}\\ \in C^1(S^3, \Z_2)}} e^{i \int \pi \left(\widehat{a}\delta a + \widehat{b}\delta b+ \widehat{c}\delta c + abc + \frac{1}{2} a \beta a  \right)} \sim \sum_{\substack{a,b,c\\ \in Z^1(S^3, \Z_2)}} e^{i \int \pi \left(abc + \frac{1}{2} a \beta a \right)} \sim 1~.
\end{split}
\end{eqnarray}
Note that we suppressed all the real positive normalization constants. In the second $\sim$ we summed over $\widehat{a}, \widehat{b}, \widehat{c}$, which enforced $a,b,c$ to be cocycles. In the third $\sim$ we used the fact that the spacetime is $S^3$, and hence that all the flat connections on $S^3$ are gauge equivalent to the trivial connection.

We then compute the correlation function $\braket{\widehat{U}_a(M_1) \widehat{U}_{b}(M'_1) \widehat{U}_c(M''_1)}$, 
\begin{equation}
    \begin{split}
        &\braket{\widehat{U}_a(M_1) \widehat{U}_b(M_1') \widehat{U}_c(M_1'')}=\frac{1}{Z(S^3)} Z(S^3, \widehat{U}_a(M_1), \widehat{U}_{b}(M'_1) , \widehat{U}_c(M''_1))\\
        &\qquad\sim \sum_{\substack{a,b,c,\widehat{a},\widehat{b},\widehat{c}\\
\in C^1(S^3, \Z_2)
}}  \sum_{\substack{\phi_2, \phi_3\\\in C^0(M_1, \Z_2)}} \sum_{\substack{\phi'_2, \phi'_3\\\in C^0(M_1', \Z_2)}} \sum_{\substack{\phi''_2, \phi''_3\\\in C^0(M_1'', \Z_2)}}   e^{i \int_{S^3} \pi \left(\widehat{a}\delta a + \widehat{b}\delta b+ \widehat{c}\delta c + abc + \frac{1}{2} a \delta a \right)} e^{i\pi \oint_{M_1} \widehat{a} + i\pi \oint_{M_1} (-\phi_2 c + \phi_3 b + \phi_2 \delta \phi_3)} \\& \hspace{1.5cm} \times e^{i\pi \oint_{M_1'} \widehat{b} + i\pi \oint_{M_1'} (-\phi'_3 a + \phi'_1 c + \phi'_3 \delta \phi'_1)} e^{i\pi \oint_{M_1''} \widehat{c} + i\pi \oint_{M_1''} (-\phi''_1 b + \phi''_2 a + \phi''_1 \delta \phi''_2)}~.
    \end{split}
\end{equation}
Summing over $\widehat{a}, \widehat{b}, \widehat{c}$ enforces that $\delta a = -  \delta^{\perp}(M_1)$,  $\delta b = -\delta^{\perp}(M_1')$, and $\delta c = -\delta^{\perp}(M_1'')$. These also imply that up to gauge transformation  $a= - \delta^{\perp}(M_2)$, $b= -\delta^{\perp}(M_2')$, and $c= -\delta^{\perp}(M_2'')$, where $M_1=\partial M_2$, $M_1'=\partial M_2'$, and $M_1''=\partial M_2''$.  The correlation function then simplifies to 
\begin{equation}
\begin{split}
&\braket{\widehat{U}_a(M_1) \widehat{U}_{b}(M_1') \widehat{U}_c(M_1'')}\\ 
&\qquad\sim \sum_{\substack{a,b,c\\
		\in C^1(S^3, \Z_2)
}}  \sum_{\substack{\phi_2, \phi_3\\\in C^0(M_1, \Z_2)}} \sum_{\substack{\phi_3', \phi_1'\\ \in C^0(M_1', \Z_2)}} \sum_{\substack{\phi_1'', \phi_2''\\\in C^0(M_1'', \Z_2)}} e^{i \int_{S^3} \pi \left(abc + \frac{1}{2} a \delta a \right)}e^{ i\pi \oint_{M_1} (-\phi_2 c + \phi_3 b + \phi_2 \delta \phi_3)} \\
&\hspace{1.5cm} \times e^{i\pi \oint_{M_1'} (-\phi'_3 a + \phi'_1 c + \phi'_3 \delta \phi'_1)}
 e^{i\pi \oint_{M_1''} (-\phi''_1 b + \phi''_2 a + \phi''_1 \delta \phi''_2)}|_{a = -  \delta^{\perp}(M_2),  b = -  \delta^{\perp}(M_2'),  c = -  \delta^{\perp}(M_2'')}\\
&\qquad\sim \sum_{\substack{\phi_2, \phi_3\\\in C^0(M_1, \Z_2)}} \sum_{\substack{\phi_3', \phi_1'\\ \in C^0(M_1', \Z_2)}} \sum_{\substack{\phi_1'', \phi_2''\\\in C^0(M_1'', \Z_2)}} e^{i \int_{S^3} \pi \delta^{\perp}(M_2\cap M_2' \cap M_2'')  }\\
&\hspace{1.5cm} \times e^{i \pi \int_{S^3} \phi_2 \delta^{\perp}(M_1\cap M_2'') - \phi_3 \delta^{\perp}(M_1\cap M_2') + \phi_2 \delta \phi_3 \delta^{\perp}(M_1)} e^{ i \pi \int_{S^3} \phi_3' \delta^{\perp}(M_1'\cap M_2) - \phi_1' \delta^{\perp}(M_1'\cap M_2'')+\phi_3' \delta \phi_1'  \delta^{\perp}(M_1')} \\
&\hspace{1.5cm} \times e^{i\pi \int_{S^3} \phi_1'' \delta^{\perp}(M_1''\cap M_2')- \phi_2'' \delta^{\perp}(M_1''\cap M_2)+ \phi_1'' \delta \phi_2''  \delta^{\perp}(M_1'')}~.
\end{split}
\end{equation}
We now simplify the sum over $\phi_2, \phi_3$. The relevant portion is 
\begin{equation}
\begin{split}
&\sum_{\substack{\phi_2, \phi_3\in C^0(M_1, \Z_2)}} e^{i \pi \int_{S^3} \phi_2 \delta^{\perp}(M_1\cap M_2'') - \phi_3 \delta^{\perp}(M_1\cap M_2') + \phi_2 \delta \phi_3 \delta^{\perp}(M_1)}
\\&\hspace{0.7 in} = \sum_{\substack{\phi_3\in C^0(M_1, \Z_2)}} e^{i \pi \int_{S^3} \phi_3 \delta^{\perp}(M_1\cap M_2') }|_{\delta \phi_3 \delta^{\perp}(M_1)= \delta^{\perp}(M_1\cap M_2'')}~.
\end{split}
\end{equation}
The constraint means $\delta \phi_3 = \delta^{\perp}(M_2'')$. However, since $\delta \delta^{\perp}(M_2'') = \delta^{\perp}(M_1'')$, $\delta^{\perp}(M_2'')$ itself can not be an exact form. Hence the constraint is solvable only when  $M_1$ and $M_1''$ do not form a Hopf link, i.e. $\delta^{\perp}(M_1\cap M_2'')=0\mod 2$. Hence $\delta \phi_3=0$, i.e. $\phi_3=0,1$ is a constant over $M_1$. Further summing over $\phi_3$ constrains $\delta^{\perp}(M_1\cap M_2')=0\mod 2$, so we need both $M_1'$ and $M_1''$ to have trivial linking number with $M_1$ in order to obtain a non-vanishing correlation function.  Further summing over $\phi_2', \phi_3'$ and $\phi_2'', \phi_3''$, we find that the correlation function is non-vanishing only when $M_1, M_1'$, and $M_1''$ have trivial linking number between any pair, and in particular they are not mutually Hopf linked. Moreover, the phase of the correlation function is non-trivial when $M_1, M_1'$, and $M_1''$ form a non-trivial Borromean ring. The final result of the correlation function is
\begin{eqnarray}
\braket{\widehat{U}_a(M_1) \widehat{U}_{b}(M_1') \widehat{U}_c(M_1'')}\sim (-1)^{ \int_{S^3} \delta^{\perp}(M_2\cap M_2' \cap M_2'') }\sim  (-1)^{\text{Link}(M_1,M_1',M_1'')_{0}}~,
\end{eqnarray}
as quoted in (\ref{eq:ABClink}).

\section{$\Z_{2MN}^{(0)}\times \Z_{N}^{(1)}$ Dijkgraaf-Witten model in 5d}
\label{app:5d}

In this appendix we study the properties of the 5d Dijkgraaf-Witten model given in (\ref{eq:AnomTFT5d}). For convenience, we reproduce the action here
\begin{eqnarray}\label{eq:5dSymTFT1}
\int_{X_5} \left( \frac{2\pi}{2MN} \widehat{a}^{(3)} \delta a^{(1)} + \frac{2\pi}{N} \widehat{b}^{(2)}\delta b^{(2)} + \frac{2\pi}{2N} a^{(1)} b^{(2)} b^{(2)} + \frac{\pi(N^2-1)}{6 N} a^{(1)} \frac{\delta a^{(1)}}{2 MN} \frac{\delta a^{(1)}}{2 MN}\right)~,
\end{eqnarray}
where we take $N$ to be even. 
All the gauge fields are either $\Z_{2MN}$ cochains (labeled by $a$ or $\widehat{a}$) or $\Z_N$ cochains (labeled by $b$ or $\widehat{b}$) with appropriate form degree indicated in their superscripts. 
For simplicity, we will assume that $X_5$ is a spin manifold.
The action is invariant under the following gauge transformation, 
\begin{eqnarray}
\begin{split}
    a^{(1)} &\to  a^{(1)} + \delta \alpha^{(0)}~,\\
    b^{(2)} &\to b^{(2)} + \delta \beta^{(1)}~,\\
    \widehat{a}^{(3)} &\to \widehat{a}^{(3)} + \delta \widehat{\alpha}^{(2)} - 2 
 M \beta^{(1)} b^{(2)} - M\beta^{(1)} \delta \beta^{(1)}~,\\
    \widehat{b}^{(2)} &\to \widehat{b}^{(2)} + \delta \widehat{\beta}^{(1)} + \alpha^{(0)} b^{(2)} + \beta^{(1)} a^{(1)} + \alpha^{(0)} \delta \beta^{(1)}~.
\end{split}
\end{eqnarray}

\subsection{Extended operators and fusion rules}

The obvious gauge invariant operators are the Wilson lines of $a^{(1)}$ and the Wilson surfaces of $b^{(2)}$, 
\begin{eqnarray}
U_a(M_1)= e^{\frac{2\pi i}{2 MN}\oint_{M_1} a^{(1)}}, \hspace{1cm} U_{b}(M_2) = e^{ \frac{2\pi i}{N} \oint_{M_2} b^{(2)}}~,
\end{eqnarray}
which satisfy the obvious fusion rules
\begin{eqnarray}
U_a^{2 MN} =1~, \hspace{1cm} U_b^{N}=1~.
\end{eqnarray}
The magnetic operators are more interesting. Because of the non-trivial  gauge transformations of $\widehat{a}^{(3)}$ and $\widehat{b}^{(2)}$, their naive Wilson operators are not gauge invariant. Instead, we should attach a 3d TQFT to $e^{\frac{2\pi}{2MN} \oint_{M_3} \widehat{a}^{(3)}}$, and a 2d TQFT to $e^{i \frac{2\pi}{N} \oint_{M_2} \widehat{b}^{(2)}}$ to maintain gauge invariance. This can be achieved by taking the gauge invariant magnetic operators to be
\begin{eqnarray}\label{eq:magneticop5d}
\begin{split}
    \widehat{U}_{a}(M_3) &\sim \sum_{\phi^{(1)}\in C^1(M_3, \Z_{N})} e^{\frac{2\pi i}{2MN}\oint_{M_3} \left( \widehat{a}^{(3)} - M\phi^{(1)} \delta \phi^{(1)} + 2 M\phi^{(1)} b^{(2)} \right)}~,\\
    \widehat{U}_{b}(M_2) &\sim \sum_{\substack{\phi^{(0)}\in C^0(M_2, \Z_{2MN})\\ \phi^{(1)}\in C^1(M_2, \Z_{N})}} e^{{2\pi i \over N}\oint_{M_2} \left( \widehat{b}^{(2)} - \phi^{(0)}b^{(2)} -  \phi^{(1)} a^{(1)} + \phi^{(1)} \delta \phi^{(0)} \right)}~,\\
\end{split}
\end{eqnarray}
where the fields $\phi^{(0)}$ and $\phi^{(1)}$ live only on the defect worldvolumes and transform as
\begin{eqnarray}
\phi^{(0)}\to \phi^{(0)} + \alpha^{(0)}~, \hspace{1cm} \phi^{(1)}\to \phi^{(1)} + \beta^{(1)}~.
\end{eqnarray}
Because of the non-trivial TQFT on the defect worldvolumes, the defects \eqref{eq:magneticop5d} are non-invertible.

We proceed to consider fusion rules involving the non-invertible operators. First, fusing $\widehat{U}_{a}(M_3)$ with $U_a(M_1)$ produces a new operator, which is a line on $M_1$ living within the 3-volume $M_3$, i.e. $\widehat{U}_{a}(M_3)\times U_a(M_1)$. One can also insert multiple $U_a$ operators within $M_3$, or even construct condensation surface and 3-volume defects of $U_a$ and fuse them with $\widehat{U}_{a}(M_3)$. All of these will give rise to new defects. However, when computing the correlation functions, we will only consider the simple topology without non-contractible cycles $M_3=S^3$, and hence the condensation defects are trivial.  This significantly simplifies the computation.

On the other hand, fusing $U_b(M_2)$ with $\widehat{U}_{a}(M_3)$ with $M_2 \subset M_3$ trivializes $U_b$. To see this, we consider 
\begin{equation}
\begin{split}
    \widehat{U}_{a}(M_3) \times U_b(M_2) &\sim \sum_{\phi^{(1)}\in C^1(M_3, \Z_{N})} e^{\frac{2\pi i}{2MN}\oint_{M_3} \left( \widehat{a}^{(3)} - M \phi^{(1)} \delta \phi^{(1)} + 2M (\phi^{(1)} + \delta^{\perp}(M_2)) b^{(2)} \right)}\\
    &\sim  \sum_{\phi^{(1)}\in C^1(M_3, \Z_{N})} e^{\frac{2\pi i}{2 MN}\oint_{M_3} \left( \widehat{a}^{(3)} - M(\phi^{(1)}- \delta^{\perp}(M_2)) \delta (\phi^{(1)}- \delta^{\perp}(M_2)) + 2M \phi^{(1)} b^{(2)} \right)}\\
    &\sim  \sum_{\phi^{(1)}\in C^1(M_3, \Z_{N})} e^{\frac{2\pi i}{2 MN}\oint_{M_3} \left( \widehat{a}^{(3)} - M\phi^{(1)}\delta \phi^{(1)} + 2 M\phi^{(1)} b^{(2)} \right)} e^{{2\pi i \over N} \oint_{M_2} \delta \phi^{(1)}}\\
    &\sim \widehat{U}_{a}(M_3)~.
\end{split}
\end{equation}
In the first line we used $e^{{2 \pi i \over N}\oint_{M_2} b^{(2)}}= e^{{2 \pi i \over N} \int_{M_3} \delta^{\perp}(M_2) b^{(2)}}$. 
In the second line we performed a change of variable $\phi^{(1)}\to \phi^{(1)}- \delta^{\perp}(M_2)$. In the last line we used $e^{{2\pi i \over N} \oint_{M_2} \delta \phi^{(1)}}=1$ since $M_2$ is closed.

Similarly, we find that $\widehat{U}_b(M_2)\times U_a(M_1)$ gives rise to a new operator, which is a line living inside a surface. On the other hand, fusing $U_a^{2M}(M_1)$ with $\widehat{U}_b(M_2)$ does not produce a new operator, because by the same discussion as in the previous paragraph, multiplying by $U_a^{2M}(M_1)$ amounts to shifting $\phi^{(1)}\to \phi^{(1)}- \delta^{\perp}(M_1)$, and by field redefinition amounts to multiplying by $e^{{2 \pi i \over N_c}\oint_{M_1} \delta \phi^{(0)}}=1$. Hence $U_a$ becomes a $\Z_{2M}$ operator within the worldvolume of $\widehat{U}_b(M_2)$. Finally, a similar discussion shows that $\widehat{U}_b(M_2)\times U_b(M_2)= \widehat{U}_b(M_2)$.

We finally consider the fusion rules between the non-invertible magnetic operators.  Let us first consider the fusion rule $\widehat{U}_a\times \overline{\widehat{U}}_a$, 
\begin{equation}
    \begin{split}
        \widehat{U}_a(M_3)\times \overline{\widehat{U}}_a(M_3) & \sim  \sum_{ \phi^{(1)}, \phi'^{(1)}\in C^1(M_3, \Z_{N})} e^{\frac{2\pi i}{{2 N }} \int_{M_3} - \phi^{(1)} \delta \phi^{(1)} + 2 \phi^{(1)} b^{(2)} + \phi'^{(1)} \delta \phi'^{(1)} - 2 \phi'^{(1)} b^{(2)}}\\
        &\sim  \sum_{ \phi^{(1)}, \phi'^{(1)}\in C^1(M_3, \Z_{N})} e^{\frac{2\pi i}{{2 N }} \int_{M_3}  \phi'^{(1)} \delta \phi'^{(1)} - 2 \phi'^{(1)} b^{(2)} + 2 \phi'^{(1)} \delta \phi^{(1)}}\\
        &\sim \sum_{ \phi'^{(1)}\in Z^1(M_3, \Z_{N})} e^{\frac{2\pi i}{{2 N}} \int_{M_3}  \phi'^{(1)} \delta \phi'^{(1)} - 2 \phi'^{(1)} b^{(2)} }\\
        &\sim  \sum_{\phi'^{(1)}\in H^1(M_3, \Z_{N})} e^{{i \pi}Q(\text{PD}(\phi'^{(1)}))}  e^{{2 \pi i \over N} \oint_{\text{PD}(\phi'^{(1)})} b^{(2)}}~.
    \end{split}
\end{equation}
In the fourth line, $\text{PD}(\phi'^{(1)})$ is the Poincare dual of $\phi^{(1)}$, and we also defined the triple intersection number $Q(\text{PD}(\phi'^{(1)})) := \frac{1}{N}\int \phi^{(1)} \delta \phi^{(1)}$ mod $N$ which is trivial for odd $N$ and can be non-trivial for even $N$. 
Because the right-hand-side of the fusion rule is a sum of multiple terms,  $\widehat{U}_a$ is non-invertible. A similar calculation yields
\begin{eqnarray}
\begin{split}
    \widehat{U}_b\times \overline{\widehat{U}}_b  \sim \sum_{\substack{\phi^{(0)}\in H^0(M_2, \Z_{2MN})\\ \phi^{(1)}\in H^1(M_2, \Z_{N})}} e^{{2\pi i \over N} \oint_{\text{PD}(\phi'^{(0)})} b^{(2)} + {2\pi i \over N} \oint_{\text{PD}(\phi'^{(1)})} a^{(1)}}
\end{split}
\end{eqnarray}
which implies that $\widehat{U}_b$ is also non-invertible.

\subsection{Linking numbers between extended operators}

We now describe the linking numbers between the extended operators. 
The non-trivial linking numbers are as follows: 
\begin{enumerate}
    \item  Linking number between the electric and magnetic surfaces, i.e. 
    \begin{eqnarray}
    \braket{U_b(M_2) \widehat{U}_b(M'_2)} \sim e^{-{2\pi i \over N}\text{Link}(M_2, M'_2)}~.
    \end{eqnarray}
    This is a consequence of the BF coupling ${2\pi\over N} \widehat{b}\delta b$ in \eqref{eq:5dSymTFT1}. Moreover, the overall normalization is nonzero only when $b^{(2)}$ is pure gauge on $M_2'$. 
    \item  Linking number between the electric line and the magnetic 3-volume operator, i.e. 
    \begin{eqnarray}
    \braket{U_a(M_1) \widehat{U}_a(M'_3)} \sim e^{-\frac{2\pi i}{2 MN}\text{Link}(M_1, M_3')}~.
    \end{eqnarray}
    This is a consequence of the BF coupling $\frac{2\pi}{2 MN} \widehat{a}\delta a$ in \eqref{eq:5dSymTFT1}. 
    \item  Linking number of type 0 between two $\widehat{U}_b$  magnetic surface operators and one $\widehat{U}_a$ magnetic 3-volume operator, i.e.
    \begin{eqnarray}\label{eq:trip0}
    \braket{\widehat{U}_a(M_3) \widehat{U}_b (M'_2) \widehat{U}_b(M''_2)} \sim e^{-\frac{2\pi i}{N} \text{Link}(M_3, M'_2, M''_2)_{0}}~.
    \end{eqnarray}
    This is a consequence of the DW twist term $\frac{2 \pi}{ 2 N} a^{(1)} b^{(2)} b^{(2)}$ in 
    \eqref{eq:5dSymTFT1}.  
    
    \item  Linking number of type 2 between three $\widehat{U}_a$  magnetic 3-volume operators, i.e. 
    \begin{eqnarray}\label{eq:trip2}
    \braket{\widehat{U}_a(M_3) \widehat{U}_a (M'_3) \widehat{U}_a(M''_3)} \sim e^{-{ i \pi (N^2 - 1) \over 4 N^3 M^2}\text{Link}(M_3, M'_3, M''_3)_{2}}~.
    \end{eqnarray}
    This is the consequence of the final term in 
    \eqref{eq:5dSymTFT1}. Note that the two triple linking invariants 
\eqref{eq:trip0} and \eqref{eq:trip2} are different: one is between a 3-volume and two surface operators, while the other one is between three 3-volume operators. 
\end{enumerate}
We finally note that when the defect worldvolume has non-trivial topology, the TQFTs on the defect worldvolumes can contribute additional phases, hence contaminating the linking numbers found above.

\bibliographystyle{ytphys}
\baselineskip=.95\baselineskip
\bibliography{bib.bib}

\providecommand{\href}[2]{#2}\begingroup\raggedright\begin{thebibliography}{100}

\bibitem{verlinde1988fusion}
E.~Verlinde, ``Fusion rules and modular transformations in 2d conformal field
  theory,'' {\em Nuclear Physics B} {\bfseries 300} (1988) 360--376.

\bibitem{Petkova:2000ip}
V.~B. Petkova and J.~B. Zuber, ``{Generalized twisted partition functions},''
  \href{http://dx.doi.org/10.1016/S0370-2693(01)00276-3}{{\em Phys. Lett. B}
  {\bfseries 504} (2001) 157--164},
  \href{http://arxiv.org/abs/hep-th/0011021}{{\ttfamily arXiv:hep-th/0011021}}.

\bibitem{Fuchs:2002cm}
J.~Fuchs, I.~Runkel, and C.~Schweigert, ``{TFT construction of RCFT correlators
  1. Partition functions},''
  \href{http://dx.doi.org/10.1016/S0550-3213(02)00744-7}{{\em Nucl. Phys. B}
  {\bfseries 646} (2002) 353--497},
  \href{http://arxiv.org/abs/hep-th/0204148}{{\ttfamily arXiv:hep-th/0204148}}.

\bibitem{Bhardwaj:2017xup}
L.~Bhardwaj and Y.~Tachikawa, ``{On finite symmetries and their gauging in two
  dimensions},'' \href{http://dx.doi.org/10.1007/JHEP03(2018)189}{{\em JHEP}
  {\bfseries 03} (2018) 189}, \href{http://arxiv.org/abs/1704.02330}{{\ttfamily
  arXiv:1704.02330 [hep-th]}}.

\bibitem{Chang:2018iay}
C.-M. Chang, Y.-H. Lin, S.-H. Shao, Y.~Wang, and X.~Yin, ``{Topological Defect
  Lines and Renormalization Group Flows in Two Dimensions},''
  \href{http://dx.doi.org/10.1007/JHEP01(2019)026}{{\em JHEP} {\bfseries 01}
  (2019) 026}, \href{http://arxiv.org/abs/1802.04445}{{\ttfamily
  arXiv:1802.04445 [hep-th]}}.

\bibitem{Lin:2022dhv}
Y.-H. Lin, M.~Okada, S.~Seifnashri, and Y.~Tachikawa, ``{Asymptotic density of
  states in 2d CFTs with non-invertible symmetries},''
  \href{http://arxiv.org/abs/2208.05495}{{\ttfamily arXiv:2208.05495
  [hep-th]}}.

\bibitem{Komargodski:2020mxz}
Z.~Komargodski, K.~Ohmori, K.~Roumpedakis, and S.~Seifnashri, ``{Symmetries and
  strings of adjoint QCD$_{2}$},''
  \href{http://dx.doi.org/10.1007/JHEP03(2021)103}{{\em JHEP} {\bfseries 03}
  (2021) 103}, \href{http://arxiv.org/abs/2008.07567}{{\ttfamily
  arXiv:2008.07567 [hep-th]}}.

\bibitem{Tachikawa:2017gyf}
Y.~Tachikawa, ``{On gauging finite subgroups},''
  \href{http://dx.doi.org/10.21468/SciPostPhys.8.1.015}{{\em SciPost Phys.}
  {\bfseries 8} no.~1, (2020) 015},
  \href{http://arxiv.org/abs/1712.09542}{{\ttfamily arXiv:1712.09542
  [hep-th]}}.

\bibitem{Frohlich:2004ef}
J.~Frohlich, J.~Fuchs, I.~Runkel, and C.~Schweigert, ``{Kramers-Wannier duality
  from conformal defects},''
  \href{http://dx.doi.org/10.1103/PhysRevLett.93.070601}{{\em Phys. Rev. Lett.}
  {\bfseries 93} (2004) 070601},
  \href{http://arxiv.org/abs/cond-mat/0404051}{{\ttfamily
  arXiv:cond-mat/0404051}}.

\bibitem{Frohlich:2006ch}
J.~Frohlich, J.~Fuchs, I.~Runkel, and C.~Schweigert, ``{Duality and defects in
  rational conformal field theory},''
  \href{http://dx.doi.org/10.1016/j.nuclphysb.2006.11.017}{{\em Nucl. Phys. B}
  {\bfseries 763} (2007) 354--430},
  \href{http://arxiv.org/abs/hep-th/0607247}{{\ttfamily arXiv:hep-th/0607247}}.

\bibitem{Frohlich:2009gb}
J.~Frohlich, J.~Fuchs, I.~Runkel, and C.~Schweigert,
  \href{http://dx.doi.org/10.1142/9789814304634_0056}{``{Defect lines,
  dualities, and generalised orbifolds},''} in {\em {16th International
  Congress on Mathematical Physics}}.
\newblock 9, 2009.
\newblock \href{http://arxiv.org/abs/0909.5013}{{\ttfamily arXiv:0909.5013
  [math-ph]}}.

\bibitem{Carqueville:2012dk}
N.~Carqueville and I.~Runkel, ``{Orbifold completion of defect bicategories},''
  \href{http://dx.doi.org/10.4171/qt/76}{{\em Quantum Topol.} {\bfseries 7}
  no.~2, (2016) 203--279}, \href{http://arxiv.org/abs/1210.6363}{{\ttfamily
  arXiv:1210.6363 [math.QA]}}.

\bibitem{Brunner:2013xna}
I.~Brunner, N.~Carqueville, and D.~Plencner, ``{A quick guide to defect
  orbifolds},'' \href{http://dx.doi.org/10.1090/pspum/088/01456}{{\em Proc.
  Symp. Pure Math.} {\bfseries 88} (2014) 231--242},
  \href{http://arxiv.org/abs/1310.0062}{{\ttfamily arXiv:1310.0062 [hep-th]}}.

\bibitem{Huang:2021zvu}
T.-C. Huang, Y.-H. Lin, and S.~Seifnashri, ``{Construction of two-dimensional
  topological field theories with non-invertible symmetries},''
  \href{http://dx.doi.org/10.1007/JHEP12(2021)028}{{\em JHEP} {\bfseries 12}
  (2021) 028}, \href{http://arxiv.org/abs/2110.02958}{{\ttfamily
  arXiv:2110.02958 [hep-th]}}.

\bibitem{Thorngren:2019iar}
R.~Thorngren and Y.~Wang, ``{Fusion Category Symmetry I: Anomaly In-Flow and
  Gapped Phases},'' \href{http://arxiv.org/abs/1912.02817}{{\ttfamily
  arXiv:1912.02817 [hep-th]}}.

\bibitem{Thorngren:2021yso}
R.~Thorngren and Y.~Wang, ``{Fusion Category Symmetry II: Categoriosities at
  $c$ = 1 and Beyond},'' \href{http://arxiv.org/abs/2106.12577}{{\ttfamily
  arXiv:2106.12577 [hep-th]}}.

\bibitem{Lootens:2021tet}
L.~Lootens, C.~Delcamp, G.~Ortiz, and F.~Verstraete, ``{Dualities in
  one-dimensional quantum lattice models: symmetric Hamiltonians and matrix
  product operator intertwiners},''
  \href{http://arxiv.org/abs/2112.09091}{{\ttfamily arXiv:2112.09091
  [quant-ph]}}.

\bibitem{Huang:2021nvb}
T.-C. Huang, Y.-H. Lin, K.~Ohmori, Y.~Tachikawa, and M.~Tezuka, ``{Numerical
  Evidence for a Haagerup Conformal Field Theory},''
  \href{http://dx.doi.org/10.1103/PhysRevLett.128.231603}{{\em Phys. Rev.
  Lett.} {\bfseries 128} no.~23, (2022) 231603},
  \href{http://arxiv.org/abs/2110.03008}{{\ttfamily arXiv:2110.03008
  [cond-mat.stat-mech]}}.

\bibitem{Inamura:2022lun}
K.~Inamura, ``{Fermionization of fusion category symmetries in 1+1
  dimensions},'' \href{http://arxiv.org/abs/2206.13159}{{\ttfamily
  arXiv:2206.13159 [cond-mat.str-el]}}.

\bibitem{Ji:2019jhk}
W.~Ji and X.-G. Wen, ``{Categorical symmetry and noninvertible anomaly in
  symmetry-breaking and topological phase transitions},''
  \href{http://dx.doi.org/10.1103/PhysRevResearch.2.033417}{{\em Phys. Rev.
  Res.} {\bfseries 2} no.~3, (2020) 033417},
  \href{http://arxiv.org/abs/1912.13492}{{\ttfamily arXiv:1912.13492
  [cond-mat.str-el]}}.

\bibitem{Kong:2020cie}
L.~Kong, T.~Lan, X.-G. Wen, Z.-H. Zhang, and H.~Zheng, ``{Algebraic higher
  symmetry and categorical symmetry -- a holographic and entanglement view of
  symmetry},'' \href{http://dx.doi.org/10.1103/PhysRevResearch.2.043086}{{\em
  Phys. Rev. Res.} {\bfseries 2} no.~4, (2020) 043086},
  \href{http://arxiv.org/abs/2005.14178}{{\ttfamily arXiv:2005.14178
  [cond-mat.str-el]}}.

\bibitem{Ji:2021esj}
W.~Ji and X.-G. Wen, ``{A unified view on symmetry, anomalous symmetry and
  non-invertible gravitational anomaly},''
  \href{http://arxiv.org/abs/2106.02069}{{\ttfamily arXiv:2106.02069
  [cond-mat.str-el]}}.

\bibitem{Chatterjee:2022kxb}
A.~Chatterjee and X.-G. Wen, ``{Algebra of local symmetric operators and
  braided fusion $n$-category -- symmetry is a shadow of topological order},''
  \href{http://arxiv.org/abs/2203.03596}{{\ttfamily arXiv:2203.03596
  [cond-mat.str-el]}}.

\bibitem{Chatterjee:2022tyg}
A.~Chatterjee and X.-G. Wen, ``{Holographic theory for the emergence and the
  symmetry protection of gaplessness and for continuous phase transitions},''
  \href{http://arxiv.org/abs/2205.06244}{{\ttfamily arXiv:2205.06244
  [cond-mat.str-el]}}.

\bibitem{Moradi:2022lqp}
H.~Moradi, S.~F. Moosavian, and A.~Tiwari, ``{Topological holography: Towards a
  unification of Landau and beyond-Landau physics},''
  \href{http://arxiv.org/abs/2207.10712}{{\ttfamily arXiv:2207.10712
  [cond-mat.str-el]}}.

\bibitem{Kaidi:2021xfk}
J.~Kaidi, K.~Ohmori, and Y.~Zheng, ``{Kramers-Wannier-like Duality Defects in
  (3+1)D Gauge Theories},''
  \href{http://dx.doi.org/10.1103/PhysRevLett.128.111601}{{\em Phys. Rev.
  Lett.} {\bfseries 128} no.~11, (2022) 111601},
  \href{http://arxiv.org/abs/2111.01141}{{\ttfamily arXiv:2111.01141
  [hep-th]}}.

\bibitem{Choi:2021kmx}
Y.~Choi, C.~Cordova, P.-S. Hsin, H.~T. Lam, and S.-H. Shao, ``{Non-Invertible
  Duality Defects in 3+1 Dimensions},''
  \href{http://arxiv.org/abs/2111.01139}{{\ttfamily arXiv:2111.01139
  [hep-th]}}.

\bibitem{Koide:2021zxj}
M.~Koide, Y.~Nagoya, and S.~Yamaguchi, ``{Non-invertible topological defects in
  4-dimensional $\mathbb{Z}_2$ pure lattice gauge theory},''
  \href{http://arxiv.org/abs/2109.05992}{{\ttfamily arXiv:2109.05992
  [hep-th]}}.

\bibitem{Choi:2022zal}
Y.~Choi, C.~Cordova, P.-S. Hsin, H.~T. Lam, and S.-H. Shao, ``{Non-invertible
  Condensation, Duality, and Triality Defects in 3+1 Dimensions},''
  \href{http://arxiv.org/abs/2204.09025}{{\ttfamily arXiv:2204.09025
  [hep-th]}}.

\bibitem{Apruzzi:2021nmk}
F.~Apruzzi, F.~Bonetti, I.~n.~G. Etxebarria, S.~S. Hosseini, and
  S.~Schafer-Nameki, ``{Symmetry TFTs from String Theory},''
  \href{http://arxiv.org/abs/2112.02092}{{\ttfamily arXiv:2112.02092
  [hep-th]}}.

\bibitem{Arias-Tamargo:2022nlf}
G.~Arias-Tamargo and D.~Rodriguez-Gomez, ``{Non-Invertible Symmetries from
  Discrete Gauging and Completeness of the Spectrum},''
  \href{http://arxiv.org/abs/2204.07523}{{\ttfamily arXiv:2204.07523
  [hep-th]}}.

\bibitem{Hayashi:2022fkw}
Y.~Hayashi and Y.~Tanizaki, ``{Non-invertible self-duality defects of
  Cardy-Rabinovici model and mixed gravitational anomaly},''
  \href{http://arxiv.org/abs/2204.07440}{{\ttfamily arXiv:2204.07440
  [hep-th]}}.

\bibitem{Roumpedakis:2022aik}
K.~Roumpedakis, S.~Seifnashri, and S.-H. Shao, ``{Higher Gauging and
  Non-invertible Condensation Defects},''
  \href{http://arxiv.org/abs/2204.02407}{{\ttfamily arXiv:2204.02407
  [hep-th]}}.

\bibitem{Bhardwaj:2022yxj}
L.~Bhardwaj, L.~Bottini, S.~Schafer-Nameki, and A.~Tiwari, ``{Non-Invertible
  Higher-Categorical Symmetries},''
  \href{http://arxiv.org/abs/2204.06564}{{\ttfamily arXiv:2204.06564
  [hep-th]}}.

\bibitem{Kaidi:2022uux}
J.~Kaidi, G.~Zafrir, and Y.~Zheng, ``{Non-invertible symmetries of $
  \mathcal{N} $ = 4 SYM and twisted compactification},''
  \href{http://dx.doi.org/10.1007/JHEP08(2022)053}{{\em JHEP} {\bfseries 08}
  (2022) 053}, \href{http://arxiv.org/abs/2205.01104}{{\ttfamily
  arXiv:2205.01104 [hep-th]}}.

\bibitem{Choi:2022jqy}
Y.~Choi, H.~T. Lam, and S.-H. Shao, ``{Non-invertible Global Symmetries in the
  Standard Model},'' \href{http://arxiv.org/abs/2205.05086}{{\ttfamily
  arXiv:2205.05086 [hep-th]}}.

\bibitem{Cordova:2022ieu}
C.~Cordova and K.~Ohmori, ``{Non-Invertible Chiral Symmetry and Exponential
  Hierarchies},'' \href{http://arxiv.org/abs/2205.06243}{{\ttfamily
  arXiv:2205.06243 [hep-th]}}.

\bibitem{Antinucci:2022eat}
A.~Antinucci, G.~Galati, and G.~Rizi, ``{On Continuous 2-Category Symmetries
  and Yang-Mills Theory},'' \href{http://arxiv.org/abs/2206.05646}{{\ttfamily
  arXiv:2206.05646 [hep-th]}}.

\bibitem{Bashmakov:2022jtl}
V.~Bashmakov, M.~Del~Zotto, and A.~Hasan, ``{On the 6d Origin of Non-invertible
  Symmetries in 4d},'' \href{http://arxiv.org/abs/2206.07073}{{\ttfamily
  arXiv:2206.07073 [hep-th]}}.

\bibitem{Damia:2022rxw}
J.~A. Damia, R.~Argurio, and L.~Tizzano, ``{Continuous Generalized Symmetries
  in Three Dimensions},'' \href{http://arxiv.org/abs/2206.14093}{{\ttfamily
  arXiv:2206.14093 [hep-th]}}.

\bibitem{Damia:2022bcd}
J.~A. Damia, R.~Argurio, and E.~Garcia-Valdecasas, ``{Non-Invertible Defects in
  5d, Boundaries and Holography},''
  \href{http://arxiv.org/abs/2207.02831}{{\ttfamily arXiv:2207.02831
  [hep-th]}}.

\bibitem{Choi:2022rfe}
Y.~Choi, H.~T. Lam, and S.-H. Shao, ``{Non-invertible Time-reversal
  Symmetry},'' \href{http://arxiv.org/abs/2208.04331}{{\ttfamily
  arXiv:2208.04331 [hep-th]}}.

\bibitem{Lu:2022ver}
D.-C. Lu and Z.~Sun, ``{On Triality Defects in 2d CFT},''
  \href{http://arxiv.org/abs/2208.06077}{{\ttfamily arXiv:2208.06077
  [hep-th]}}.

\bibitem{Bhardwaj:2022lsg}
L.~Bhardwaj, S.~Schafer-Nameki, and J.~Wu, ``{Universal Non-Invertible
  Symmetries},'' \href{http://arxiv.org/abs/2208.05973}{{\ttfamily
  arXiv:2208.05973 [hep-th]}}.

\bibitem{Bartsch:2022mpm}
T.~Bartsch, M.~Bullimore, A.~E.~V. Ferrari, and J.~Pearson, ``{Non-invertible
  Symmetries and Higher Representation Theory I},''
  \href{http://arxiv.org/abs/2208.05993}{{\ttfamily arXiv:2208.05993
  [hep-th]}}.

\bibitem{Lin:2022xod}
L.~Lin, D.~G. Robbins, and E.~Sharpe, ``{Decomposition, condensation defects,
  and fusion},'' \href{http://arxiv.org/abs/2208.05982}{{\ttfamily
  arXiv:2208.05982 [hep-th]}}.

\bibitem{Apruzzi:2022rei}
F.~Apruzzi, I.~Bah, F.~Bonetti, and S.~Schafer-Nameki, ``{Non-Invertible
  Symmetries from Holography and Branes},''
  \href{http://arxiv.org/abs/2208.07373}{{\ttfamily arXiv:2208.07373
  [hep-th]}}.

\bibitem{GarciaEtxebarria:2022vzq}
I.~n. Garc\'\i{}a~Etxebarria, ``{Branes and Non-Invertible Symmetries},''
  \href{http://arxiv.org/abs/2208.07508}{{\ttfamily arXiv:2208.07508
  [hep-th]}}.

\bibitem{Benini:2022hzx}
F.~Benini, C.~Copetti, and L.~Di~Pietro, ``{Factorization and global symmetries
  in holography},'' \href{http://arxiv.org/abs/2203.09537}{{\ttfamily
  arXiv:2203.09537 [hep-th]}}.

\bibitem{Wang:2021vki}
J.~Wang and Y.-Z. You, ``{Gauge Enhanced Quantum Criticality Between Grand
  Unifications: Categorical Higher Symmetry Retraction},''
  \href{http://arxiv.org/abs/2111.10369}{{\ttfamily arXiv:2111.10369
  [hep-th]}}.

\bibitem{Chen:2021xuc}
X.~Chen, A.~Dua, P.-S. Hsin, C.-M. Jian, W.~Shirley, and C.~Xu, ``{Loops in
  4+1d Topological Phases},'' \href{http://arxiv.org/abs/2112.02137}{{\ttfamily
  arXiv:2112.02137 [cond-mat.str-el]}}.

\bibitem{DelZotto:2022ras}
M.~Del~Zotto and I.~n. Garc\'\i{}a~Etxebarria, ``{Global Structures from the
  Infrared},'' \href{http://arxiv.org/abs/2204.06495}{{\ttfamily
  arXiv:2204.06495 [hep-th]}}.

\bibitem{Bhardwaj:2022dyt}
L.~Bhardwaj, M.~Bullimore, A.~E.~V. Ferrari, and S.~Schafer-Nameki,
  ``{Anomalies of Generalized Symmetries from Solitonic Defects},''
  \href{http://arxiv.org/abs/2205.15330}{{\ttfamily arXiv:2205.15330
  [hep-th]}}.

\bibitem{Brennan:2022tyl}
T.~D. Brennan, C.~Cordova, and T.~T. Dumitrescu, ``{Line Defect Quantum Numbers
  \& Anomalies},'' \href{http://arxiv.org/abs/2206.15401}{{\ttfamily
  arXiv:2206.15401 [hep-th]}}.

\bibitem{Delmastro:2022pfo}
D.~Delmastro, J.~Gomis, P.-S. Hsin, and Z.~Komargodski, ``{Anomalies and
  Symmetry Fractionalization},''
  \href{http://arxiv.org/abs/2206.15118}{{\ttfamily arXiv:2206.15118
  [hep-th]}}.

\bibitem{Heckman:2022muc}
J.~J. Heckman, M.~H\"ubner, E.~Torres, and H.~Y. Zhang, ``{The Branes Behind
  Generalized Symmetry Operators},''
  \href{http://arxiv.org/abs/2209.03343}{{\ttfamily arXiv:2209.03343
  [hep-th]}}.

\bibitem{Freed:2022qnc}
D.~S. Freed, G.~W. Moore, and C.~Teleman, ``{Topological symmetry in quantum
  field theory},'' \href{http://arxiv.org/abs/2209.07471}{{\ttfamily
  arXiv:2209.07471 [hep-th]}}.

\bibitem{Freed:2022iao}
D.~S. Freed, ``{Introduction to topological symmetry in QFT},''
  \href{http://arxiv.org/abs/2212.00195}{{\ttfamily arXiv:2212.00195
  [hep-th]}}.

\bibitem{Niro:2022ctq}
P.~Niro, K.~Roumpedakis, and O.~Sela, ``{Exploring Non-Invertible Symmetries in
  Free Theories},'' \href{http://arxiv.org/abs/2209.11166}{{\ttfamily
  arXiv:2209.11166 [hep-th]}}.

\bibitem{Kaidi:2022cpf}
J.~Kaidi, K.~Ohmori, and Y.~Zheng, ``{Symmetry TFTs for Non-Invertible
  Defects},'' \href{http://arxiv.org/abs/2209.11062}{{\ttfamily
  arXiv:2209.11062 [hep-th]}}.

\bibitem{Mekareeya:2022spm}
N.~Mekareeya and M.~Sacchi, ``{Mixed Anomalies, Two-groups, Non-Invertible
  Symmetries, and 3d Superconformal Indices},''
  \href{http://arxiv.org/abs/2210.02466}{{\ttfamily arXiv:2210.02466
  [hep-th]}}.

\bibitem{vanBeest:2022fss}
M.~van Beest, D.~S.~W. Gould, S.~Schafer-Nameki, and Y.-N. Wang, ``{Symmetry
  TFTs for 3d QFTs from M-theory},''
  \href{http://arxiv.org/abs/2210.03703}{{\ttfamily arXiv:2210.03703
  [hep-th]}}.

\bibitem{Antinucci:2022vyk}
A.~Antinucci, F.~Benini, C.~Copetti, G.~Galati, and G.~Rizi, ``{The holography
  of non-invertible self-duality symmetries},''
  \href{http://arxiv.org/abs/2210.09146}{{\ttfamily arXiv:2210.09146
  [hep-th]}}.

\bibitem{Chen:2022cyw}
S.~Chen and Y.~Tanizaki, ``{Solitonic symmetry beyond homotopy: invertibility
  from bordism and non-invertibility from TQFT},''
  \href{http://arxiv.org/abs/2210.13780}{{\ttfamily arXiv:2210.13780
  [hep-th]}}.

\bibitem{Bashmakov:2022uek}
V.~Bashmakov, M.~Del~Zotto, A.~Hasan, and J.~Kaidi, ``{Non-invertible
  Symmetries of Class $\mathcal{S}$ Theories},''
  \href{http://arxiv.org/abs/2211.05138}{{\ttfamily arXiv:2211.05138
  [hep-th]}}.

\bibitem{Karasik:2022kkq}
A.~Karasik, ``{On anomalies and gauging of U(1) non-invertible symmetries in 4d
  QED},'' \href{http://arxiv.org/abs/2211.05802}{{\ttfamily arXiv:2211.05802
  [hep-th]}}.

\bibitem{Cordova:2022fhg}
C.~Cordova, S.~Hong, S.~Koren, and K.~Ohmori, ``{Neutrino Masses from
  Generalized Symmetry Breaking},''
  \href{http://arxiv.org/abs/2211.07639}{{\ttfamily arXiv:2211.07639
  [hep-ph]}}.

\bibitem{Decoppet:2022dnz}
T.~D. D\'ecoppet and M.~Yu, ``{Gauging Noninvertible Defects: A 2-Categorical
  Perspective},'' \href{http://arxiv.org/abs/2211.08436}{{\ttfamily
  arXiv:2211.08436 [math.CT]}}.

\bibitem{GarciaEtxebarria:2022jky}
I.~n. Garc\'\i{}a~Etxebarria and N.~Iqbal, ``{A Goldstone theorem for
  continuous non-invertible symmetries},''
  \href{http://arxiv.org/abs/2211.09570}{{\ttfamily arXiv:2211.09570
  [hep-th]}}.

\bibitem{Choi:2022fgx}
Y.~Choi, H.~T. Lam, and S.-H. Shao, ``{Non-invertible Gauss Law and Axions},''
  \href{http://arxiv.org/abs/2212.04499}{{\ttfamily arXiv:2212.04499
  [hep-th]}}.

\bibitem{Yokokura:2022alv}
R.~Yokokura, ``{Non-invertible symmetries in axion electrodynamics},''
  \href{http://arxiv.org/abs/2212.05001}{{\ttfamily arXiv:2212.05001
  [hep-th]}}.

\bibitem{Bhardwaj:2022kot}
L.~Bhardwaj, S.~Schafer-Nameki, and A.~Tiwari, ``{Unifying Constructions of
  Non-Invertible Symmetries},''
  \href{http://arxiv.org/abs/2212.06159}{{\ttfamily arXiv:2212.06159
  [hep-th]}}.

\bibitem{Bhardwaj:2022maz}
L.~Bhardwaj, L.~E. Bottini, S.~Schafer-Nameki, and A.~Tiwari, ``{Non-Invertible
  Symmetry Webs},'' \href{http://arxiv.org/abs/2212.06842}{{\ttfamily
  arXiv:2212.06842 [hep-th]}}.

\bibitem{Bartsch:2022ytj}
T.~Bartsch, M.~Bullimore, A.~E.~V. Ferrari, and J.~Pearson, ``{Non-invertible
  Symmetries and Higher Representation Theory II},''
  \href{http://arxiv.org/abs/2212.07393}{{\ttfamily arXiv:2212.07393
  [hep-th]}}.

\bibitem{Hsin:2022heo}
P.-S. Hsin, ``{Non-Invertible Defects in Nonlinear Sigma Models and Coupling to
  Topological Orders},'' \href{http://arxiv.org/abs/2212.08608}{{\ttfamily
  arXiv:2212.08608 [cond-mat.str-el]}}.

\bibitem{Heckman:2022xgu}
J.~J. Heckman, M.~Hubner, E.~Torres, X.~Yu, and H.~Y. Zhang, ``{Top Down
  Approach to Topological Duality Defects},''
  \href{http://arxiv.org/abs/2212.09743}{{\ttfamily arXiv:2212.09743
  [hep-th]}}.

\bibitem{Antinucci:2022cdi}
A.~Antinucci, C.~Copetti, G.~Galati, and G.~Rizi, ``{''Zoology'' of
  non-invertible duality defects: the view from class $\mathcal{S}$},''
  \href{http://arxiv.org/abs/2212.09549}{{\ttfamily arXiv:2212.09549
  [hep-th]}}.

\bibitem{Apte:2022xtu}
A.~Apte, C.~Cordova, and H.~T. Lam, ``{Obstructions to Gapped Phases from
  Non-Invertible Symmetries},''
  \href{http://arxiv.org/abs/2212.14605}{{\ttfamily arXiv:2212.14605
  [hep-th]}}.

\bibitem{Garcia-Valdecasas:2023mis}
E.~Garc\'\i{}a-Valdecasas, ``{Non-Invertible Symmetries in Supergravity},''
  \href{http://arxiv.org/abs/2301.00777}{{\ttfamily arXiv:2301.00777
  [hep-th]}}.

\bibitem{Delcamp:2023kew}
C.~Delcamp and A.~Tiwari, ``{Higher categorical symmetries and gauging in
  two-dimensional spin systems},''
  \href{http://arxiv.org/abs/2301.01259}{{\ttfamily arXiv:2301.01259
  [hep-th]}}.

\bibitem{Bhardwaj:2023zix}
L.~Bhardwaj, M.~Bullimore, A.~E.~V. Ferrari, and S.~Schafer-Nameki,
  ``{Generalized Symmetries and Anomalies of 3d N=4 SCFTs},''
  \href{http://arxiv.org/abs/2301.02249}{{\ttfamily arXiv:2301.02249
  [hep-th]}}.

\bibitem{Cordova:2019jnf}
C.~C\'ordova, D.~S. Freed, H.~T. Lam, and N.~Seiberg, ``{Anomalies in the Space
  of Coupling Constants and Their Dynamical Applications I},''
  \href{http://dx.doi.org/10.21468/SciPostPhys.8.1.001}{{\em SciPost Phys.}
  {\bfseries 8} no.~1, (2020) 001},
  \href{http://arxiv.org/abs/1905.09315}{{\ttfamily arXiv:1905.09315
  [hep-th]}}.

\bibitem{Cordova:2019bsd}
C.~C\'ordova and K.~Ohmori, ``{Anomaly Obstructions to Symmetry Preserving
  Gapped Phases},'' \href{http://arxiv.org/abs/1910.04962}{{\ttfamily
  arXiv:1910.04962 [hep-th]}}.

\bibitem{Freed:2014iua}
D.~S. Freed, ``{Anomalies and Invertible Field Theories},''
  \href{http://dx.doi.org/10.1090/pspum/088/01462}{{\em Proc. Symp. Pure Math.}
  {\bfseries 88} (2014) 25--46},
  \href{http://arxiv.org/abs/1404.7224}{{\ttfamily arXiv:1404.7224 [hep-th]}}.

\bibitem{Monnier:2019ytc}
S.~Monnier, ``{A Modern Point of View on Anomalies},''
  \href{http://dx.doi.org/10.1002/prop.201910012}{{\em Fortsch. Phys.}
  {\bfseries 67} no.~8-9, (2019) 1910012},
  \href{http://arxiv.org/abs/1903.02828}{{\ttfamily arXiv:1903.02828
  [hep-th]}}.

\bibitem{Callan:1984sa}
C.~G. Callan, Jr. and J.~A. Harvey, ``{Anomalies and Fermion Zero Modes on
  Strings and Domain Walls},''
  \href{http://dx.doi.org/10.1016/0550-3213(85)90489-4}{{\em Nucl. Phys. B}
  {\bfseries 250} (1985) 427--436}.

\bibitem{Freed:2012bs}
D.~S. Freed and C.~Teleman, ``{Relative quantum field theory},''
  \href{http://dx.doi.org/10.1007/s00220-013-1880-1}{{\em Commun. Math. Phys.}
  {\bfseries 326} (2014) 459--476},
  \href{http://arxiv.org/abs/1212.1692}{{\ttfamily arXiv:1212.1692 [hep-th]}}.

\bibitem{2015arXiv150201690K}
L.~{Kong}, X.-G. {Wen}, and H.~{Zheng}, ``{Boundary-bulk relation for
  topological orders as the functor mapping higher categories to their
  centers},'' {\em arXiv e-prints} (Feb., 2015) arXiv:1502.01690,
  \href{http://arxiv.org/abs/1502.01690}{{\ttfamily arXiv:1502.01690
  [cond-mat.str-el]}}.

\bibitem{Freed:2018cec}
D.~S. Freed and C.~Teleman, ``{Topological dualities in the Ising model},''
  \href{http://arxiv.org/abs/1806.00008}{{\ttfamily arXiv:1806.00008
  [math.AT]}}.

\bibitem{Gaiotto:2020iye}
D.~Gaiotto and J.~Kulp, ``{Orbifold groupoids},''
  \href{http://dx.doi.org/10.1007/JHEP02(2021)132}{{\em JHEP} {\bfseries 02}
  (2021) 132}, \href{http://arxiv.org/abs/2008.05960}{{\ttfamily
  arXiv:2008.05960 [hep-th]}}.

\bibitem{Apruzzi:2022dlm}
F.~Apruzzi, ``{Higher Form Symmetries TFT in 6d},''
  \href{http://arxiv.org/abs/2203.10063}{{\ttfamily arXiv:2203.10063
  [hep-th]}}.

\bibitem{Burbano:2021loy}
I.~M. Burbano, J.~Kulp, and J.~Neuser, ``{Duality Defects in $E_8$},''
  \href{http://arxiv.org/abs/2112.14323}{{\ttfamily arXiv:2112.14323
  [hep-th]}}.

\bibitem{Kitaev:2011dxc}
A.~Kitaev and L.~Kong, ``{Models for Gapped Boundaries and Domain Walls},''
  \href{http://dx.doi.org/10.1007/s00220-012-1500-5}{{\em Commun. Math. Phys.}
  {\bfseries 313} no.~2, (2012) 351--373},
  \href{http://arxiv.org/abs/1104.5047}{{\ttfamily arXiv:1104.5047
  [cond-mat.str-el]}}.

\bibitem{2017arXiv170401447K}
L.~{Kong} and H.~{Zheng}, ``{Drinfeld center of enriched monoidal
  categories},'' {\em arXiv e-prints} (Apr., 2017) arXiv:1704.01447,
  \href{http://arxiv.org/abs/1704.01447}{{\ttfamily arXiv:1704.01447
  [math.CT]}}.

\bibitem{2021arXiv210403121K}
L.~{Kong}, W.~{Yuan}, Z.-H. {Zhang}, and H.~{Zheng}, ``{Enriched monoidal
  categories I: centers},'' {\em arXiv e-prints} (Apr., 2021) arXiv:2104.03121,
  \href{http://arxiv.org/abs/2104.03121}{{\ttfamily arXiv:2104.03121
  [math.CT]}}.

\bibitem{Hsin:2018vcg}
P.-S. Hsin, H.~T. Lam, and N.~Seiberg, ``{Comments on One-Form Global
  Symmetries and Their Gauging in 3d and 4d},''
  \href{http://dx.doi.org/10.21468/SciPostPhys.6.3.039}{{\em SciPost Phys.}
  {\bfseries 6} no.~3, (2019) 039},
  \href{http://arxiv.org/abs/1812.04716}{{\ttfamily arXiv:1812.04716
  [hep-th]}}.

\bibitem{2012PhRvB..86k5109L}
M.~{Levin} and Z.-C. {Gu}, ``{Braiding statistics approach to
  symmetry-protected topological phases},''
  \href{http://dx.doi.org/10.1103/PhysRevB.86.115109}{{\em prb} {\bfseries 86}
  no.~11, (Sept., 2012) 115109},
  \href{http://arxiv.org/abs/1202.3120}{{\ttfamily arXiv:1202.3120
  [cond-mat.str-el]}}.

\bibitem{Freed:1998tg}
D.~Freed, J.~A. Harvey, R.~Minasian, and G.~W. Moore, ``{Gravitational anomaly
  cancellation for M theory five-branes},''
  \href{http://dx.doi.org/10.4310/ATMP.1998.v2.n3.a8}{{\em Adv. Theor. Math.
  Phys.} {\bfseries 2} (1998) 601--618},
  \href{http://arxiv.org/abs/hep-th/9803205}{{\ttfamily arXiv:hep-th/9803205}}.

\bibitem{Harvey:1998bx}
J.~A. Harvey, R.~Minasian, and G.~W. Moore, ``{NonAbelian tensor multiplet
  anomalies},'' \href{http://dx.doi.org/10.1088/1126-6708/1998/09/004}{{\em
  JHEP} {\bfseries 09} (1998) 004},
  \href{http://arxiv.org/abs/hep-th/9808060}{{\ttfamily arXiv:hep-th/9808060}}.

\bibitem{Bah:2019rgq}
I.~Bah, F.~Bonetti, R.~Minasian, and E.~Nardoni, ``{Anomalies of QFTs from
  M-theory and Holography},''
  \href{http://dx.doi.org/10.1007/JHEP01(2020)125}{{\em JHEP} {\bfseries 01}
  (2020) 125}, \href{http://arxiv.org/abs/1910.04166}{{\ttfamily
  arXiv:1910.04166 [hep-th]}}.

\bibitem{Bah:2020jas}
I.~Bah, F.~Bonetti, R.~Minasian, and P.~Weck, ``{Anomaly Inflow Methods for
  SCFT Constructions in Type IIB},''
  \href{http://dx.doi.org/10.1007/JHEP02(2021)116}{{\em JHEP} {\bfseries 02}
  (2021) 116}, \href{http://arxiv.org/abs/2002.10466}{{\ttfamily
  arXiv:2002.10466 [hep-th]}}.

\bibitem{Bah:2020uev}
I.~Bah, F.~Bonetti, and R.~Minasian, ``{Discrete and higher-form symmetries in
  SCFTs from wrapped M5-branes},''
  \href{http://dx.doi.org/10.1007/JHEP03(2021)196}{{\em JHEP} {\bfseries 03}
  (2021) 196}, \href{http://arxiv.org/abs/2007.15003}{{\ttfamily
  arXiv:2007.15003 [hep-th]}}.

\bibitem{He:2016xpi}
H.~He, Y.~Zheng, and C.~von Keyserlingk, ``{Field theories for gauged
  symmetry-protected topological phases: Non-Abelian anyons with Abelian gauge
  group $\mathbb Z_2^{\otimes 3}$},''
  \href{http://dx.doi.org/10.1103/PhysRevB.95.035131}{{\em Phys. Rev. B}
  {\bfseries 95} no.~3, (2017) 035131},
  \href{http://arxiv.org/abs/1608.05393}{{\ttfamily arXiv:1608.05393
  [cond-mat.str-el]}}.

\bibitem{deWildPropitius:1995cf}
M.~D.~F. de~Wild~Propitius, {\em {Topological interactions in broken gauge
  theories}}.
\newblock PhD thesis, Amsterdam U., 1995.
\newblock \href{http://arxiv.org/abs/hep-th/9511195}{{\ttfamily
  arXiv:hep-th/9511195}}.

\bibitem{Sulejmanpasic:2018upi}
T.~Sulejmanpasic and Y.~Tanizaki, ``{C-P-T anomaly matching in bosonic quantum
  field theory and spin chains},''
  \href{http://dx.doi.org/10.1103/PhysRevB.97.144201}{{\em Phys. Rev. B}
  {\bfseries 97} no.~14, (2018) 144201},
  \href{http://arxiv.org/abs/1802.02153}{{\ttfamily arXiv:1802.02153
  [hep-th]}}.

\bibitem{Wan:2018zql}
Z.~Wan, J.~Wang, and Y.~Zheng, ``{New higher anomalies, SU(N)
  Yang\textendash{}Mills gauge theory and $\mathbb{CP}^{\mathrm{N}-1}$ sigma
  model},'' \href{http://dx.doi.org/10.1016/j.aop.2020.168074}{{\em Annals
  Phys.} {\bfseries 414} (2020) 168074},
  \href{http://arxiv.org/abs/1812.11968}{{\ttfamily arXiv:1812.11968
  [hep-th]}}.

\bibitem{Komargodski:2017dmc}
Z.~Komargodski, A.~Sharon, R.~Thorngren, and X.~Zhou, ``{Comments on Abelian
  Higgs Models and Persistent Order},''
  \href{http://dx.doi.org/10.21468/SciPostPhys.6.1.003}{{\em SciPost Phys.}
  {\bfseries 6} no.~1, (2019) 003},
  \href{http://arxiv.org/abs/1705.04786}{{\ttfamily arXiv:1705.04786
  [hep-th]}}.

\bibitem{Metlitski:2017fmd}
M.~A. Metlitski and R.~Thorngren, ``{Intrinsic and emergent anomalies at
  deconfined critical points},''
  \href{http://dx.doi.org/10.1103/PhysRevB.98.085140}{{\em Phys. Rev. B}
  {\bfseries 98} no.~8, (2018) 085140},
  \href{http://arxiv.org/abs/1707.07686}{{\ttfamily arXiv:1707.07686
  [cond-mat.str-el]}}.

\bibitem{Cheng:2022sgb}
M.~Cheng and N.~Seiberg, ``{Lieb-Schultz-Mattis, Luttinger, and 't Hooft --
  anomaly matching in lattice systems},''
  \href{http://arxiv.org/abs/2211.12543}{{\ttfamily arXiv:2211.12543
  [cond-mat.str-el]}}.

\bibitem{Cordova:2018acb}
C.~C\'ordova and T.~T. Dumitrescu, ``{Candidate Phases for SU(2) Adjoint
  QCD$_4$ with Two Flavors from $\mathcal{N}=2$ Supersymmetric Yang-Mills
  Theory},'' \href{http://arxiv.org/abs/1806.09592}{{\ttfamily arXiv:1806.09592
  [hep-th]}}.

\bibitem{Gaiotto:2017yup}
D.~Gaiotto, A.~Kapustin, Z.~Komargodski, and N.~Seiberg, ``{Theta, Time
  Reversal, and Temperature},''
  \href{http://dx.doi.org/10.1007/JHEP05(2017)091}{{\em JHEP} {\bfseries 05}
  (2017) 091}, \href{http://arxiv.org/abs/1703.00501}{{\ttfamily
  arXiv:1703.00501 [hep-th]}}.

\bibitem{Witten:1995gf}
E.~Witten, ``{On S duality in Abelian gauge theory},''
  \href{http://dx.doi.org/10.1007/BF01671570}{{\em Selecta Math.} {\bfseries 1}
  (1995) 383}, \href{http://arxiv.org/abs/hep-th/9505186}{{\ttfamily
  arXiv:hep-th/9505186}}.

\bibitem{Kravec:2014aza}
S.~M. Kravec, J.~McGreevy, and B.~Swingle, ``{All-fermion electrodynamics and
  fermion number anomaly inflow},''
  \href{http://dx.doi.org/10.1103/PhysRevD.92.085024}{{\em Phys. Rev. D}
  {\bfseries 92} no.~8, (2015) 085024},
  \href{http://arxiv.org/abs/1409.8339}{{\ttfamily arXiv:1409.8339 [hep-th]}}.

\bibitem{Seiberg:2018ntt}
N.~Seiberg, Y.~Tachikawa, and K.~Yonekura, ``{Anomalies of Duality Groups and
  Extended Conformal Manifolds},''
  \href{http://dx.doi.org/10.1093/ptep/pty069}{{\em PTEP} {\bfseries 2018}
  no.~7, (2018) 073B04}, \href{http://arxiv.org/abs/1803.07366}{{\ttfamily
  arXiv:1803.07366 [hep-th]}}.

\bibitem{Hsieh:2019iba}
C.-T. Hsieh, Y.~Tachikawa, and K.~Yonekura, ``{Anomaly of the Electromagnetic
  Duality of Maxwell Theory},''
  \href{http://dx.doi.org/10.1103/PhysRevLett.123.161601}{{\em Phys. Rev.
  Lett.} {\bfseries 123} no.~16, (2019) 161601},
  \href{http://arxiv.org/abs/1905.08943}{{\ttfamily arXiv:1905.08943
  [hep-th]}}.

\bibitem{Hsieh:2018ifc}
C.-T. Hsieh, ``{Discrete gauge anomalies revisited},''
  \href{http://arxiv.org/abs/1808.02881}{{\ttfamily arXiv:1808.02881
  [hep-th]}}.

\bibitem{2014PhRvL.113h0403W}
C.~{Wang} and M.~{Levin}, ``{Braiding Statistics of Loop Excitations in Three
  Dimensions},'' \href{http://dx.doi.org/10.1103/PhysRevLett.113.080403}{{\em
  Phys. Rev. Lett.} {\bfseries 113} no.~8, (Aug., 2014) 080403},
  \href{http://arxiv.org/abs/1403.7437}{{\ttfamily arXiv:1403.7437
  [cond-mat.str-el]}}.

\bibitem{Wang:2014oya}
J.~Wang and X.-G. Wen, ``{Non-Abelian string and particle braiding in
  topological order: Modular SL(3,$\mathbb{Z}$) representation and (3+1)
  -dimensional twisted gauge theory},''
  \href{http://dx.doi.org/10.1103/PhysRevB.91.035134}{{\em Phys. Rev. B}
  {\bfseries 91} no.~3, (2015) 035134},
  \href{http://arxiv.org/abs/1404.7854}{{\ttfamily arXiv:1404.7854
  [cond-mat.str-el]}}.

\end{thebibliography}\endgroup

\end{document}